\PassOptionsToPackage{twoside=false}{geometry}

\documentclass[manuscript]{acmart}
\AtBeginDocument{%
  }

\usepackage{subcaption}
\usepackage{xspace}
\usepackage{graphicx}
\usepackage{amsmath}
\usepackage{comment}
\usepackage{booktabs}
\usepackage{multirow}
\usepackage{adjustbox}
\usepackage{pifont}

\makeatletter
\g@addto@macro\caption@subtypehook{%
  \let\ltx@label\subcaption@label
}
\makeatother

\input{dfn.tex}
\copyrightyear{YYYY}
\acmYear{YYYY}
\acmDOI{XX.XXXX/XXXXXXX.XXXXXXX}

\acmJournal{CSUR}
\acmVolume{00}
\acmNumber{0}
\acmArticle{000}
\acmMonth{0}

\begin{document}

\title[A Survey on Centrality and Importance Measures in Hypergraphs]{A Survey on Centrality and Importance Measures in Hypergraphs: Categorization and Empirical Insights}

\author{Jaewan Chun}
\authornote{Both authors contributed equally to this survey.}
\email{jjwpalace@kaist.ac.kr}
\affiliation{%
  \institution{Kim Jaechul Graduate School of AI, KAIST}
  \city{Seoul}
  \country{South Korea}
}

\author{Fanchen Bu}
\authornotemark[1]
\affiliation{%
  \institution{School of Electrical Engineering, KAIST}
  \city{Daejeon}
  \country{South Korea}
}

\author{Yeongho Kim}
\email{yeongho@kaist.ac.kr}
\affiliation{%
  \institution{Kim Jaechul Graduate School of AI, KAIST}
  \city{Seoul}
  \country{South Korea}
}

\author{Atsushi Miyauchi}
\email{atsushi.miyauchi@intesasanpaolo.com}
\affiliation{%
  \institution{Intesa Sanpaolo}
  \city{Turin}
  \country{Italy}
}

\author{Francesco Bonchi}
\email{francesco.bonchi@intesasanpaolo.com}
\affiliation{%
  \institution{Intesa Sanpaolo AI Research}
  \city{Turin}
  \country{Italy}
}
\author{Kijung Shin}
\email{kijungs@kaist.ac.kr}
\affiliation{%
  \institution{Kim Jaechul Graduate School of AI, KAIST}
  \city{Seoul}
  \country{South Korea}
}

\renewcommand{\shortauthors}{Chun \& Bu et al.}

\begin{abstract}

Identifying central entities and interactions is a fundamental problem in network science. While well-studied for graphs (pairwise relations), many biological and social systems exhibit higher-order interactions best modeled by hypergraphs. This has led to a proliferation of specialized hypergraph centrality measures, but the field remains fragmented and lacks a unifying framework.

This paper addresses this gap by providing the first systematic survey of 39 distinct measures. We introduce a novel taxonomy classifying them as (1) structural (topology-based), (2) functional (impact on system dynamics), or (3) contextual (incorporating external features). We also present an experimental assessment comparing their empirical similarity and computation time. Finally, we 
discuss applications, establishing a coherent roadmap for future research in this area.

\end{abstract}



\received{DD Month YYYY}
\received[revised]{DD Month YYYY}
\received[accepted]{DD Month YYYY}

\maketitle

\section{Introduction}\label{sec:intro}

Identifying key entities or interactions is a foundational task in network science~\citep{lu2016vital}.
Historically, this problem has been studied through the lens of \emph{centrality}~\citep{das2018study}, a concept rooted in network science to quantify the \emph{structural prominence} of a node or edge based on its position in a graph. Classical measures --- such as degree~\citep{shaw1954group}, closeness~\citep{bavelas1948mathematical}, and betweenness~\citep{freeman1977set} centrality --- are cornerstones of this tradition. Over time, researchers have also sought to move beyond purely structural prominence, developing measures that capture the \textit{importance}~\citep{liu2016evaluating} of each entity in shaping system-level behaviors such as network robustness~\citep{freitas2022graph}, information diffusion~\citep{myers2012information}, or collective dynamics~\citep{watts1998collective}.

\vspace{-2mm}
\begin{discussion}[Centrality vs. Importance]
Originated in network science, the notion of \textit{centrality} traditionally
quantified the structural prominence of nodes or edges within a graph. In parallel,
the notion of \textit{importance} has often been used more broadly to capture functional criticality. While there is a part of the literature that uses the two terms interchangeably as if they were synonyms, recognizing the evolution over time of the term ``centrality'' beyond its original structural connotation,  others consider centrality as a structure-only type of importance, thus with centrality being a narrower and specialized notion of importance. Acknowledging the existence of this ambiguity, in this survey, we do not take a stance in the debate or aim to fix the ambiguity; instead we fully embrace it: We adopt an inclusive perspective and comprehensively review all measures that assign scores of centrality or importance in hypergraphs, without attempting to draw the borders dividing the two notions.
\end{discussion}
\vspace{-2mm}

While these concepts were originally formulated for \emph{pairwise graphs}, where interactions occur between exactly two entities, many real-world systems---from scientific collaborations and communication platforms to biochemical and ecological processes---exhibit \emph{higher-order interactions} that involve groups of entities simultaneously~\citep{benson2016higher,bianconi2021higher,bick2023higher}.
Overlooking these higher-order interactions can lead to incomplete or misleading conclusions because doing so loses the shared context and simultaneous nature of group activities.
For example, modeling a group email as separate messages misses the shared context of a single communication; reducing a multi-author paper to a set of two-person partnerships hides the team's synergy; and analyzing a complex biochemical process in pairs overlooks the key catalytic event that involves several molecules at the same time.

The shift from dyadic to higher-order structures requires extending classical notions of centrality and importance to \emph{hypergraphs}, which naturally represent multiway relations by allowing hyperedges to connect an arbitrary number of nodes~\citep{battiston2021physics}.
This transition raises unique challenges, e.g., multiple non-equivalent ways to define paths and distances~\citep{vasilyeva2023distances}, and ambiguity in how influence or connectivity propagates across multiway links~\citep{chitra2019random}.
On one hand, these complexities hinder straightforward generalizations of graph-based measures.
On the other hand, they also open opportunities for designing novel frameworks tailored to higher-order systems.

\begin{figure}[t]
    \begin{subfigure}{0.95\linewidth}
        \centering
        \includegraphics[width=\linewidth]{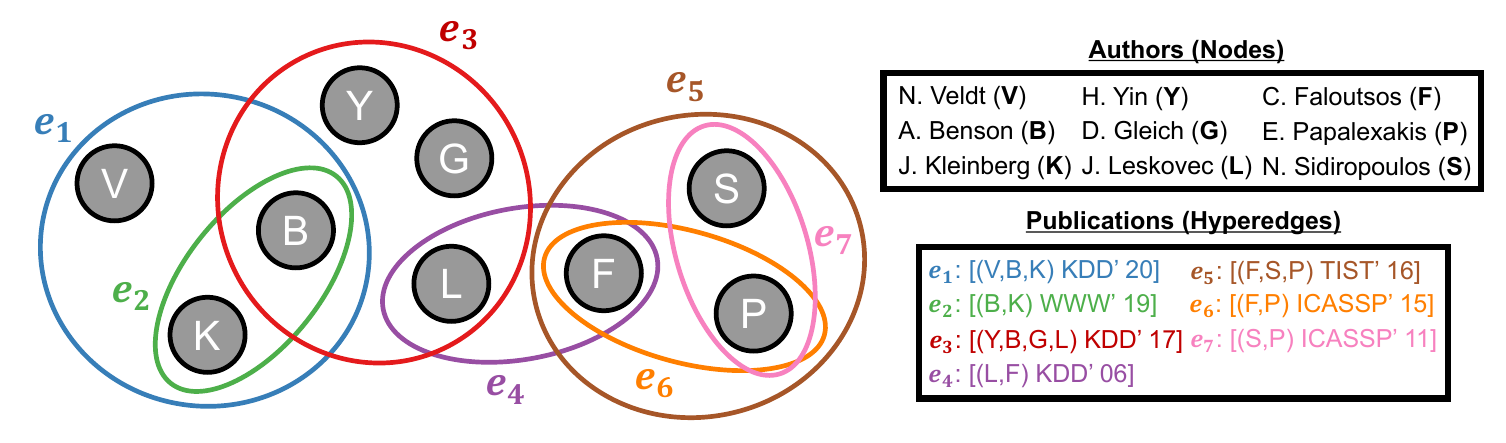}
        \caption{Example hypergraph constructed from a co-authorship network. Nodes represent authors and hyperedges represent publications containing the corresponding sets of authors. \label{fig:example:hypergraph}}
    \end{subfigure}

    \vspace{0.5em}

    \begin{subfigure}{0.95\linewidth}
        \centering
        \begin{tabular}{c|ccccccccc}
        \hline
        Measure $\backslash$ Node 
            & \textbf{V} & \textbf{B} & \textbf{K} & \textbf{Y} & \textbf{G} 
            & \textbf{L} & \textbf{F} & \textbf{P} & \textbf{S} \bigstrut \\
        \hline
        Degree (\cref{cent:degree}) 
            & 0.056 & \textbf{0.167} & 0.111 & 0.056 & 0.056 
            & 0.111 & \textbf{0.167} & \textbf{0.167} & 0.111 \bigstrut[t]\\
        Neighbor-degree (\cref{cent:neighbor_deg}) 
            & 0.077 & \textbf{0.192} & 0.077 & 0.115 & 0.115 
            & 0.154 & 0.115 & 0.077 & 0.077 \\
        Closeness (\cref{cent:closeness}) 
            & 0.088 & 0.131 & 0.090 & 0.112 & 0.112 
            & \textbf{0.149} & 0.131 & 0.095 & 0.093 \\
        Betweenness (\cref{cent:betweenness}) 
            & 0.000 & 0.269 & 0.006 & 0.000 & 0.000 
            & \textbf{0.335} & 0.304 & 0.060 & 0.025 \\
        Eigenvector (\cref{cent:eigenvector}) 
            & 0.029 & 0.100 & 0.053 & 0.046 & 0.046 
            & 0.098 & 0.226 & \textbf{0.238} & 0.164 \\
        Hypercoreness (\cref{cent:core}) 
            & 0.083 & 0.083 & 0.083 & 0.083 & 0.083 
            & 0.083 & \textbf{0.167} & \textbf{0.167} & \textbf{0.167} \bigstrut[b]\\
        \hline
        \end{tabular}
        \caption{Node centrality values in the example hypergraph, with the highest-scoring node(s) under each measure shown in \textbf{bold}. \label{fig:example:measure}}
    \end{subfigure}
    \vspace{-3mm}
    \caption{(a) Example co-authorship hypergraph and (b) corresponding node centrality and importance values. 
    Different measures emphasize different structural roles, resulting in distinct rankings. 
    For example, betweenness centrality (\cref{cent:betweenness}) highlights author L due to its bridging position across multiple publication groups, 
    whereas hypercoreness (\cref{cent:core}) assigns the highest scores to F, P, and S, reflecting the tightly connected triadic region they form.}
    \Description{A coauthorship hypergraph whose nodes are authors and hyperedges are papers, paired with a table showing how six centrality measures rank nine authors differently.}
    \label{fig:example}
    
\end{figure}

\cref{fig:example} offers a concrete example of a real-world hypergraph, illustrating how higher-order modeling naturally arises in practical systems, e.g., co-authorship networks here. In this representation, each hyperedge corresponds to a publication and can include any number of nodes (i.e., authors), demonstrating the size flexibility that distinguishes hypergraphs from pairwise graphs where interactions are strictly dyadic.
This flexibility enables the model to preserve group-level structure that would otherwise be lost under pairwise reductions.
\cref{fig:example:measure} also previews a key theme of this survey: different centrality and importance measures often emphasize distinct characteristics and perspectives. As shown in the centrality table (\cref{fig:example}b), different measures, such as degree, closeness, and eigenvector centralities, may highlight different nodes as most central. This diversity motivates our systematic examination of the many measures developed for hypergraphs and the distinct analytical perspectives they encode.

As foreshadowed in \cref{fig:example}, researchers have proposed a wide variety of centrality and importance measures tailored specifically for hypergraphs. These range from extensions of classical measures---such as degree, closeness, and betweenness centralities---to fundamentally new frameworks leveraging spectral methods~\citep{shen2011spectral}, perturbation analyses~\citep{jeong2001lethality}, and cooperative game theory~\citep{slikker2012social}.
Despite rapid progress, the literature remains fragmented, lacking a unified framework that systematically categorizes measures based on underlying principles.
Such unification would reveal common high-level goals and assumptions of the measures, facilitate meaningful empirical comparisons, and guide future researchers in selecting or developing measures based on such goals and assumptions.

This survey aims to fill that gap by providing the first dedicated and systematic overview of centrality and importance measures in hypergraphs.
We introduce a taxonomy that distinguishes three broad categories---\emph{structural}, \emph{functional}, and \emph{contextual} measures---each reflecting a different perspective on what makes an entity ``central'' or ``important.''
By analyzing underlying intuitions and mathematical foundations, we aim to establish a coherent framework for understanding existing measures and to highlight opportunities for future research.
Additionally, we conduct empirical analyses across diverse real-world datasets to provide practical insights.

\subsection{Scope and Contributions}

Our contributions can be summarized as follows:
\begin{itemize}
    \item (\cref{sec:measures:structural,sec:measures:functional,sec:measures:contextual}) We propose a systematic taxonomy of hypergraph centrality and importance measures that organizes 39 measures into three principled categories:
    (1) \textbf{structural measures} that capture centrality based purely on topology;
    (2) \textbf{functional measures} that evaluate importance via system-level behavior such as perturbation or coalition value; and
    (3) \textbf{contextual measures} that incorporate external features, labels, or learned representations.
    \item (\cref{sec:insight}) 
    We conduct a comprehensive empirical study across diverse real-world hypergraphs to compare representative measures. Our analyses examine (1) their \textbf{empirical similarity}, identifying when different methodological choices lead to similar or divergent rankings, and (2) their \textbf{computational behavior}, including run-time complexity and the practicality of heavier path-based or spectral variants. These results provide guidance on selecting measures under accuracy-efficiency tradeoffs.    
    \item (\cref{sec:applications})
    We examine \textbf{key application domains} where hypergraph centrality and importance play a critical role, including computational biology, social and communication systems, and infrastructure networks. Across these domains, we highlight how different measures capture domain-specific higher-order structures, linking methodological choices to practical modeling needs.
\end{itemize}
Together, these contributions provide the first comprehensive map of the field, enabling researchers to position existing measures within a unified framework and to design new measures aligned with their analytical goals.

\subsection{Related Surveys}

The literature on identifying important entities in pairwise graphs is extensive, with numerous surveys covering two related themes. On one hand, many reviews have cataloged the vast family of centrality measures, from classical formulations to modern machine learning-based approaches~\citep{das2018study, saxena2020centrality, wan2021survey, grando2018machine, bloch2023centrality}. On the other hand, the closely related problem of identifying key, critical, or influential nodes has also been thoroughly surveyed, often with a focus on specific domains such as information diffusion in social networks~\citep{bian2019identifying, hafiene2020influential, sarkar2016survey, ait2023influential}, network robustness~\citep{chen2025critical, zhao2023survey}, or biological systems~\citep{wang2022mini}.

Simultaneously, while the study of higher-order models has spurred a significant number of surveys on hypergraphs, these works have predominantly focused on topics other than centrality. A large body of literature is dedicated to hypergraph machine learning, with dedicated reviews on representation learning~\citep{gao2020hypergraph,antelmi2023survey}, hypergraph neural networks~\citep{kim2024survey, yang2025recent, wang2022survey}, and applications in recommender systems~\citep{liu2022survey}. Other topics that have received comprehensive treatment include hypergraph patterns and generators~\citep{lee2025survey}, partitioning algorithms~\citep{ccatalyurek2023more}, visualization techniques~\citep{fischer2021towards}, and hyperedge prediction~\citep{chen2023survey}.

Despite this wealth of research in parallel fields, a systematic and dedicated review of centrality and importance measures in hypergraphs has been notably absent. The most relevant works are broad surveys on critical node identification in graphs that only briefly touch upon higher-order networks (e.g., \citep{chen2025critical}). This survey aims to fill this critical gap by providing the first unified and systematic roadmap of this emerging field.

\subsection{Paper Outline}

The remainder of this paper is structured as follows. 
\Cref{sec:prelim} introduces preliminaries on graphs and hypergraphs, as well as classical centrality measures on pairwise graphs. 
Sections \ref{sec:measures:structural}, \ref{sec:measures:functional}, and \ref{sec:measures:contextual} 
survey structural, functional, and contextual measures of centrality and importance in hypergraphs, respectively.
\Cref{sec:insight} reports empirical comparisons and discusses insights on representative centrality and importance measures. \Cref{sec:applications} surveys real-world applications of hypergraph centrality and importance measures across diverse domains, including social, biological, communication, and infrastructure systems. \Cref{sec:conclusion} contains our concluding remarks outlining the road ahead and future research opportunities.   \cref{fig:taxonomy_elements} presents the taxonomy of measures covered in this survey, which is also a navigable table of contents for Sections \ref{sec:measures:structural}, \ref{sec:measures:functional}, and \ref{sec:measures:contextual}.  
Complementing this taxonomy, \Cref{tab:summary} summarizes, for each measure, any extra input information it incorporates, its output types, and available time and space complexity bounds.

\definecolor{mygreen}{HTML}{D5E8D4}
\definecolor{myblue}{HTML}{DAE8FC}
\definecolor{mypink}{HTML}{F8CECC}
\definecolor{mygrey}{HTML}{F5F5F5}
\definecolor{myborder}{HTML}{6A879E}

\begin{figure}[H]
    \centering
    \tikzset{
        basic/.style  = {
            draw=myborder,
            align=center, 
            font=\sffamily, 
            rectangle,         
            inner sep=1.4pt,                   
            blur shadow={shadow opacity=20} 
        },
        root/.style = {
            basic, 
            rounded corners=2pt, 
            thin, 
            align=center, 
            fill=mygreen, 
            text width=7cm,
            rotate=90,                   
            anchor=center,               
        },
        cat/.style = {
            basic, thin, rounded corners=2pt, fill=myblue,
            align=center, 
            inner sep=5pt,
            anchor=center, 
            child anchor=north,
            parent anchor=south,
      },
        tnode/.style  = {basic, thin, rounded corners=2pt, align=center, fill=mypink, text width=5.2cm}, 
        onode/.style  = {basic, thin, rounded corners=2pt, align=left, fill=mygrey, 
        text width=6cm,        
        },          
    }
    \begin{adjustbox}{max width=\textwidth}
    \begin{forest}
        for tree={
            s sep=2.5pt,       
            grow'=0,
            anchor=west,
            child anchor=west,
            tier/.option=level, 
            forked edges,         
            fork sep=2mm,              
            edge={semithick, ->, draw=myborder},   
        }
        [Centrality and importance measures, root,
        calign=center, 
        parent anchor=south,  
            [Structural measures\\(\cref{sec:measures:structural}), cat,
            child anchor=west,
            parent anchor=east,
                [Degree-based measures\\ (\cref{sec:measures:structural:degree}), tnode
                     [\cref{cent:degree}: Degree centrality, onode]
                     [\cref{cent:neighbor_deg}: Neighbor-degree centrality, onode]
                     [\cref{cent:isolating}: Isolating centrality, onode]
                     [\cref{cent:line_exp_deg}: Line-expansion degree centrality, onode]                     
                ]
                [Path-based measures\\ (\cref{sec:measures:structural:path}), tnode
                     [\cref{cent:closeness}: Closeness centrality, onode]
                     [\cref{cent:betweenness}: Betweenness centrality, onode]
                     [\cref{cent:harmonic}: Harmonic centrality, onode]
                     [\cref{cent:neighborhood}: Neighborhood centrality, onode]
                     [\cref{cent:fuzzy_collective_influence}: Distance-based fuzzy centrality, onode]
                ]
                [Walk-based measures\\ (\cref{sec:measures:structural:walk}), tnode
                     [\cref{cent:eigenvector}: Eigenvector centrality, onode]
                     [\cref{cent:rw_edge_dep}: Random-walk centrality, onode]
                     [\cref{cent:pagerank}: Personalized PageRank, onode]
                     [\cref{cent:vector}: Vector centrality, onode]
                     [\cref{cent:threeeigenvector}: Z/H-eigenvector centrality, onode]
                     [\cref{cent:genralized_eigenvector}: Gen. Z/H-eigenvector centrality, onode]
                     [\cref{cent:node_edge_nonlinear}: Nonlinear eigenvector centrality, onode]
                     [\cref{cent:sub_hypergraph}: Sub-hypergraph centrality, onode]
                ]
                [Subhypergraph-based measures\\ (\cref{sec:measures:structural:subgraph}), tnode
                    [\cref{cent:core}: Hypercoreness, onode]
                    [\cref{cent:truss}: Hypertrussness, onode]
                    [\cref{cent:hitting_set}: Hitting-set scores, onode]
                    [\cref{cent:core_periphery}: Core-periphery scores, onode]
                    [\cref{cent:motif_counts}: Motif counts, onode]
                    [\cref{cent:clustering_coef}: Clustering coefficients, onode]
                ]
                [Hybrid measures\\ (\cref{sec:measures:structural:hybrid}), tnode
                    [\cref{cent:gravity}: Gravity centrality, onode]
                    [\cref{cent:coreness_plus_degree}: Complex centrality, onode]
                    [\cref{cent:entropy_based_multi_cent}: Multi-centrality, onode]
                    [\cref{cent:multi_criteria}: Multi-criteria centrality, onode]
                    [\cref{cent:sprie}: Improved PageRank, onode]
                ]
            ]
            [Functional measures\\(\cref{sec:measures:functional}), cat,
            child anchor=west,
            parent anchor=east,
                [Perturbation-based measures\\ (\cref{sec:measures:functional:perturbation}), tnode
                    [\cref{cent:higher_order_von_neumann}: Higher-order von Neumann entropy, onode]
                    [\cref{cent:total_loss}: Total loss, onode]
                    [\cref{cent:grounded_laplacian}: Grounded-Laplacian eigenvalue, onode]
                ]
                [Coalition-based measures\\ (\cref{sec:measures:functional:coalition}), tnode
                    [\cref{cent:clique_shapley}: Clique-induced Shapley value, onode]
                    [\cref{cent:line_exp_shapley}: Line-expansion Shapley value, onode]
                ]
            ]
            [Contextual measures\\(\cref{sec:measures:contextual}), cat,
            child anchor=west,
            parent anchor=east,
                [Feature-based measures\\ (\cref{sec:measures:functional:feature}), tnode
                    [\cref{cent:struc_and_embed}: Structure-and-embedding score, onode]
                    [\cref{cent:AHGA}: Representation-based influence score, onode]
                ]
                [Label-based measures\\ (\cref{sec:measures:functional:label}), tnode
                    [\cref{cent:WHATSNet}: Edge-dependent role score, onode]
                    [\cref{cent:matrix_centrality}: Matrix centrality, onode]
                ]
                [Hybrid measures\\ (\cref{sec:measures:contextual:hybrid}), tnode
                    [\cref{cent:hypergraph_conv_att}: Hypergraph attention score, onode]
                    [\cref{cent:ID_HAN}: Identity-aware score, onode]
                ]
            ]
        ]
    \end{forest}
    \end{adjustbox}
    \caption{\textbf{Taxonomy of hypergraph centrality and importance measures.} The framework distinguishes three major categories: structural measures, which rely on the combinatorial structure of the hypergraph; functional measures, which evaluate importance in terms of system behavior; and contextual measures, which integrate information beyond structure. Each category is further divided into sub-categories that capture specific methodological choices.}
    \Description{A three-level taxonomy organizing 39 hypergraph centrality and importance measures into structural, functional, and contextual categories and their subcategories.}
    \label{fig:taxonomy_elements}
\end{figure}
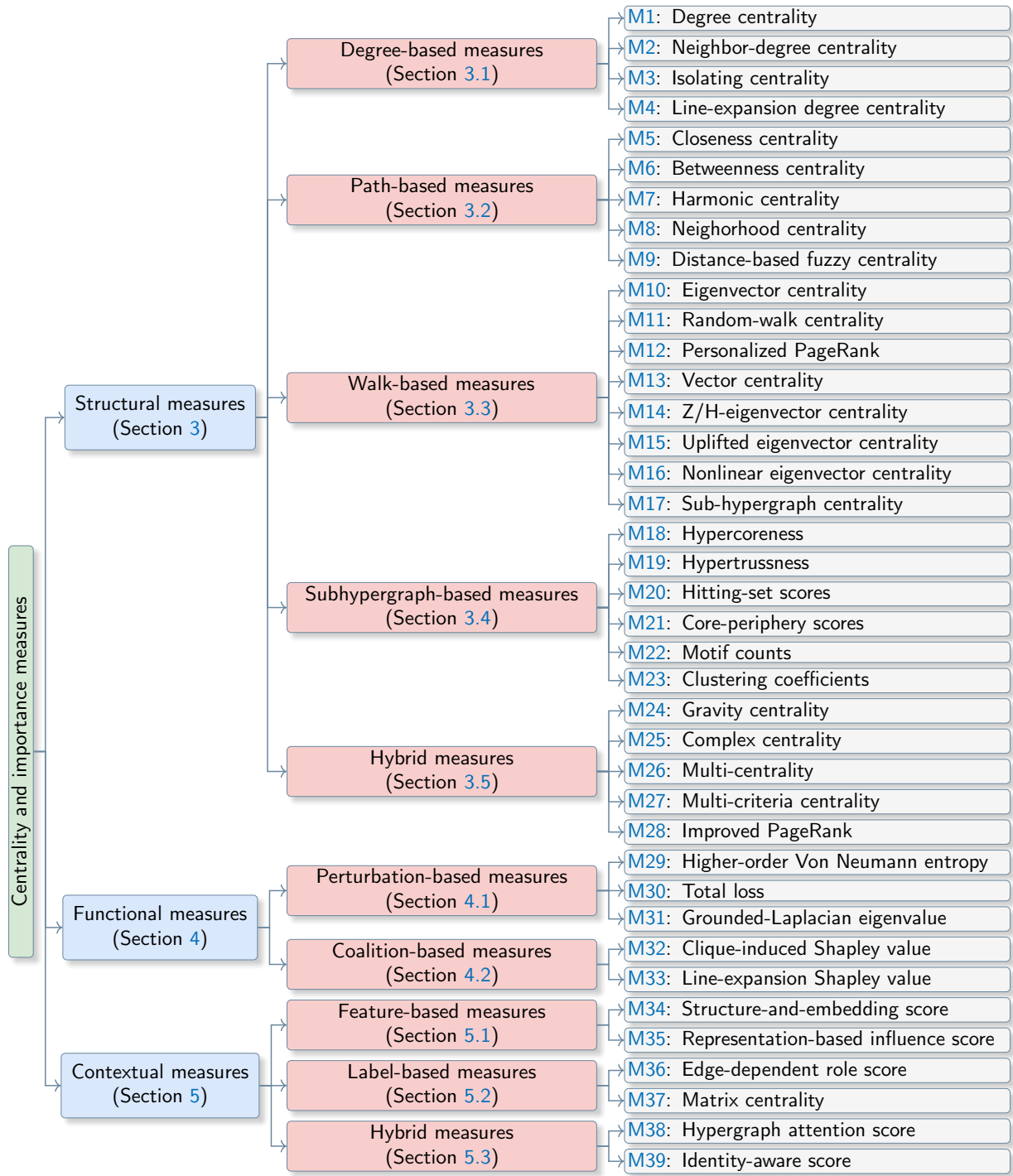

\begin{table*}[!ht]
\centering
\caption{
    Additional inputs, output types, and available time and space complexity bounds for the reviewed measures.
    Under ``Input'', a check mark indicates that at least one formulation of the measure explicitly incorporates the corresponding additional information.
    \textbf{Wts.} denotes weights;
    \textbf{Dir.}, directedness;
    \textbf{Feat.}, {features} or embeddings; and
    \textbf{Lab.}, {labels}.
    Under ``Output'', \textbf{N} and \textbf{E} denote node- and hyperedge-level scores, respectively, and \textbf{NS} denotes non-scalar outputs.
    Complexity bounds are included only when explicitly reported in a cited source or readily derived from the stated formulation; a dash (--) indicates otherwise.
    Space complexities include working memory and output storage but exclude storage of the input hypergraph. Complexity notations are defined below the table.}
\label{tab:summary}
{
\setlength{\tabcolsep}{4pt}
\renewcommand{\arraystretch}{0.7}
\begin{adjustbox}{max width=\textwidth,center}
\begin{tabular}{>{\collectcell\ScaleCell}l<{\endcollectcell}|cccc|ccc|>{\collectcell\ScaleCell}l<{\endcollectcell}>{\collectcell\ScaleCell}l<{\endcollectcell}}
\toprule
&
\multicolumn{4}{c}{\textbf{Input}}
&
\multicolumn{3}{c}{\textbf{Output}}
&
\multicolumn{2}{c}{\normalsize\textbf{Complexity}}
\\
\cmidrule(lr){2-5}
\cmidrule(lr){6-8}
\cmidrule(lr){9-10}
\multicolumn{1}{l|}{\normalsize\textbf{Measure}}
& \rotatebox[origin=c]{90}{\textbf{Wts.}}
& \rotatebox[origin=c]{90}{\textbf{Dir.}}
& \rotatebox[origin=c]{90}{\textbf{Feat.}}
& \rotatebox[origin=c]{90}{\textbf{Lab.}}
& \textbf{N}
& \textbf{E}
& \textbf{NS}
& \multicolumn{1}{l}{\normalsize\textbf{Time}}
& \multicolumn{1}{l}{\normalsize\textbf{Space}}
\\
\midrule
\cref{cent:degree}: Degree centrality
& \cmark & & &
& \cmark & \cmark &
& $O(p)$~\citep{choe2023classification}
& $O(n)$ (N); $O(m)$ (E) \\
\cref{cent:neighbor_deg}: Neighbor-degree centrality& \cmark & & &
& \cmark & &
& $O(q_n)$
& $O(n)$ \\
\cref{cent:isolating}: Isolating centrality& & & &
& \cmark & &
& $O(q_n + n \log n)$
& $O(n)$ \\
\cref{cent:line_exp_deg}: Line-expansion degree centrality& & & &
& & \cmark &
& $O(p + q_m)$
& $O(m)$ \\

\midrule

\cref{cent:closeness}: Closeness centrality
& & & &
& \cmark & \cmark &
& $O((n+m)p)$~\citep{faust1997centrality}
& $O(n+m)$~\citep{faust1997centrality} \\
\cref{cent:betweenness}: Betweenness centrality
& & & &
& \cmark & \cmark &
& $O((n+m)p)$~\citep{faust1997centrality,brandes2001faster}
& $O(n+m+p)$~\citep{faust1997centrality,brandes2001faster} \\
\cref{cent:harmonic}: Harmonic centrality
& & & &
& \cmark & \cmark &
& $O((n+m)p)$~\citep{esposito2022venture}
& $O(n+m)$~\citep{esposito2022venture} \\
\cref{cent:neighborhood}: Neighborhood centrality
& & & &
& \cmark & &
& --
& -- \\
\cref{cent:fuzzy_collective_influence}: Distance-based fuzzy centrality
& & & &
& \cmark & &
& --
& -- \\

\midrule

\cref{cent:eigenvector}: Eigenvector centrality
& \cmark & & &
& \cmark & \cmark &
& $O(Kp)$~\citep{choe2023classification}
& -- \\
\cref{cent:rw_edge_dep}: Random-walk centrality
& \cmark & & &
& \cmark & & 
& $O(m^3+m^2n)$~\citep{chitra2019random}
& -- \\
\cref{cent:pagerank}: Personalized PageRank
& \cmark & \cmark  & &
& \cmark & &
& $O(p/\Delta_{\mathrm{PR}})$~\citep{takai2020hypergraph}
& -- \\
\cref{cent:vector}: Vector centrality
& & & &
& \cmark & & \cmark
& --
& -- \\
\cref{cent:threeeigenvector}: Z/H-eigenvector centrality
& & & &
& \cmark & &
& --
& -- \\
\cref{cent:genralized_eigenvector}: Gen. Z/H-eigenvector centrality
& \cmark & & &
& \cmark & &
& $O\!\left(K\sum\limits_{e\in\mathcal E}r_{\max}|e|\binom{r_{\max}-1}{r_{\max}-|e|}\right)$~\citep{aksoy2024scalable}
& -- \\
\cref{cent:node_edge_nonlinear}: Nonlinear eigenvector centrality
& \cmark & & &
& \cmark & \cmark &
& --
& -- \\
\cref{cent:sub_hypergraph}: Sub-hypergraph centrality
& & & &
& \cmark & &
& --
& -- \\

\midrule

\cref{cent:core}: Hypercoreness
& & & &
& \cmark & \cmark &
& $O(p)$~\citep{choe2023classification}
& -- \\
\cref{cent:truss}: Hypertrussness
& & & &
& \cmark & \cmark &
& $O(T_{\triangle}+md_{\mathrm Q}\log d_{\mathrm Q})$~\citep{qin2025truss}
& -- \\
\cref{cent:hitting_set}: Hitting-set scores
& & & &
& \cmark & &
& $O(Kr_{\max}^2k^\star m)$~\citep{amburg2021planted}
& -- \\
\cref{cent:core_periphery}: Core-periphery scores
& \cmark & & &
& \cmark & &
& $O(Kp)$~\citep{tudisco2023core}
& -- \\
\cref{cent:motif_counts}: Motif counts
& & & &
& \cmark & \cmark & \cmark
& $O_{\mathrm{exp}}\!\left(
  \sum_{e\in\mathcal E}
  |e|\,|\mathcal N_{\mathcal E}(e)|^2
  \right)$~\citep{lee2024hypergraph}
& -- \\
\cref{cent:clustering_coef}: Clustering coefficients
& & & &
& \cmark & \cmark &
& --
& -- \\

\midrule

\cref{cent:gravity}: Gravity centrality
& & & &
& \cmark & &
& --
& -- \\
\cref{cent:coreness_plus_degree}: Complex centrality
& & & &
& \cmark & &
& $O(p)$~\citep{zhou2021identification,choe2023classification}
& -- \\
\cref{cent:entropy_based_multi_cent}: Multi-centrality
& & & &
& \cmark &  &
& --
& -- \\
\cref{cent:multi_criteria}: Multi-criteria centrality
& & & &
& \cmark & &
& --
& -- \\
\cref{cent:sprie}: Improved PageRank
& \cmark & & &
& \cmark & &
& --
& -- \\

\midrule

\cref{cent:higher_order_von_neumann}: Higher-order von Neumann entropy
& & & &
& \cmark & \cmark &
& $O(\max\{nm^2,m^3\})$~\citep{hu2023identifying}
& -- \\
\cref{cent:total_loss}: Total loss
& & & &
& \cmark & &
& --
& -- \\
\cref{cent:grounded_laplacian}: Grounded-Laplacian eigenvalue 
& & & &
& & \cmark &
& --
& -- \\

\midrule

\cref{cent:clique_shapley}: Clique-induced Shapley value
& & & &
& \cmark & &
& --
& -- \\
\cref{cent:line_exp_shapley}: Line-expansion Shapley value
& \cmark & & &
& \cmark & \cmark &
& $O(p+q_{m}+mq_{m}+m^2\log m)$~\citep{roy2015measuring}
& -- \\

\midrule

\cref{cent:struc_and_embed}: Structure-and-embedding score
& & & \cmark  &
& \cmark & &
& --
& -- \\
\cref{cent:AHGA}: Representation-based influence score
& & & \cmark  &
& \cmark & &
& --
& -- \\

\midrule

\cref{cent:WHATSNet}: Edge-dependent role score
& & & & \cmark 
& \cmark & &
& --
& -- \\

\cref{cent:matrix_centrality}: Matrix centrality
& & & & \cmark 
& \cmark & & \cmark
& --
& -- \\

\midrule

\cref{cent:hypergraph_conv_att}: Hypergraph attention score
& & \cmark & \cmark  & \cmark 
& \cmark & \cmark &
& --
& -- \\
\cref{cent:ID_HAN}: Identity-aware score
& & & \cmark  & \cmark 
& & \cmark &
& --
& -- \\
\bottomrule
\end{tabular}
\end{adjustbox}

\vspace{1mm}
\parbox{\textwidth}{\footnotesize
\textit{Complexity notations.}
For a hypergraph $H=(\mathcal V,\mathcal E)$, $n:=|\mathcal V|$, $m:=|\mathcal E|$,  $p:=\sum_{e\in\mathcal E}|e|$, $q_n:=\sum_{e\in\mathcal E}\binom{|e|}{2}$, $q_m:=\sum_{v\in\mathcal V}\binom{\deg(v)}{2}$, $d_{\mathrm Q}:=\max_{e\in\mathcal E} |\mathcal N_{\mathcal E}(e)|$, and $r_{\max}:=\max_{e\in\mathcal E}|e|$.
$K$ is the number of iterations or repeated runs. 
$\Delta_{\mathrm{PR}}$ denotes the time span of one Euler step in the personalized PageRank algorithm by \citep{takai2020hypergraph}.
$T_{\triangle}$ is the cost of the hyperedge-triangle-counting subroutine by \citet{qin2025truss}.
$k^\star$ is the minimum hitting-set size used by \citet{amburg2021planted}.
$O_{\mathrm{exp}}$ denotes expected time under the expected constant-time hash-table operations assumed by \citet{lee2024hypergraph}.
}

}
\end{table*}

\section{Preliminaries}\label{sec:prelim}

In this section, we present the preliminaries and notations used throughout the paper. 
We first define hypergraphs and their expansions to pairwise graphs, and then briefly review centrality measures in pairwise graphs as a foundation for their generalizations to hypergraphs.

\subsection{Hypergraphs}\label{sec:prelim:HGs}

First, we introduce the concept of hypergraphs, together with some related concepts.

\prelim{Hypergraphs.}\label{prelim:HGs}
A \textit{hypergraph} is defined as $H = (\mathcal{V}, \mathcal{E})$, where $\mathcal{V} = \{v_1, \ldots, v_n\}$ is the set of nodes and $\mathcal{E} = \{e_1, \ldots, e_m\}$ is the set of hyperedges, with $n = \lvert \mathcal{V} \rvert$ being the number of nodes and $m = \lvert \mathcal{E} \rvert$ the number of hyperedges.
Each hyperedge $e \in \mathcal{E}$ is a non-empty subset of $\mathcal{V}$, i.e., $e \subseteq \mathcal{V}$ and $e \neq \emptyset$.
Refer to Fig.~\ref{fig:example:hypergraph} for an example.

\prelim{Incident hyperedges and constituent nodes.}\label{prelim:incident_edge_n_const_nodes}
For a node $v$ and a hyperedge $e$, if $v \in e$, we call $e$ an \textit{incident hyperedge} of $v$, and $v$ a \textit{constituent node} of $e$.

\prelim{Node degrees.}\label{prelim:degree}
The \textit{degree} of a node $v$, $\deg(v) := \vert \{ e\in \mathcal{E} \mid v \in e\} \vert$, is the number of its incident hyperedges.

\prelim{Neighbors and neighbor-degrees.}\label{prelim:neighbor}
The set of \textit{neighbors} of a node $v$, $\mathcal{N}(v) := \{v \neq u \in \mathcal{V} \mid \exists e \in \mathcal{E}, \{u, v\} \subseteq e\}$, consists of nodes that co-occur in at least one hyperedge with $v$.
The number of neighbors, $|\mathcal{N}(v)|$, is called the \textit{neighbor-degree} of $v$.

\prelim{Hyperedge sizes.}\label{prelim:edge_size}
The \textit{size} of a hyperedge $e$, $s_e := |e|$, is the number of its constituent nodes, which is also sometimes referred to as the \textit{degree} of the hyperedge.
In this work, we consider hypergraphs with each hyperedge of size at least two, i.e., $|e| \geq 2, \forall e \in \mathcal{E}$.

\prelim{Subhypergraphs.}\label{prelim:subhg}
A hypergraph ${H}'=(\mathcal{V}', \mathcal{E}')$ is a \textit{subhypergraph} of another hypergraph ${H}=(\mathcal{V}, \mathcal{E})$ if the node and hyperedge sets of ${H}'$ are subsets of those of $H$, respectively, i.e., $\mathcal{V}'\subseteq \mathcal{V}$ and $\mathcal{E}' \subseteq \mathcal{E}$.

\prelim{Incidence matrix.}\label{prelim:incidence_mat}
The \textit{incidence matrix} of a hypergraph ${H}$, $\mat{H} = \mat{H}({H}) \in \mathbb{R}^{n \times m}$, summarizes the membership information.
For the commonly considered unweighted case, the entries are binary, where $\mat{H}_{ij} = 1$ if node $v_i$ is in hyperedge $e_j$, i.e., $v_i \in e_j$, and $\mat{H}_{ij} = 0$ otherwise.
This can be extended to weighted hypergraphs, where the entries of $\mat{H}$ can take non-binary values to represent the strengths or weights of membership.

\prelim{Uniform hypergraphs.}\label{prelim:uniform_HGs}
A hypergraph ${H} = (\mathcal{V}, \mathcal{E})$ is $r$-\textit{uniform} if all hyperedges are of size $r$, i.e., $\vert e \vert = r, \forall e \in \mathcal{E}$.

\prelim{Pairwise graphs.}\label{prelim:graphs}
A $2$-uniform hypergraph is specifically called a \textit{pairwise graph}, where each edge consists of exactly two nodes.
When the context is clear, we may simply use \textit{graphs} to refer to pairwise graphs.

\subsection{Connectivity-Related Concepts}\label{sec:prelim:connectivity}
We now introduce concepts related to connectivity, e.g., walks, paths, and distances.

\prelim{Adjacent hyperedges.}\label{prelim:adj_edges}
Given a pairwise graph $G = (\mathcal{V}, \mathcal{E})$,
two different edges $e \neq e' \in \mathcal{E}$ are \textit{adjacent} if $e \cap e' \neq \emptyset$.
One can straightforwardly extend this concept to a hypergraph ${H} = (\mathcal{V}, \mathcal{E})$, where two different hyperedges $e \neq e' \in \mathcal{E}$ are \textit{adjacent} if $e \cap e' \neq \emptyset$.
Recall that each edge in a pairwise graph contains exactly two nodes (see \cref{prelim:graphs}), so if two (non-identical) edges are adjacent, they must have exactly one node in common.
By contrast, in hypergraphs, a hyperedge can contain an arbitrary number of nodes, which implies that two hyperedges may intersect at multiple nodes, allowing greater flexibility in the definition of \textit{adjacent hyperedges}.
Indeed, researchers~\citep{battiston2020networks,tian2025representing,aksoy2020hypernetwork,lu2013high} have considered the definition of \textit{$s$-adjacent} hyperedges: Two hyperedges $e$ and $e'$ are $s$-adjacent for some $s \in \mathbb{N}$ if they have at least $s$ nodes in common, i.e., $\lvert e \cap e' \rvert \geq s$.
When $s = 1$, this definition reduces to the aforementioned straightforward generalization.
Researchers~\citep{larock2023encapsulation,lotito2024hyperlink,zhou2022topological} have also considered using a proportional threshold (e.g., two hyperedges $e$ and $e'$ are considered adjacent if $|e \cap e'|/\min(|e|, |e'|) \geq p$ for some proportional threshold $p \in [0, 1]$).
Rather than using the size of intersection solely as a threshold, some researchers~\citep{vasilyeva2023distances,carletti2020random,carletti2021random,nortier2025higher,lee2021hyperedges} have also incorporated it as additional information, e.g., as the weight of the adjacency
between hyperedges.

\vspace{-2mm}
\begin{discussion}[On adjacent hyperedges]
    Existing connectivity-related concepts and centrality measures use different definitions of \textit{adjacent hyperedges}.
    These concepts and measures usually remain well-defined and meaningful regardless of the specific adjacency definition.
    Therefore, in subsequent discussions of related concepts and centrality measures, we will use the term ``adjacent hyperedges'' without explicitly specifying the adjacency criterion, unless otherwise noted.
\end{discussion}
\vspace{-2mm}

\prelim{Walks.}\label{prelim:walk}
Given a hypergraph ${H} = (\mathcal{V}, \mathcal{E})$, a \emph{walk} of length $k$ is a sequence of nodes $(v_{0},v_{1},\ldots,v_{k})$ such that there exists a sequence of hyperedges $(e_{1}, e_{2}, \ldots, e_{k})$ satisfying:
(1) each pair of consecutive nodes $v_{i-1}$ and $v_{i}$ coexists in the hyperedge $e_i$, i.e., $\{v_{i-1}, v_i\} \subseteq e_i$ for each $i \in \{1,\ldots,k\}$, and
(2) every pair of consecutive hyperedges is adjacent, i.e., $e_{i - 1}$ and $e_{i}$ are adjacent, for each $i \in \{2,\ldots, k\}$.
In a walk, the same node can appear multiple times, i.e., it is allowed that $v_i = v_j$ for some $i \neq j$.

\prelim{Paths.}\label{prelim:path}
A \emph{path} of length $k$ is a walk $(v_{0},v_{1},\dots,v_{k})$ where all nodes are distinct, i.e., $v_i \neq v_j, \forall i \neq j$.

\prelim{Shortest paths and distances.}\label{prelim:distance}
For two nodes $u, v\in \mathcal{V}$, a \emph{shortest path} between them is a path $(v_{0},v_{1},\dots,v_{k})$ of minimum length $k$ among all paths connecting $u$ and $v$, i.e., where $v_0 = u$ and $v_k = v$.
The minimum length $k$ is called the \emph{distance} $\delta(u,v)$ between $u$ and $v$.
There are possibly multiple shortest paths of the same length between the same node pair.
If no path exists between $u$ and $v$, we set $\delta(u,v)=\infty$.

\subsection{Expansions from Hypergraphs to Pairwise Graphs}\label{sec:prelim:expansion}

Researchers have considered several \textit{expansions} of hypergraphs, i.e., projections of hypergraphs into pairwise graphs, to borrow results from pairwise-graph analysis for hypergraph analysis.

\prelim{Clique expansion.}\label{prelim:cliq_exps}
Given a hypergraph ${H} = (\mathcal{V}, \mathcal{E})$, its \textit{clique expansion} \cite{sun2008hypergraph} $G_C=(\mathcal{V}_C, \mathcal{E}_C)$ is generated by replacing each hyperedge by a clique over its constituent 
nodes, i.e., $\mathcal{V}_C=\mathcal{V}, \mathcal{E}_C=\{ \{u,v\} \mid u, v \in e,\, e \in \mathcal{E} \}$. 
This reduces ${H}$ to a weighted or unweighted pairwise graph, depending on whether edge multiplicities are considered.

\prelim{Star expansion.}\label{prelim:star_exps}
Given a hypergraph ${H} = (\mathcal{V}, \mathcal{E})$, its \textit{star expansion} \cite{zien2002multilevel} $G_S=(\mathcal{V}_S, \mathcal{E}_S)$ is generated by combining sets of nodes and hyperedges as a set of nodes (i.e., $\mathcal{V}_S=\mathcal{V} \cup \mathcal{E}$) and putting an edge between a hyperedge and each of its constituent nodes (i.e., $\mathcal{E}_S=\{ \{v,e\} \mid v \in e,\, v \in \mathcal{V},\, e \in \mathcal{E} \}$).

\prelim{Line expansion.}\label{prelim:line_exps}
Given a hypergraph ${H} = (\mathcal{V}, \mathcal{E})$, its \textit{line expansion} \cite{whitney1992congruent,bermond1977line} is a pairwise graph $G_L=(\mathcal{V}_L, \mathcal{E}_L)$ generated by treating original hyperedges as nodes, i.e., $\mathcal{V}_L = \mathcal{E}$ and putting an edge between two nodes (corresponding to two hyperedges) if and only if they are adjacent in the original hypergraph. 
As mentioned above in \cref{sec:prelim:connectivity}, there exist various definitions of ``adjacent hyperedges'', and they give different definitions of line expansion.

\subsection{Basic Centrality and Importance Measures in Pairwise Graphs}\label{sec:prelim:pairwise_cent}

Given a pairwise graph $G = (\mathcal{V}, \mathcal{E})$, a centrality or importance measure assigns scores to nodes or edges to quantify their importance in the network.
Following~\citet{boldi2014axioms}, we briefly review several classical graph centrality measures that serve as the conceptual basis for hypergraph generalizations.
See also, e.g., \cite{freeman1978centrality, koschutzki2005centrality, saxena2020centrality,bian2019identifying,sarkar2016survey}, for comprehensive surveys on centrality and importance measures in pairwise graphs.

\prelim{Degree centrality.}\label{prelim:degree_cent}
\textit{Degree centrality} \cite{shaw1954group} measures the immediate connectedness of a node: $C_{\text{deg}}(v) = \deg(v)$, where $\deg(v)$ is the number of edges incident to $v$ (see \cref{prelim:degree}). Sometimes, this measure is normalized by the number of nodes in the entire graph, i.e., $\tilde{C}_{\text{deg}}(v) = {\deg(v)} / {\lvert \mathcal{V} \rvert}$, which does not affect the relative node ranking in the same graph.

\prelim{Closeness centrality.}\label{prelim:closeness_cent}
\textit{Closeness centrality} \cite{bavelas1948mathematical} measures how close a node is to the other nodes, specifically, the inverse of the sum of distances (see \cref{prelim:distance}) to all other nodes: $C_{\text{cls}}(v) = \left (\sum_{u \in \mathcal{V} \setminus \{v\}} \delta(v,u) \right )^{-1}$.
A limitation of this measure is that the scores of all nodes become zero if the graph is disconnected.

\prelim{Harmonic centrality.}\label{prelim:harmonic_cent}
\textit{Harmonic centrality} \cite{boldi2014axioms} is similar to closeness centrality, but it measures the sum of the inverses instead of the inverse of the sum, so that it overcomes the above limitation:
$C_{\text{hmn}}(v) = \sum_{u \in \mathcal{V} \setminus \{v\}} {\delta(v,u)}^{-1}$.

\prelim{Betweenness centrality.}\label{prelim:betweenness_cent}
\textit{Betweenness centrality} \cite{freeman1977set} measures how often a node appears on the shortest paths (see \cref{prelim:distance}).
Specifically, it is the sum of the fractions of shortest paths that pass through the node, taken over all pairs of other nodes:
$C_{\text{btw}}(v) = \sum_{\substack{s, t \in \mathcal{V} \setminus\{v\}}} {\sigma_{st}(v)} / {\sigma_{st}}$,
where $\sigma_{st}$ is the number of all shortest paths between $s$ and $t$, and $\sigma_{st}(v)$ is the number of those passing through $v$. 

\prelim{Current-flow centralities.}\label{prelim:current_flow_cent}
\textit{Current-flow betweenness and closeness centralities}
\cite{brandes2005centrality} replace the shortest-path model of information
spread with an electrical-current model, in which flow may split across multiple
routes.
Viewing each edge as a unit resistor, current-flow betweenness aggregates the
amount $\tau_{st}(v)$ of unit current between each pair $(s,t)$ that passes
through $v$:
$C_{\text{cf-btw}}(v)=\sum_{s,t\in\mathcal{V}}\tau_{st}(v)$, up to normalization.
Current-flow closeness instead replaces shortest-path distance with effective
resistance:
$C_{\text{cf-cls}}(v)
=
\left(\sum_{u\in\mathcal{V}\setminus\{v\}}
R_{\mathrm{eff}}(v,u)\right)^{-1}$.

\prelim{Eigenvector centrality.}\label{prelim:eigenvec_cent}
\textit{Eigenvector centrality} \cite{seeley1949net} is a spectral measure of a node's influence within a network. It is calculated using the graph's adjacency matrix $\mat{A} \in \mathbb{R}^{n \times n}$, where the entry $\mat{A}_{ij}$ represents the weight of the edge between nodes $v_i$ and $v_j$ (for unweighted cases, $\mat{A}_{ij} = 1$ if an edge exists and $0$ otherwise).
The centrality score of each node $v_i$ is given by the $i$-th component of the \textit{principal eigenvector} $\vect{c}$ of $\mat{A}$, which is the eigenvector corresponding to the largest eigenvalue. This formulation defines a node's score as being proportional to the sum of the scores of its neighbors, i.e., a node is considered more important if it is connected to other important nodes.

\prelim{Katz centrality.}\label{prelim:katz_cent}
\textit{Katz centrality} \cite{katz1953new} extends the core idea of eigenvector centrality.
Instead of only considering a node's immediate neighbors, it defines a node's importance by counting the total number of walks of all lengths that terminate at that node, while attenuating longer walks with exponential decay:
$C_{\text{katz}}(v_i) = \sum_{k=1}^\infty \sum_{j = 1}^n \alpha^k (\mat{A}^k)_{ji}$, where $\alpha > 0$ is a damping factor.
The term $(\mat{A}^k)_{ji}$ is the number of walks of length $k$ from node $v_j$ to node $v_i$, and $\sum_{j = 1}^n \alpha^k (\mat{A}^k)_{ji}$ thus counts the total number of walks of length $k$ ending at node $v_i$.

\prelim{PageRank centrality.}\label{prelim:pagerank_cent}
\textit{PageRank} \cite{page1999pagerank} defines a node's importance as its stationary probability under a ``random surfer'' model.
This surfer either follows a random edge from its current node (with probability $\alpha$) or ``teleports'' to a new node in the graph (with probability $1-\alpha$).
The vector of PageRank scores, $\vect{c}$, is the stationary distribution of this process, satisfying the recursive equation:
$\vect{c} = \alpha \mat{P}^\top \vect{c} + (1 - \alpha) \vect{v}$,
where $\mat{P}$ is the row-stochastic transition matrix,\footnote{The entry $P_{ij}$ of the transition matrix $\mat{P}$ specifies the probability of the random surfer moving from node $i$ to node $j$. The matrix is row-stochastic (i.e., each of its rows sums to one) because the sum of probabilities for all possible destinations from each node $i$ must be $1$.}
and $\vect{v}$ is the teleportation distribution, which is typically uniform.
If $\vect{v}$ is biased toward a specific set of nodes, the measure is referred to as \textit{Personalized PageRank} \citep{jeh2003scaling}.

\begin{discussion}[On applying pairwise-graph measures to hypergraphs]
    Expansions (see~\cref{sec:prelim:expansion}) convert a hypergraph into a pairwise graph, allowing standard graph centrality and importance measures to be applied.
    However, the resulting scores characterize the chosen pairwise representation rather than the full hypergraph, because each expansion preserves some aspects of the original higher-order structure while discarding others.
    Clique expansion (see~\cref{prelim:cliq_exps}) replaces each hyperedge with pairwise connections among all of its nodes. It preserves which nodes co-occur but loses information about which pairwise connections belong to the same group interaction; consequently, distinct hypergraphs can produce the same projected graph~\citep{aksoy2020hypernetwork}. The resulting scores may also vary depending on how repeated co-occurrences are weighted.
    Star expansion (see~\cref{prelim:star_exps}) represents both the original nodes and hyperedges as nodes in a bipartite graph, thereby preserving every node--hyperedge incidence. However, standard pairwise-graph measures do not account for the different meanings of these two entity types. The scores assigned to original nodes and hyperedges may therefore have different interpretations and may not be directly comparable~\citep{faust1997centrality}.
    Line expansion (see~\cref{prelim:line_exps}) represents hyperedges as nodes and connects them according to their overlaps. Because the original nodes do not appear explicitly in this representation, directly applying a pairwise-graph measure yields scores only for hyperedges. Moreover, the line expansion may not retain the identities of the original nodes that produce each overlap, and the resulting scores depend on how hyperedge adjacency and edge weights are defined.
    Projection-based scores should therefore be interpreted as centrality or importance measures of the chosen pairwise representation rather than as projection-independent properties of the original hypergraph.
    Projection-based measures are appropriate when the relationships preserved by the expansion match the analytical goal; measures defined directly on the hypergraph may be preferable when hyperedge identity, interaction size, or node roles within individual hyperedges are important.
\end{discussion}

\section{Structural Measures}\label{sec:measures:structural}

\textit{Structural measures} form the most fundamental class of hypergraph centrality and importance measures, relying solely on the static topology of hypergraphs. They determine importance from \emph{primary structural elements} such as node degrees, hyperedge sizes, paths, walks, or subhypergraphs (see \cref{sec:prelim}), and capture how the combinatorial arrangement of higher-order connections shapes the prominence of individual nodes or hyperedges.

\subsection{Degree-based Measures}\label{sec:measures:structural:degree}

The most straightforward and intuitive structural elements that can be used for calculating measures are \textit{degrees}, including node degrees (see \cref{prelim:degree}) and neighbor-degrees (see \cref{prelim:neighbor}).\footnote{Note that node degrees and neighbor-degrees are equivalent for pairwise graphs (more precisely, they are equivalent for pairwise graphs that are undirected, unweighted, and without self-loops), but they are different in general for hypergraphs.}
Degree-based measures typically assume that nodes/hyperedges that are \textit{connected to more nodes/hyperedges} are more central and important, and thus are assigned higher centrality scores.

\centrality{Degree centrality.}\label{cent:degree}
\citet{faust1997centrality} extended degree centrality (see \cref{prelim:degree_cent}) to hypergraphs for both nodes and hyperedges.
The \textit{degree centrality} of each node $v$ is its degree $\deg(v)$ (see \cref{prelim:degree}), and that of each hyperedge $e$ is its size $|e|$ (see \cref{prelim:edge_size}).
This measure is equivalent to computing the degrees in the star expansion (see \cref{prelim:star_exps}).
Weights on nodes and/or hyperedges can be naturally included.
Specifically, the degree centrality of each node can be the sum of weights of its incident hyperedges (see \cref{prelim:incident_edge_n_const_nodes}), and that of each hyperedge can be the sum of weights of its constituent nodes (see \cref{prelim:incident_edge_n_const_nodes}).
For example, \citet{kapoor2013weighted} considered degree centrality for nodes while allowing different weights on hyperedges.

\centrality{Neighbor-degree centrality.}\label{cent:neighbor_deg}
Similarly, we can define the \textit{neighbor-degree centrality} for each node as its neighbor-degree $|\mathcal{N}(v)|$ (see \cref{prelim:neighbor}).
This measure is equivalent to the degrees in the unweighted clique expansion (see \cref{prelim:cliq_exps}).
We can incorporate weights on hyperedges into the definition by considering weighted clique expansion (see, e.g., \citep{dubey2025influential,kim2024survey}).

\centrality{Isolating centrality.}\label{cent:isolating}
\citet{tejaswi2024identifying} proposed \textit{isolating centrality} to measure the importance of nodes.
Let $\mathcal{V}_\text{min}$ denote a set of nodes with the lowest degree in the hypergraph (how many nodes to choose to put in $\mathcal{V}_\text{min}$ is a hyperparameter). The isolating centrality of each node $v$ is the product of (1) the number of its neighbors in $\mathcal{V}_\text{min}$ and (2) the total number of its neighbors.
The intuition is that a node is considered important if it is not only connected to many nodes, but also safeguards many low-degree nodes.
Accordingly, isolating centrality is well suited to analyses of network vulnerability and fragmentation, where the goal is to identify nodes connecting many low-degree peripheral nodes to the rest of the hypergraph, but less suited to notions of importance based on global accessibility, shortest-path brokerage, or spectral influence.
This measure can be seen as a variant of neighbor-degree centrality (see \cref{cent:neighbor_deg}) where we assign higher weights to the lowest-degree nodes.

\centrality{Line-expansion degree centrality.}\label{cent:line_exp_deg}
After obtaining the line expansion (see \cref{prelim:line_exps}), the degrees of nodes in the line expansion (which correspond to hyperedges in the original hypergraph) give a centrality measure for hyperedges.
Specifically, the \textit{line-expansion degree centrality} of each hyperedge $e$ is the number of other hyperedges $e'$ that are adjacent (see \cref{prelim:adj_edges}) to $e$.
As discussed in \cref{sec:prelim:connectivity}, we can have different variants of this measure by considering different definitions of adjacent hyperedges.

\subsection{Path-based Measures}\label{sec:measures:structural:path}
Another structural element commonly used to define centrality and importance measures is \textit{paths} (see \cref{prelim:path}).
{These measures typically evaluate} either the quality or the quantity of paths associated with an entity.
From a quality perspective, nodes/hyperedges are considered more central and important if they lie on \textit{shorter} paths to others, a concept typically captured using distances (see \cref{prelim:distance}).
From a quantity perspective, nodes/hyperedges are considered more central and important if they lie on \textit{a greater number} of paths.

\vspace{-2mm}
\begin{discussion}[Paths and distances in hypergraphs]
    The key challenge in defining path-based measures on hypergraphs lies in generalizing the notions of paths and distances (see \cref{prelim:path} and \cref{prelim:distance}).
    As discussed in \cref{sec:prelim:connectivity},
    the higher-order nature of hypergraphs allows various definitions of paths and their length.
    Typically, we consider distances between nodes, and there are two major categories of common approaches in existing literature.
    The first category uses clique expansion (see \cref{prelim:cliq_exps}) or star expansion (see \cref{prelim:star_exps}) to reduce hypergraphs into pairwise graphs (see \cref{sec:prelim:expansion}), and then considers distances between nodes in the reduced graphs.
    The second category considers adjacent hyperedges (see \cref{prelim:adj_edges}), obtains distances between hyperedges (which can be interpreted as computing distances between nodes in the line expansion; see \cref{prelim:line_exps}), and then sets the distance between two nodes as the minimum distance between their respective incident hyperedges, as described in \cref{sec:prelim:connectivity} (see \cref{prelim:adj_edges,prelim:walk,prelim:path,prelim:distance}).
\end{discussion}
\vspace{-2mm}

\centrality{Closeness centrality.}\label{cent:closeness}
\citet{faust1997centrality} extended closeness centrality (see \cref{prelim:closeness_cent}) for both nodes and hyperedges using star expansion (see \cref{prelim:star_exps}).
Both original nodes and hyperedges become nodes in the star expansion, and the \textit{closeness centrality} of each original node/hyperedge is defined as the closeness centrality of its corresponding node in the star expansion.
As mentioned in \cref{sec:prelim:connectivity} and above, we can obtain various definitions of closeness centrality in hypergraphs by considering different definitions of paths and distances.
Closeness centrality has also been extended to hypergraphs using clique expansion~\citep{vasilyeva2023distances} (see \cref{prelim:cliq_exps}) and line expansion~\citep{vasilyeva2023distances,aksoy2020hypernetwork} (see \cref{prelim:line_exps}).
Similarly to the closeness centrality on graphs (see \cref{prelim:closeness_cent}), this measure is meaningful only when the hypergraph is connected, i.e., when all node pairs have finite distances under the corresponding definition of adjacent hyperedges.

\centrality{Betweenness centrality.}\label{cent:betweenness}
As another centrality measure based on paths, betweenness centrality (see \cref{prelim:betweenness_cent}) can also be extended to hypergraphs in various ways by considering different definitions of paths.
Researchers have also defined \textit{betweenness centrality} for hypergraphs in various ways using
star expansion~\citep{faust1997centrality},
clique expansion~\citep{vasilyeva2023distances}, and
line expansion~\citep{vasilyeva2023distances,aksoy2020hypernetwork}.
\citet{puzis2013augmented} proposed a variant of betweenness centrality for pairwise graphs with weights on each node pair, which can be naturally extended to hypergraphs.
A well-known limitation of betweenness centrality is its high computational complexity since the computation involves calculating the shortest paths between all node pairs.
To this end, there are existing techniques for efficient computation or approximation of betweenness centrality in graphs (see, e.g., \cite{bader2007approximating,baglioni2012fast}) and hypergraphs (see, e.g., \cite{puzis2013betweenness}).

\centrality{Harmonic centrality.}\label{cent:harmonic}
Similarly to closeness centrality (see \cref{cent:closeness}), harmonic centrality (see \cref{prelim:harmonic_cent}) is also defined by distances, and has been extended to hypergraphs in various ways by considering different definitions of distances.
Again, researchers have considered
star expansion~\citep{esposito2022venture},
clique expansion~\citep{aksoy2020hypernetwork}, and
line expansion~\citep{aksoy2020hypernetwork}.
Similar to the harmonic centrality on pairwise graphs (see \cref{prelim:harmonic_cent}), it has the merit that it is meaningful even on disconnected hypergraphs.
Harmonic centrality essentially gives each distance $\delta$ a weight $w_\delta = \delta^{-1}$.
One can consider more general weights $w_\delta$ on different distances.
Such ideas have been considered for pairwise graphs (see, e.g., \cite{inariba2017random} and \cite{dangalchev2006residual}), and they can be straightforwardly generalized to hypergraphs.

\centrality{Neighborhood centrality.}\label{cent:neighborhood}
\citet{amato2018centrality} proposed \emph{neighborhood centrality}, which quantifies the importance of each node as the fraction of nodes in the hypergraph within a distance threshold (which is a hyperparameter).
The authors specifically used the distances in line expansion, while in principle, as discussed in \cref{sec:prelim:connectivity} and above, any definition of distances would suffice to give a reasonable definition.
This measure can be interpreted as a generalization of neighbor-degree centrality (see \cref{cent:neighbor_deg}), extending beyond immediate neighbors to those within a multi-hop radius.

\centrality{Distance-based fuzzy centrality.}\label{cent:fuzzy_collective_influence}
\citet{zhang2025locating} proposed \emph{higher-order distance-based fuzzy centrality}, which quantifies node importance using the entropy of the distance distribution to other nodes.
For each node $v$, the other nodes are grouped by their distance to $v$; the counts at each distance are weighted using distance-dependent fuzzy decays, normalized into a probability distribution, and summarized by its entropy.
A node is considered important when it is close to many others while maintaining balanced influence across multiple distance layers, capturing both local impact and broader spreading potential.
The authors specifically used distances in the $s$-line expansion (see \cref{prelim:line_exps}), although other distance definitions could also be adopted, as discussed in \cref{sec:prelim:connectivity}.
Similar ideas were considered by \citet{liu2024influential}.

\vspace{-2mm}
\begin{discussion}[Why usually only shortest paths?]
    As we can see above, path-based measures typically rely exclusively on \textit{shortest} paths, rather than general paths.
    But why is this preference prevalent?
    {In our view,} the main reason lies in the computational complexity.
    Calculating all paths is computationally intensive and often infeasible, as it involves enumerating and analyzing a combinatorially large set of possible paths.
    On the other hand, shortest paths can be computed efficiently using algorithms such as Dijkstra's, due to their well-defined and limited nature.
    Specifically, on pairwise graphs, the time complexity of finding all paths between a single node pair is $O(|V|!)$ in the worst case,
    while that of finding all shortest paths between a node pair is at most $O(|V|^2)$.
\end{discussion}
\vspace{-2mm}

\subsection{Walk-based Measures}\label{sec:measures:structural:walk}
A broader and more expressive class of centrality and importance measures is based on \textit{walks} (see \cref{prelim:walk}).
Such measures typically rely on matrix or tensor representations of hypergraphs and extract importance scores as eigenvectors or stationary distributions of associated operators.
With such spectral operations, walk-based measures typically assume that a node is important if it is frequently visited by random walks on a hypergraph. This perspective implies that the centrality of a node/hyperedge is reinforced by the importance of the other nodes/hyperedges that constitute the walks leading to it.

\centrality{Eigenvector centrality.}\label{cent:eigenvector}
In pairwise graphs, eigenvector centrality (see \cref{prelim:eigenvec_cent}) is one of the most straightforward walk-based measures.
Generalizations to hypergraphs have defined centrality for both \textbf{nodes} and \textbf{hyperedges}. For node centrality, \citet{zhou2006learning} proposed computing eigenvector centrality on the clique expansion (see \cref{prelim:cliq_exps}), while \citet{faust1997centrality} considered the star expansion (see \cref{prelim:star_exps}). To measure the importance of hyperedges, \citet{kovalenko2022vector} considered the line expansion (see \cref{prelim:line_exps}), where the centrality of a node in the expansion corresponds to the centrality of a hyperedge in the original hypergraph.

\centrality{Random-walk centrality.}\label{cent:rw_edge_dep}
\citet{chitra2019random} proposed \textit{random-walk centrality} for hypergraphs, where each node has distinct, edge-dependent weights reflecting its importance within specific hyperedges.
This measure is defined via a random walk process: starting from a node, the walker selects the next hyperedge with probability proportional to the hyperedge's weight, and subsequently selects the next node from that hyperedge with probability proportional to its edge-dependent node weight.
The final centrality {score} of each node corresponds to {its stationary probability in} this random walk.
\citet{chitra2019random} also showed that without edge-dependent node weights, random walks are equivalent to those on clique expansions.
\citet{chun2024random} extended this by further considering random walks \textit{with restart} on hypergraphs.
\citet{stephan2024sparse} and \citet{chodrow2023nonbacktracking} considered \textit{non-backtracking walks} to mitigate localization, where walkers tend to get trapped in short loops or high-degree neighborhoods.

\centrality{Personalized PageRank.}\label{cent:pagerank}
PageRank (see \cref{prelim:pagerank_cent}) is a walk-based measure that evaluates node importance using the stationary distribution of a random walk with teleportation, where the walker randomly jumps to a preferred set of nodes with some probability.
\citet{takai2020hypergraph} extended \emph{personalized PageRank} to hypergraphs by defining the random walk directly on the hypergraph structure, rather than relying on a pairwise projection.
In this formulation, the teleportation distribution can be customized to prioritize certain nodes, enabling personalization of the ranking process.
\citet{piao2025identifying} considered PageRank for \textit{directed hypergraphs}, where each hyperedge has explicitly defined head and tail node sets inside it.

\centrality{Vector centrality.}\label{cent:vector}
\citet{kovalenko2022vector} proposed \emph{vector centrality}, which assigns each node a \emph{vector} (instead of a \emph{scalar} in usual cases) of centrality scores.
The dimension of this vector corresponds to the number of hyperedge sizes of interest, with each component of the vector quantifying the node's involvement in interactions of a specific size.
Specifically, after computing the eigenvector centrality of all hyperedges (see \cref{cent:eigenvector}), for each node, the element in the vector corresponding to each hyperedge size $k$ is the sum of the eigenvector centrality values of all its incident hyperedges (see \cref{prelim:incident_edge_n_const_nodes}) of size $k$.
The vector centrality thus captures a node's importance depending on the size of the interactions (i.e., hyperedges).

\centrality{Z/H-eigenvector centrality.}\label{cent:threeeigenvector}
\citet{benson2019three} defined three variants of eigenvector centrality measures: C/Z/H-eigenvector centrality.
Among them, C-eigenvector centrality
is equivalent to \emph{eigenvector centrality} computed on the clique expansion (see \cref{cent:eigenvector}).
The other two variants,
\emph{Z-eigenvector centrality} (ZEC) and \emph{H-eigenvector centrality} (HEC), are directly computed on $r$-uniform hypergraphs (see \cref{prelim:uniform_HGs}) using the \emph{hypergraph adjacency tensor}.
ZEC and HEC adapt the idea of eigenvector centrality to hypergraphs by working directly with their multi-way connections, using tensor-based spectral definitions that differ in how the eigenvalues are defined.
See also, e.g., \citep{kolda2006tophits,kolda2009tensor}, for general tensor-based methods for identifying important nodes in hypergraphs.
A key limitation of both ZEC and HEC is that they are defined only for uniform hypergraphs.

\centrality{Generalized Z/H-eigenvector centrality.}\label{cent:genralized_eigenvector}
Although Z- and H-eigenvector centralities (see \cref{cent:threeeigenvector})  were originally defined for uniform hypergraphs, subsequent studies enabled their application to non-uniform hypergraphs.
Building on a non-uniform hypergraph adjacency tensor based on \emph{hyperedge blowups},
\citet{aksoy2024scalable} 
developed implicit tensor-times-same-vector algorithms that solve the corresponding eigenvector equations without
explicitly constructing the full high-order tensor.
\citet{contreras2023uplifting} proposed \emph{uplifted eigenvector centrality} as a tensor-based spectral approach for extending the aforementioned H-eigenvector centrality to non-uniform hypergraphs.
The method transforms a non-uniform hypergraph into a uniform one via an uplift process, in which smaller hyperedges are augmented with auxiliary nodes and larger hyperedges may be projected onto subsets of a chosen size.

\centrality{Nonlinear eigenvector centrality.}\label{cent:node_edge_nonlinear}
\citet{tudisco2021node} introduced a spectral framework that jointly ranks nodes and hyperedges in a hypergraph using a mutually reinforcing rule: Nodes are important when they belong to important hyperedges, and hyperedges are important when they contain important nodes.
The framework is flexible because it lets us choose transformation functions that specify how influence is pooled (e.g., sum‑like, product‑like, or other combinations).
With suitable choices, it can reproduce classic eigenvector centrality (see \cref{cent:eigenvector}), generalize Z-eigenvector centrality (see \cref{cent:threeeigenvector}), or create new variants for non-uniform hypergraphs.

\centrality{Sub-hypergraph centrality.}\footnote{Despite its name, this measure only uses information about walks and is thus walk-based instead of subhypergraph-based.}\label{cent:sub_hypergraph}
\citet{estrada2006subgraph} extended the notion of subgraph centrality \cite{estrada2005subgraph} from graphs to hypergraphs, defining \textit{sub-hypergraph centrality} that measures the importance of each node by counting closed walks it is involved in, with shorter walks contributing more.
A closed walk is a walk (see \cref{prelim:walk}) starting and ending at the same node.

\subsection{Subhypergraph-based Measures}\label{sec:measures:structural:subgraph}

Another important class of measures is based on \textit{subhypergraphs} (see \cref{prelim:subhg}).
These measures typically first identify central subhypergraphs, and then assume that nodes/hyperedges participating {in a larger number of, or in} more prominent, central subhypergraphs are more central and important, and are thus assigned higher centrality scores.

\centrality{Hypercoreness.}\label{cent:core}
\citet{leng2013m} generalized the notion of $k$-cores \cite{seidman1983network} to hypergraphs.
The $k$-\textit{hypercore} is the maximal subhypergraph (see \cref{prelim:subhg}) in which every node has degree (see \cref{prelim:degree}) at least $k$.
The \textit{hypercoreness} of a node or hyperedge is then the largest $k$ such that it belongs to the $k$-hypercore.
Several variants refine this basic model: \citet{luo2024hierarchical} required hyperedges of size above a threshold, \citet{arafat2023neighborhood} defined cores via neighbor-degrees (see \cref{prelim:neighbor}), and \citet{kim2023exploring} added additional conditions on node co-occurrences.
Some other researchers~\citep{limnios2021hcore,bu2023hypercore,kim2025beyond} extended the notion of subhypergraphs, proposing hypercore models in which hyperedges can be subsets of the original ones, rather than being preserved entirely as required in the typical definition of subhypergraphs (see \cref{prelim:subhg}).

\centrality{Hypertrussness.}\label{cent:truss}
\citet{wang2022efficient} and \citet{qin2025truss} generalized the notion of
$k$-trusses \cite{cohen2008trusses} to hypergraphs.
In a similar spirit to hypercoreness (see \cref{cent:core}), a $k$-\textit{hypertruss} is the maximal subhypergraph (see \cref{prelim:subhg}) in which each hyperedge participates in at least $k$ \textit{triangles}, and the \textit{hypertrussness} of a node or hyperedge is the largest $k$ for which it belongs to such a subhypergraph.
Different definitions of triangles lead to different models: \citet{wang2022efficient} defined a triangle as a triplet of nodes where each pair of nodes co-occurs in at least one hyperedge,
while \citet{qin2025truss} defined a triangle as a triplet of hyperedges that intersect pairwise but have an empty three-way intersection.
Other notions of triangles in hypergraphs
(see, e.g., \cite{benson2018simplicial,kostochka2013hypergraph,li2022chromatic})
can likewise be adopted to define hypertrussness.

\centrality{Hitting-set score.}\label{cent:hitting_set}
\citet{amburg2021planted} proposed \textit{hitting-set scores} that measure the likelihood of each node being in the hitting set.
{A hitting set} of a hypergraph is a subset of nodes that intersect every hyperedge.
To estimate which nodes belong to the hitting set,
the authors repeatedly apply a greedy set-cover heuristic, prune the output to a minimal hitting set, and take the union across many iterations.
The score of a node is defined as the proportion of runs in which it appears in these minimal hitting sets.
Nodes that are structurally indispensable for intersecting hyperedges thus accumulate higher scores.

\centrality{Core-periphery score.}\label{cent:core_periphery}
Both \citet{tudisco2023core} and \citet{papachristou2022core} proposed methods for assigning \textit{core-periphery scores} in hypergraphs, aiming to quantify how ``core'' (i.e., central) each node is.
These measures assume the existence of a subhypergraph consisting of \textit{core} nodes that frequently participate across different groups, contrasted with more peripheral nodes that interact mainly through the core.
\citet{tudisco2023core} formulated the problem as a nonlinear spectral optimization.
\citet{papachristou2022core} instead introduced a generative random hypergraph model in which the probability of each hyperedge depends on the nodes' core-periphery scores, and inference recovers the most likely scores explaining the observed hypergraph.

\centrality{Motif counts.}\label{cent:motif_counts}
The above subhypergraph-based measures assume that nodes/hyperedges participating in \textit{more prominent} central subhypergraphs are more important, while one can also consider the \textit{frequency} of participation in central subhypergraphs to quantify importance.
Following such an idea, \citet{lee2024hypergraph} proposed \textit{hypergraph motifs} and used the \textit{motif counts} of each node/hyperedge as a centrality measure.
Since there are multiple motifs, each node/hyperedge is associated with a centrality vector instead of just a scalar.
Other definitions of hypergraph motifs (see, e.g., \cite{kim2025estimating,lotito2022higher,juul2024hypergraph}) can similarly be used to define motif-based centralities.

\centrality{Clustering coefficients.}\label{cent:clustering_coef}
In pairwise graphs, the clustering coefficient (or transitivity)~\citep{watts1998collective} measures the tendency of a node's neighbors to form triangles, capturing its role in cohesive groups.
Several works have generalized this idea to hypergraphs by defining clustering coefficients for pairs of nodes~\citep{klamt2009hypergraphs,le2005clustering,pena2012bipartite,gallagher2013clustering,miyashita2025clustering} or for pairs of hyperedges~\citep{zhou2011properties,klimm2021hypergraphs,klamt2009hypergraphs,torres2021and,gallagher2013clustering,kim2023transitive}.
Aggregating these pairwise values yields centrality scores for individual nodes or hyperedges.
These measures are typically normalized as fractions, although unnormalized (count-based) versions are also used when absolute participation is of interest.

\subsection{Hybrid Measures}\label{sec:measures:structural:hybrid}
Below, we introduce \textit{hybrid structural measures} based on multiple structural elements. Such measures typically capture the importance of nodes/hyperedges from various perspectives, by aggregating multiple measures.
Under the high-level framework of hybrid measures, one can combine different measures under various aggregation functions.

\centrality{Gravity centrality.}\label{cent:gravity}
Inspired by Newton's law of universal gravitation, \citet{xie2023vital} introduced \emph{gravity centrality}
based on both \textit{degrees} and \textit{paths}.
It measures the importance of each node $v$ by considering its connections to all other nodes $u$, weighting each connection by the degrees of both nodes and inversely by the square of their distance $\delta(u,v)$, i.e., $\sum_{u\in\mathcal V\setminus\{v\}} \frac{\deg(u) \deg(v)}{\delta(u,v)^2}$.
The authors also considered a semi-local version, where for each node $v$, we set a distance threshold and only consider the other nodes $u$ within this threshold.
The authors specifically used distances in the $s$-line expansions (more precisely, a weighted sum of distances in $s$-line expansions with different $s$ values; see \cref{prelim:line_exps}), while in principle, as discussed in \cref{sec:prelim:connectivity} and above, any definition of distances would suffice to give a reasonable definition.
\citet{wang2024identification} extended this idea to further consider hyperedge-level and community-level interactions.

\centrality{Complex centrality.}\label{cent:coreness_plus_degree}
\citet{zhou2021identification} proposed \textit{complex centrality} based on both \textit{degrees} and \textit{subhypergraphs}.
It measures the importance of each node $v$ with a mixture of degree centrality (see \cref{cent:degree}) and hypercoreness (see \cref{cent:core}).
Specifically, the authors use a Euclidean aggregation.
The centrality score of each node $v$ is $\sqrt{d^2_v + k^2_v}$, where $d_v$ is the degree centrality of $v$, and $k_v$ is the hypercoreness of $v$.
\citet{guo2025method} have extended the idea to multilayer hypergraphs.

\centrality{Multi-centrality.}\label{cent:entropy_based_multi_cent}
\citet{wu2023multi} proposed \emph{multi-centrality} for nodes based on \textit{degrees}, \textit{paths}, and \textit{subhypergraphs}.
Specifically, they combine degree centrality (see \cref{cent:degree}), betweenness centrality (see \cref{cent:betweenness}), and hypercoreness (see \cref{cent:core}) into an integrated centrality measure, with an adaptive weighting scheme based on entropy.

\centrality{Multi-criteria centrality.}\label{cent:multi_criteria}
\citet{xiao2013research} proposed \emph{multi-criteria centrality} that integrates \textit{degree-based} and \textit{path-based} information.
For each node $v$, its importance is defined as a weighted sum of three components: (1) its overall degree (see \cref{prelim:degree} and \cref{cent:degree}), (2) its degree in a local subhypergraph around $v$, and (3) its betweenness (see \cref{prelim:betweenness_cent} and \cref{cent:betweenness}).

\centrality{Improved PageRank.}\label{cent:sprie}
\citet{chen2023identifying} proposed \textit{improved PageRank}, which combines \textit{walk-based} PageRank with \textit{degree-based} information.
Globally, they introduce an improved PageRank using node-similarity information.
Locally, they quantify node importance via information entropy of neighbor degree distributions, with the high-level idea that a node is more influential if its connections are not only numerous but also diverse (see similar ideas in \cref{cent:fuzzy_collective_influence}).
Those two components are then combined to give the final measure.

\section{Functional Measures}\label{sec:measures:functional}


{\textit{Functional measures} assess importance based on the role that nodes or hyperedges play in the overall \emph{functioning} of the hypergraph, rather than solely on static structural features.}
They quantify centrality in terms of system behavior: how much the network's connectivity or robustness is disrupted when a node/hyperedge is perturbed, or how much value a node/hyperedge contributes across different coalitions of multiple nodes/hyperedges.
We group these approaches into two sub-categories:
\textit{perturbation-based measures}, which evaluate structural changes caused by removing nodes or hyperedges, and \textit{coalition-based measures}, which allocate importance using cooperative game-theoretical formulations.

\subsection{Perturbation-based Measures}\label{sec:measures:functional:perturbation}
Perturbation-based measures typically assume that nodes/hyperedges whose removal leads to greater structural disruption are more central and important, and are thus assigned higher centrality scores.

\centrality{Higher-order von Neumann entropy.} \label{cent:higher_order_von_neumann}
\citet{hu2023identifying} proposed a centrality measure called \textit{higher-order von Neumann entropy}, which quantifies the importance of each hyperedge based on the change in the von Neumann entropy of the Laplacian matrix of its line expansion (see \cref{prelim:line_exps}) after removal.\footnote{The Laplacian matrix of a pairwise graph (here, the line expansion) is defined as $\mat{L} = \mat{D} - \mat{A}$, where $\mat{D}$ is a diagonal matrix of node degrees with $\mat{D}_{ii}$ being the degree (see \cref{prelim:degree}) of node $v_i$, and $A$ is the graph's adjacency matrix (see \cref{prelim:eigenvec_cent}).}
The more the von Neumann entropy decreases after a hyperedge is removed, the more central the hyperedge is considered.
The intuition is that higher von Neumann entropy often corresponds to denser structures.
The centrality value of each node is computed by aggregating the values of its incident hyperedges (see \cref{prelim:incident_edge_n_const_nodes}).

\centrality{Total loss.}\label{cent:total_loss}
\citet{xiao2016node} proposed a centrality measure called \textit{total loss}, which quantifies the importance of each node based on the decrease (i.e., ``total loss'') in the connectivity of the hypergraph.
Specifically, the connectivity here is a function of the distances between all node pairs in the hypergraph, where shorter distances give higher connectivity.

\centrality{Grounded-Laplacian eigenvalue.}\label{cent:grounded_laplacian}
\citet{li2023important} proposed \emph{grounded-Laplacian eigenvalue} to rank hyperedges by considering the effects of their removal, which combines \textit{walk-based} spectral information with \textit{perturbation-based} removal effects.
For each hyperedge $e$, they remove $e$ and all its constituent nodes (see \cref{prelim:incident_edge_n_const_nodes}),
obtain the corresponding Laplacian of the remaining hypergraph (the authors call it the \textit{grounded Laplacian} of $e$), and then compute its smallest eigenvalue.\footnote{Here, the Laplacian matrix is also defined as $\mat{L} = \mat{D} - \mat{A}$, but $\mat{D}$ and $\mat{A}$ are defined differently. Specifically, $\mat{D}$ is the diagonal matrix of neighbor-degrees (see \cref{prelim:neighbor}), and $\mat{A}$ is a weighted adjacency matrix where each entry $\mat{A}_{ij}$ is the number of hyperedges containing both nodes $v_i$ and $v_j$.}
This measure captures the structural vulnerability induced by removing each hyperedge, with larger values indicating greater importance.

\vspace{-2mm}
\begin{discussion}[Alternative metrics to quantify the structural change]
    In the above measures (see \cref{cent:higher_order_von_neumann,cent:total_loss,cent:grounded_laplacian}), the von Neumann entropy, the connectivity function, and the smallest eigenvalue of the Laplacian are the specific choices made by the authors.
    In principle, any measure that quantifies the connectivity \cite{laita2011graph} or robustness \cite{freitas2022graph} of hypergraph structure can be used instead.
\end{discussion}
\vspace{-2mm}

\subsection{Coalition-based Measures}\label{sec:measures:functional:coalition}
Similarly to perturbations, \textit{coalitions} can also be used to define centrality measures.
Rather than focusing only on the presence or absence of individual nodes/hyperedges, coalition-based methods evaluate the contribution of a node/hyperedge across all possible subsets of nodes/hyperedges.
This perspective is grounded in cooperative game theory, where the centrality of a node/hyperedge is often defined via its \textit{Shapley value}~\cite{shapley1951notes}, i.e., the expected marginal contribution it makes when joining a randomly ordered coalition.

\centrality{Clique-induced Shapley value.}\label{cent:clique_shapley}
\citet{huang2024clique} proposed \textit{clique-induced Shapley value}, a coalition-based centrality based on a cooperative game where a coalition's value comes from the cliques it induces.
Larger cliques are given higher weights, and the corresponding Shapley value then fairly splits this clique-based value among nodes.
This measure highlights nodes that tend to bridge overlapping communities.

\centrality{Line-expansion Shapley value.}\label{cent:line_exp_shapley}
\citet{roy2015measuring} proposed \textit{line-expansion Shapley value}, a coalition-based centrality for hypergraphs by first converting them into a weighted graph using line expansion (see \cref{prelim:line_exps}).
A cooperative game is then played on this weighted graph, using the Shapley value to determine each hyperedge's importance, which is finally shared equally among its constituent nodes.

\section{Contextual Measures}\label{sec:measures:contextual}

Both structural measures (see \cref{sec:measures:structural}) and functional measures (see \cref{sec:measures:functional}) define centrality and importance measures merely based on hypergraph topology. \textit{Contextual measures}, in contrast, extend beyond topology by incorporating additional information associated with nodes or hyperedges.
Such information may include \textit{features} (e.g., attributes or embeddings describing entities) or \textit{labels} (e.g., roles or annotations specifying functional positions).
These measures capture importance as a joint effect of structure and context, allowing centrality to reflect not only how elements are embedded in the hypergraph but also what characteristics they carry.
We group existing approaches into three sub-categories: \textit{feature-based}, \textit{label-based}, and \textit{hybrid} contextual measures.

\subsection{Feature-based Measures}\label{sec:measures:functional:feature}
\textit{Feature-based measures} incorporate non-structural attributes of nodes or hyperedges into the assessment of centrality and importance.
These approaches combine structural information with external descriptors such as embeddings, metadata, or other feature vectors.
The underlying assumption is that centrality or importance in real-world hypergraphs depends not only on position within the topology but also on the characteristics entities carry.
Feature-based measures therefore capture centrality as an interplay between structural prominence and attributive relevance.

\centrality{Structure-and-embedding score}\label{cent:struc_and_embed}
\citet{chang2025structure} proposed \textit{structure-and-embedding score}, which combines \textit{degree-based} structural scores and a \textit{feature-based} score using learned node embeddings.
The key idea is that central nodes are both structurally well-connected and distinctive in their learned representations.
Structurally, they define two measures: \textit{common-node centrality}, capturing how many nodes hyperedges share, and \textit{common-hyperedge centrality}, capturing how many hyperedges nodes share. Embedding-wise, a \textit{distance-based score} is computed in the learned embedding space, where nodes with large distances to others are considered more important.

\centrality{Representation-based influence score.}\label{cent:AHGA}
\citet{ni2025structural} proposed \textit{representation-based influence score}, which assigns scores to nodes by learning from diffusion outcomes.
Rather than simulating influence spread and structural disruption for every node, the method uses results from a small set of representative nodes as supervision.
A learning framework then generalizes these outcomes, producing scores that approximate each node's impact on information spreading and network connectivity.
The resulting scores align with simulation-based influence while being far more efficient and transferable across different hypergraph structures.

\subsection{Label-based Measures}\label{sec:measures:functional:label}
\textit{Label-based measures} exploit categorical annotations that specify the roles or functions of nodes within hyperedges.
Conventional node labels are fixed across hyperedges and are commonly used to characterize node- or hyperedge-level homophily and heterophily~\cite{li2025hypergraph, veldt2023combinatorial}. In contrast, the labels considered here are often \emph{edge-dependent}: the same node may occupy different roles across hyperedges.
Different from feature-based approaches, which rely on continuous attributes or embeddings, label-based measures emphasize how role assignments shape importance.
They assess centrality by considering not only where nodes or hyperedges are positioned in the structure, but also what roles they occupy in group interactions.
This perspective is particularly relevant in settings such as collaboration, communication, or Q\&A forums, where the influence of a participant depends on their position \emph{and} their functional role.

\centrality{Edge-dependent role score.}\label{cent:WHATSNet}
\citet{choe2023classification} proposed \textit{edge-dependent role score} that combines structural cues from traditional measures with role-based attributes (e.g., first/last author, asker/answerer). The model learns attention weights and role probabilities within each hyperedge, then aggregates them to produce a score capturing both structural position and role-specific relevance.
As a related effort, \citet{Bu2025Anchor} introduced the concept of \textit{group anchors}, where each hyperedge has a single but edge-dependent anchor (i.e., a node may be the anchor in one group but not in another), and modeled this through \textit{anchor scores} on node-hyperedge pairs indicating the likelihood of anchorship.

\centrality{Matrix centrality.}\label{cent:matrix_centrality}
\citet{vasilyeva2024matrix} proposed \emph{matrix centrality} for annotated hypergraphs \cite{chodrow2020annotated, choe2023classification}  where nodes may occupy different roles within hyperedges.
It extends vector centrality (see \cref{cent:vector}), which represents each node's importance as a vector indexed by hyperedge size, to a two-dimensional matrix representation in which centrality values are indexed jointly by hyperedge size and node role.
Aggregating over the role dimension recovers the original vector centrality (see \cref{cent:vector}), while aggregating over hyperedge sizes produces \emph{role centrality}, a complementary measure that captures a node's overall importance by functional position regardless of group size.

\subsection{Hybrid Measures}\label{sec:measures:contextual:hybrid}
\textit{Hybrid contextual measures} combine both node or hyperedge \textit{features} and \textit{labels} when defining importance.
By leveraging continuous attributes together with categorical role information, these approaches produce scores that reflect not only structural prominence but also how entities with particular characteristics and roles contribute to hypergraph interactions.

\centrality{Hypergraph attention score.}\label{cent:hypergraph_conv_att}
In hypergraph neural networks, attention mechanisms (see, e.g., \citep{bai2021hypergraph,li2023hypergraph,chai2024hypergraph,saifuddin2023hygnn}) are often used to weight nodes, hyperedges, or higher-order substructures during message passing.
The resulting \textit{hypergraph attention score} indicates which elements are more important for the specific learning task at hand, and can therefore be interpreted as task-specific importance.
These scores reflect importance jointly from connectivity, features, and labels, tailored to the supervised or self-supervised objective.

\centrality{Identity-aware score.}\label{cent:ID_HAN}
\citet{chen2024hyperedge} proposed \textit{identity-aware score}, a hyperedge-level measure that learns group importance from real-world features (e.g., votes, citations, and views) while accounting for members' different roles. Role-specific attention and degree-based positional context inject attributes into message passing, and iterative node-hyperedge propagation produces a score capturing both structural position and role contributions.

\section{Empirical Insights}\label{sec:insight}
In this section, we present the results of empirical analyses designed to provide insights into representative measures.
Specifically, we summarize key findings from experiments that examine (1) the empirical (dis)similarity among different measures, and (2) the empirical computation time of measures.
For reproducibility, we provide the source code and datasets utilized in our analyses at \url{https://github.com/jaewan01/hypergraph-centrality-survey}. 

\setlength{\tabcolsep}{8pt} 
\begin{table}[t]
\begin{center}
\begin{minipage}{\textwidth}
\caption{Basic statistics of the {preprocessed} real-world hypergraphs used in our empirical analyses.
The numbers of nodes and hyperedges are denoted by $n = |\mathcal{V}|$ and $m = |\mathcal{E}|$, respectively.
The average and maximum sizes of hyperedges are denoted by $\text{avg}_{e \in \mathcal{E}}\vert e \vert$ and $\text{max}_{e \in \mathcal{E}}\vert e \vert$, respectively. 
The density~\citep{hu2017maintaining} is defined as $m / n$, and the overlapness~\citep{lee2021hyperedges} is defined as $\sum_{e \in \mathcal{E}}\vert e \vert / n$.
}
\label{tab:graph_data}%
\begin{tabular}{lrrrrrr}
\hline
\toprule
\textbf{Dataset (Abbreviation)} & $n = \lvert \mathcal{V} \rvert$ & $m = \lvert \mathcal{E} \rvert$ & $\text{avg}_{e \in \mathcal{E}}\vert e \vert$ & $\text{max}_{e \in \mathcal{E}}\vert e \vert$ & Density & Overlapness\\
\midrule
\textbf{senate-bills (SB)} & 294 & 29,157 & 7.96 & 99 & 99.17 & 789.62\\
\textbf{house-bills (HB)} & 1,494 & 60,987 & 20.47 & 399 & 40.82 & 835.79\\
\textbf{email-enron (EEN)} & 143 & 10,454 & 2.53 & 37 & 73.10 & 185.20\\
\textbf{email-eu (EEU)} & 986 & 209,508 & 2.56 & 40 & 212.48 & 544.47\\
\textbf{ndc-classes (NDCC)} & 628 & 39,717 & 3.46 & 39 & 63.24 & 218.74\\
\textbf{ndc-substances (NDCS)} & 3,414 & 28,506 & 4.95 & 187 & 8.35 & 41.36\\
\textbf{contact-primary-school (CTP)} & 242 & 106,879 & 2.10 & 5 & 441.65 & 925.61\\
\textbf{contact-high-school (CTH)} & 327 & 172,035 & 2.05 & 5 & 526.10 & 1078.65\\
\textbf{tags-ask-ubuntu (TAU)} & 3,021 & 219,076 & 3.11 & 5 & 72.52 & 225.80\\
\textbf{threads-ask-ubuntu (THU)} & 82,048 & 113,575 & 2.31 & 14 & 1.38 & 3.20\\
\bottomrule\hline
\end{tabular}
\end{minipage}
\end{center}
\end{table}

\subsection{Experimental Settings}
\label{sec:insight:settings}
We describe the experimental settings, including machines, datasets, analyzed measures, and evaluation metrics.

\smallsection{Machines.}
We used a workstation with an AMD Ryzen 7 3700X processor and 128 GB of memory.

\smallsection{Datasets.}
We conduct our empirical analyses on ten real-world hypergraphs from five different domains:
\begin{itemize}
    \item \textbf{Bills domain (senate-bills, house-bills~\cite{chodrow2021generative, fowler2006connecting, fowler2006legislative}):} Each node represents a member of the U.S. Congress, and each hyperedge corresponds to a set of co-sponsors of a bill.
    \item \textbf{Email domain (email-enron, email-eu~\cite{benson2018simplicial, yin2017local, leskovec2007graph}):} Each node denotes an email address, and each hyperedge consists of the sender and all recipients of a single email.
    \item \textbf{Drug domain (ndc-classes, ndc-substances~\cite{benson2018simplicial}):} Each hyperedge represents a drug. For \textit{ndc-classes}, nodes in a hyperedge correspond to the class labels applied to the corresponding drug, and for \textit{ndc-substances}, nodes in a hyperedge represent the substances comprising the corresponding drug.
    \item \textbf{Contact domain (contact-primary-school, contact-high-school~\cite{benson2018simplicial, stehle2011high, mastrandrea2015contact}):} Each node represents an individual, and each hyperedge is a group of people who interacted within a given time interval.
    \item \textbf{Q\&A domain (tags-ask-ubuntu, threads-ask-ubuntu~\cite{benson2018simplicial}):} Each hyperedge represents a Q\&A post. For \textit{tags-ask-ubuntu}, nodes in a hyperedge correspond to the tags associated with the corresponding post, and for \textit{threads-ask-ubuntu}, nodes in a hyperedge correspond to the users participating in the corresponding post.
\end{itemize}
For all the hypergraphs, we only consider hyperedges of size greater than or equal to two, and extract the largest connected component of each hypergraph.
We report the basic statistics of the preprocessed hypergraphs in~\cref{tab:graph_data}.

\smallsection{Analyzed Measures.}
We employ ten representative hypergraph centrality and importance measures.
We exclude measures that 
(1) depend on user-specified hyperparameters (e.g., \cref{cent:isolating}, \cref{cent:neighborhood}, and \cref{cent:clique_shapley}), 
(2) produce vector-valued outputs rather than a single scalar score (e.g., \cref{cent:vector} and \cref{cent:motif_counts}), 
and (3) require external labels or features (i.e., contextual measures described in~\cref{sec:measures:contextual}).
The evaluated measures are categorized according to our proposed taxonomy as follows:
\begin{itemize}
    \item \textbf{Degree-based measures:} Degree centrality (\cref{cent:degree}), neighbor-degree centrality (\cref{cent:neighbor_deg}), and line-expansion degree centrality (\cref{cent:line_exp_deg}).
    \item \textbf{Path-based measures:} Closeness centrality (\cref{cent:closeness}), betweenness centrality (\cref{cent:betweenness}), and harmonic centrality (\cref{cent:harmonic}).
    \item \textbf{Walk-based measures:} Eigenvector centrality (\cref{cent:eigenvector}), random-walk centrality (\cref{cent:rw_edge_dep}), and uplifted eigenvector centrality (\cref{cent:genralized_eigenvector}).
    \item \textbf{Subhypergraph-based measure:} Hypercoreness (\cref{cent:core}).
\end{itemize}

We evaluate measures defined for nodes and those defined for hyperedges separately.
Among the analyzed measures, neighbor-degree centrality and uplifted eigenvector centrality are defined only for nodes, whereas line-expansion degree centrality is defined only for hyperedges.
All the other measures are defined for both nodes and hyperedges. 
All implementations are based on the XGI library \cite{landry2023xgi}. 
The source code and datasets used in this section are available at \url{https://github.com/jaewan01/hypergraph-centrality-survey}.

\smallsection{Evaluation Metrics.}
To quantitatively assess the (dis)similarity between different measures, we use four evaluation metrics from two complementary perspectives: two \emph{global metrics} and two \emph{top-$k$ metrics}.

\begin{itemize}
    \item \textbf{Global metrics:} We first assess the global (dis)similarity between each pair of measures using the \textbf{Pearson correlation coefficient} and the \textbf{Spearman rank correlation coefficient}.\footnote{Given two centrality score vectors $\mathbf{x} = (x_1, \ldots, x_n)$ and $\mathbf{y} = (y_1, \ldots, y_n)$, the Pearson correlation coefficient is defined as:
    $\frac{\sum_{i=1}^{n} (x_i - \bar{x})(y_i - \bar{y})}
    {\sqrt{\sum_{i=1}^{n} (x_i - \bar{x})^2} \sqrt{\sum_{i=1}^{n} (y_i - \bar{y})^2}}$,
    where $\bar{x}$ and $\bar{y}$ denote the mean values of $\mathbf{x}$ and $\mathbf{y}$, respectively.
    The Spearman rank correlation coefficient is computed as the Pearson correlation between the rank-transformed vectors of $\mathbf{x}$ and $\mathbf{y}$.}
    Both correlation coefficients range from $-1$ to $1$, where $-1$ indicates a perfect inverse relationship between the two measures, while $1$ indicates a perfect agreement.
    \item \textbf{Top-$k$ metrics:} To analyze the consistency among highly ranked nodes, we further employ the \textbf{Jaccard similarity} and the \textbf{top-$k$ overlap ratio}.\footnote{Let $S_x^{(k)}$ and $S_y^{(k)}$ denote the sets of top-$k$ nodes identified by two different measures. The Jaccard similarity and the top-$k$ overlap ratio are defined as 
    $\frac{\vert S_x^{(k)} \cap S_y^{(k)} \vert}{\vert S_x^{(k)} \cup S_y^{(k)} \vert}$ and 
    $\frac{2 * \vert S_x^{(k)} \cap S_y^{(k)} \vert}{\vert S_x^{(k)} \vert + \vert S_y^{(k)} \vert}$, respectively.}
    We calculate both metrics using the top $5\%$ of the ranked nodes or hyperedges.
    To handle potential ties in rankings, we follow the procedure in~\citet{webber2010similarity}, including all tied items at rank $k$ in each set.
    Both metrics take values in the range $[0, 1]$, where $0$ indicates no agreement between the two measures, while $1$ indicates complete alignment.
\end{itemize}

\begin{figure}[t]
    \centering
    \includegraphics[width=0.98\linewidth]{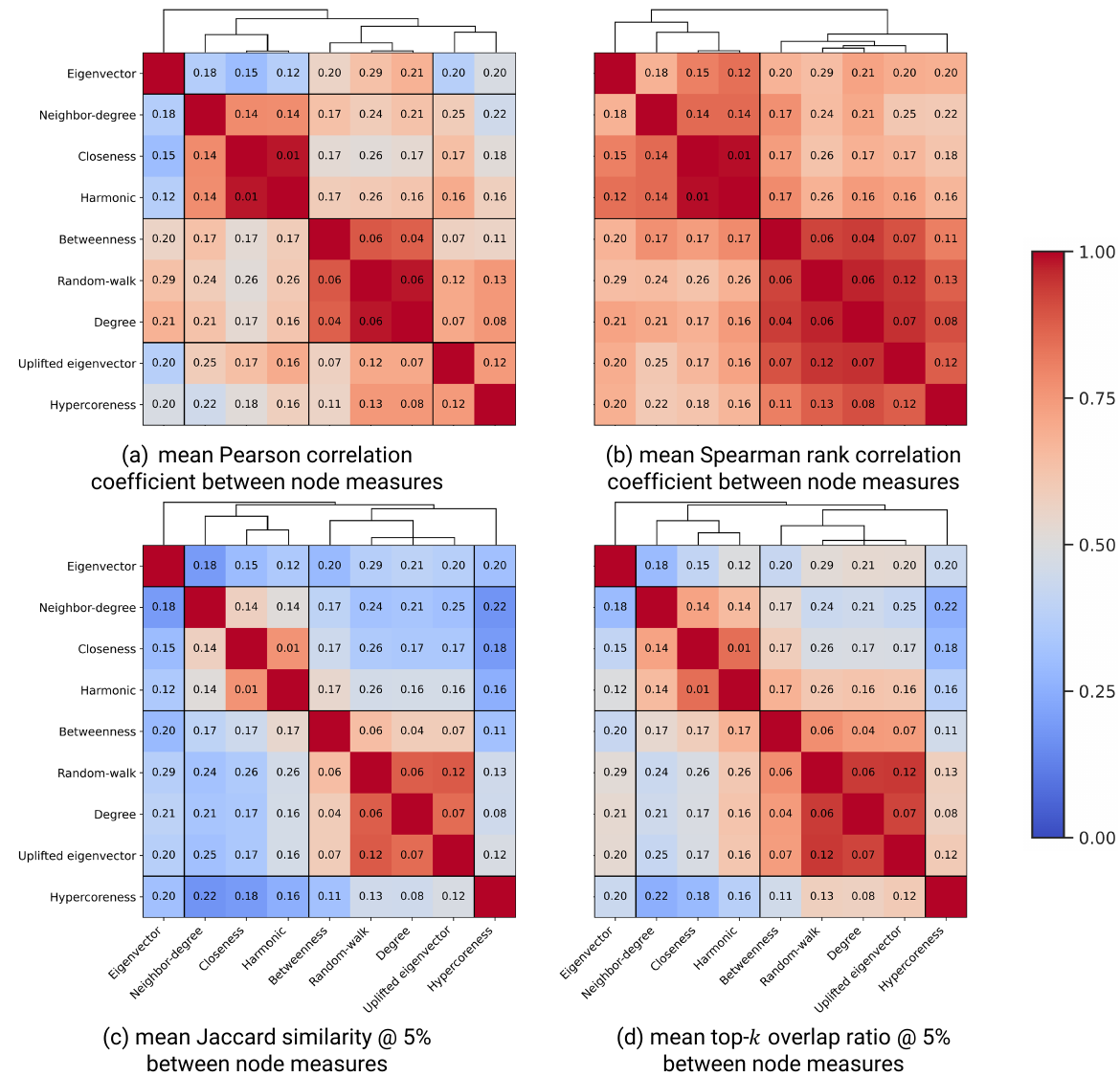}
    \vspace{-2mm}
    \caption{Correlation between node centrality and importance measures in terms of (a) Pearson correlation coefficient, (b) Spearman rank correlation coefficient, (c) Jaccard similarity @ 5\%, and (d) top-$k$ overlap ratio @ 5\%. 
    The colormap indicates the average value of each measure across all datasets, and we annotate the standard deviation across datasets in each cell.}
    \Description{Four clustered heatmaps comparing node measures by Pearson correlation, Spearman correlation, Jaccard similarity at five percent, and top-k overlap at five percent across ten datasets.}
    \label{fig:node:correlation}
\end{figure}

\begin{figure}[t]
    \centering
    \includegraphics[width=0.98\linewidth]{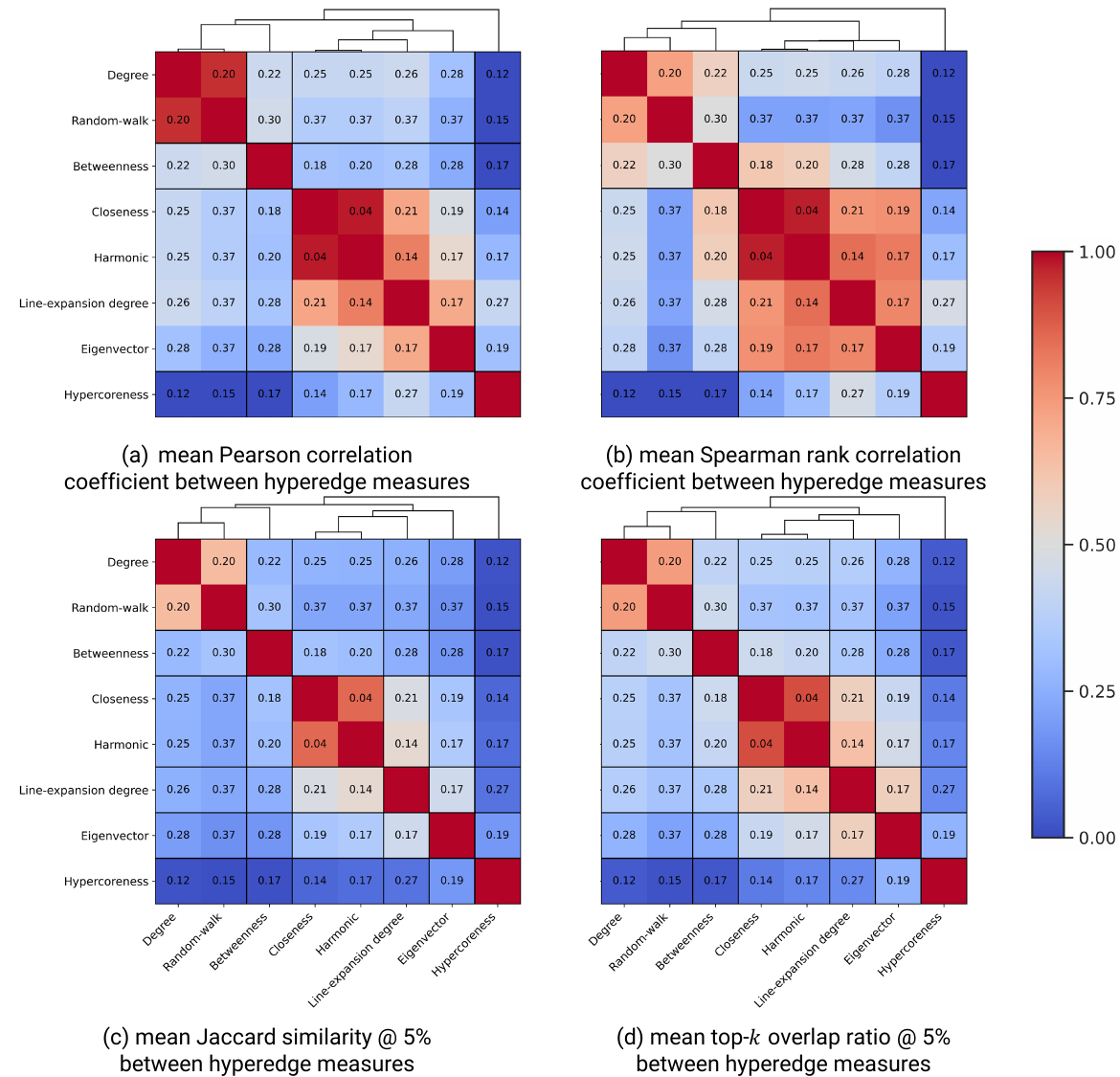}
    \vspace{-2mm}
    \caption{Correlation between hyperedge centrality and importance measures in terms of (a) Pearson correlation coefficient, (b) Spearman rank correlation coefficient, (c) Jaccard similarity @ 5\%, and (d) top-$k$ overlap ratio @ 5\%. 
    The colormap indicates the average value of each measure across all datasets, and we annotate the standard deviation across datasets in each cell.}
    \Description{Four clustered heatmaps comparing hyperedge measures by Pearson correlation, Spearman correlation, Jaccard similarity at five percent, and top-k overlap at five percent across ten datasets.}
    \vspace{-3mm}
    \label{fig:edge:correlation}
\end{figure}

\subsection{Correlation Analysis}
\label{sec:insight:correlation}
In this section, we present an empirical correlation analysis of different measures.
Specifically, we evaluate the measures described in~\cref{sec:insight:settings} across ten real-world hypergraphs, using the four evaluation metrics introduced in the same section.
In \cref{fig:node:correlation,fig:edge:correlation}, we present the mean value of each correlation metric for node and hyperedge measures, respectively, with standard deviations across datasets annotated in each cell.
Additionally, we apply hierarchical clustering to each metric, where correlation scores are linearly transformed into distances (0 indicating perfect similarity and 1 indicating complete dissimilarity). 
The resulting dendrograms are displayed above each colormap, and we further report the discrete cluster assignments that yield the highest silhouette score.
For both node and hyperedge measures, we derive three key take-home messages regarding the results:

\smallsection{High global correlation, limited top-ranked agreement.}
Across all datasets, both node and hyperedge measures exhibit strong global correlations, with average Pearson and Spearman rank correlation coefficients of 0.627 and 0.773 for nodes, and 0.400 and 0.434 for hyperedges, respectively.
Meanwhile, the agreement among highly ranked elements provides a complementary perspective: the average Jaccard similarity and top-$k$ overlap ratio for the top 5\% of elements are 0.443 and 0.568 for nodes, and 0.271 and 0.355 for hyperedges.
Note that the global correlation metrics range from $[-1,1]$, whereas the top-$k$ metrics range from $[0,1]$ and quantify a different aspect of agreement.
Together, these results suggest that while different measures capture broadly consistent overall importance trends, they often diverge in identifying the most influential nodes or hyperedges within each hypergraph.

\smallsection{Consistent clusters across metrics.}
Hierarchical clustering based on the similarity matrices consistently partitions both node and hyperedge measures into a small number of stable groups across all metrics. 
For node centralities, two clear clusters consistently appear based on the silhouette score: (1) neighbor-degree, closeness, and harmonic centralities, and (2) betweenness and degree centralities. 
Within these clusters, the average similarities are notably higher than the overall averages.  
Specifically, the average Pearson and Spearman correlation coefficients reach 0.848 and 0.898 for Cluster~1, and 0.874 and 0.934 for Cluster~2, 
while the overall averages are 0.627 and 0.773, respectively.
Similarly, for the top 5\% of nodes, Jaccard similarities are 0.616 and 0.594, and top-$k$ overlap ratios are 0.742 and 0.745 within Cluster~1 and Cluster~2, respectively, while the overall averages are 0.443 and 0.568.
{A similar pattern is observed for hyperedge centralities, 
where the overall averages are 0.400 and 0.434 for the Pearson and Spearman rank correlations, and 0.271 and 0.355 for the Jaccard similarity and top-$k$ overlap ratio, respectively.}
Here, the two clusters are (1) closeness and harmonic centralities, and (2) degree and random-walk centralities.
In terms of average Pearson and Spearman correlation coefficients, Cluster~1 achieves 0.980 and 0.976, respectively, and Cluster~2 achieves 0.954 and 0.729, respectively. 
For Jaccard similarity and top-$k$ overlap ratio, Cluster~1 achieves 0.856 and 0.909, and Cluster~2 achieves 0.649 and 0.741, respectively. 

These results confirm that certain groups of measures exhibit strong internal agreement across multiple similarity metrics, reflecting shared measurement foundations (e.g., harmonic centrality is a variant of closeness centrality designed for disconnected hypergraphs). In contrast, other measures maintain distinct computational or conceptual perspectives.

\smallsection{Distinct singleton measures.}
While most measures form consistent clusters across metrics, certain measures behave as distinct singletons, exhibiting low similarity with all others. 
For nodes, eigenvector centrality consistently forms a singleton cluster, with average correlations to the other measures of 0.470 (Pearson correlation coefficient), 0.722 (Spearman rank correlation coefficient), 0.321 (Jaccard similarity), and 0.454 (top-$k$ overlap ratio). 
For hyperedges, hypercoreness acts as a singleton across all metrics, with corresponding averages of 0.136, 0.116, 0.066, and 0.107. 
These values are substantially lower than the overall averages (Pearson correlation coefficient 0.627 and 0.400, Spearman rank correlation coefficient 0.773 and 0.434, Jaccard similarity 0.443 and 0.271, and top-$k$ overlap ratio 0.568 and 0.355 for nodes and hyperedges, respectively). 

This divergence indicates that these singleton measures capture structural characteristics that are largely independent of those reflected by other measures. 
For example, hypercoreness identifies dense subhypergraph participation, offering unique perspectives on higher-order connectivity that are not well represented by conventional degree-, path-, or walk-based measures.

\subsection{Computation Time Analysis}
\label{sec:insight:runtime}

In this section, we analyze the empirical computation time of different measures described in~\cref{sec:insight:settings}. 
We measure the total computation time required to calculate the centrality and importance scores for all nodes or hyperedges, across the ten real-world hypergraphs. 
All experiments are performed under the same computational environment, and the results are compared across measures and datasets to assess scalability and practical feasibility. 
We present the results in~\cref{fig:runtime}, evaluating the computation time with respect to three structural quantities: (1) number of nodes, (2) number of hyperedges, and (3) sum of degrees (i.e., $\sum_{v \in \mathcal{V}}\text{deg}(v)$).
From the results, we derive three key take-home messages:

{
\captionsetup[sub]{justification=centering, singlelinecheck=false}
\captionsetup[sub]{aboveskip=2pt, belowskip=0pt}

\begin{figure}[t]
\centering
\setlength{\tabcolsep}{4pt} 
\begin{tabular}{@{} m{0.34\textwidth} m{0.34\textwidth} m{0.28\textwidth} @{}}

\subcaptionbox{Scalability w.r.t. node count \\
(node measures)\label{fig:runtime:node_node}}[0.32\textwidth]{%
  \centering\includegraphics[width=\linewidth]{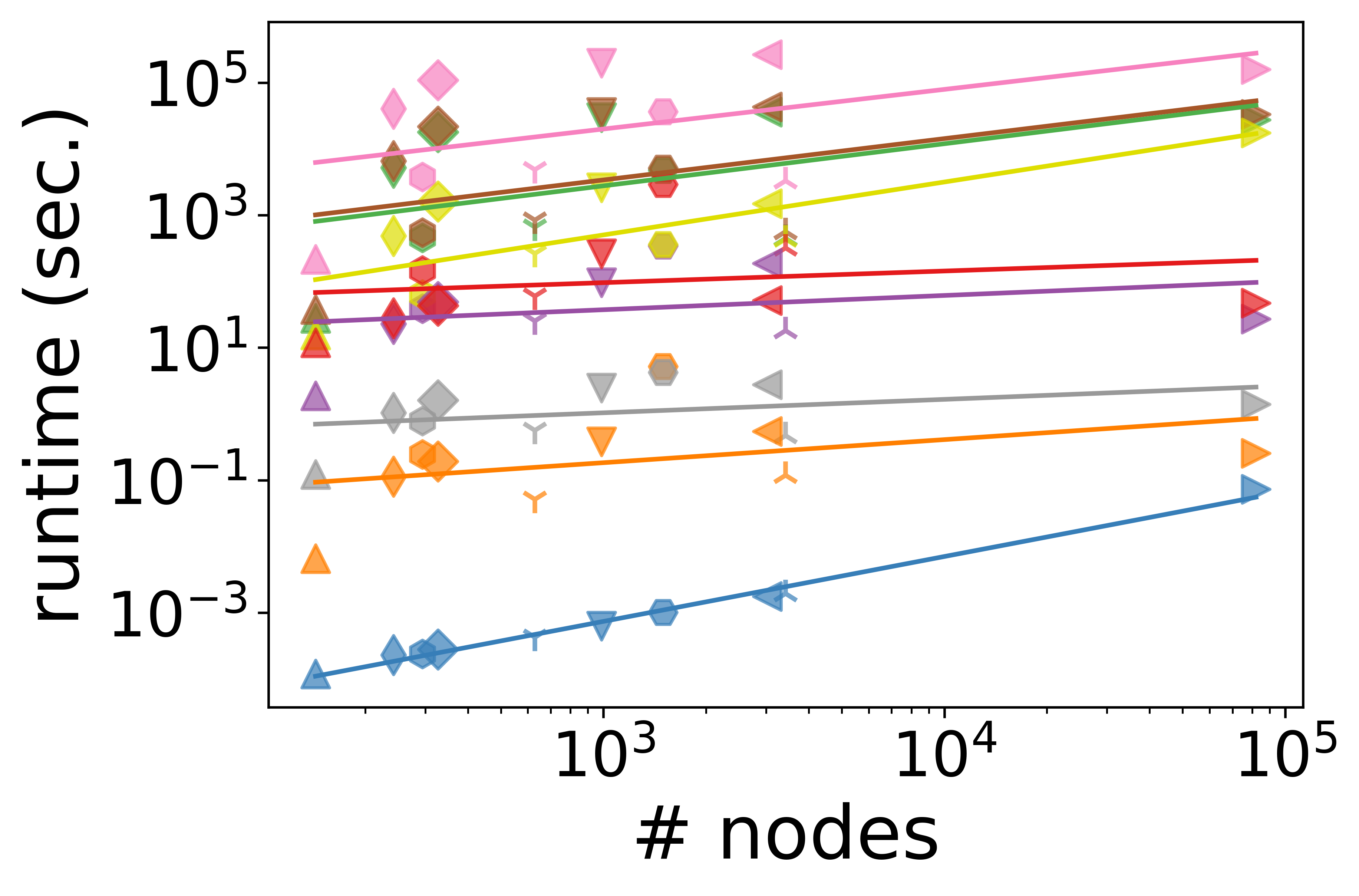}} &
\subcaptionbox{Scalability w.r.t. node count \\ (hyperedge measures)\label{fig:runtime:edge_node}}[0.32\textwidth]{%
  \centering\includegraphics[width=\linewidth]{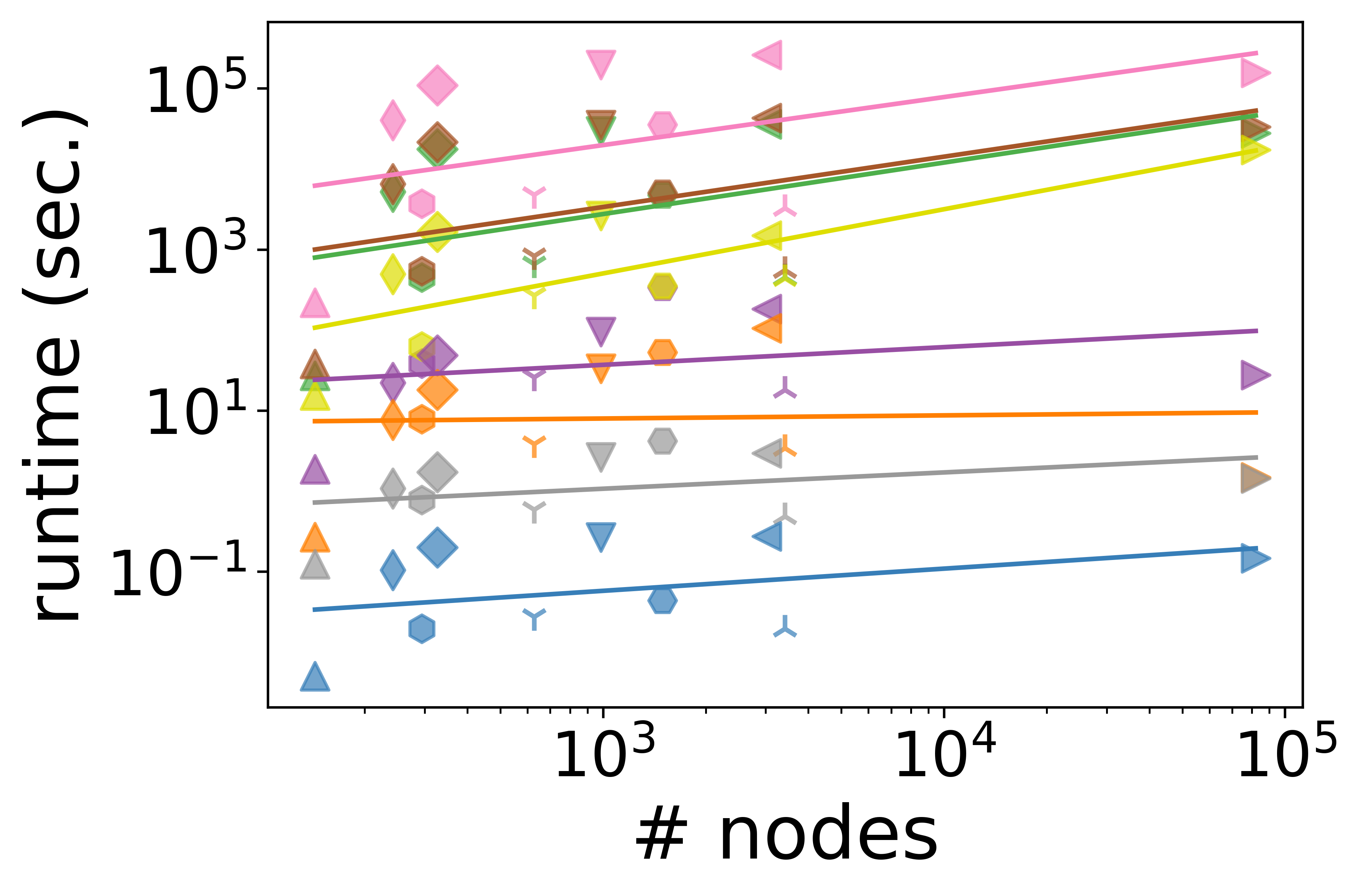}} &
\multirow{3}{*}[2.0em]{{%
  \includegraphics[width=0.85\linewidth]{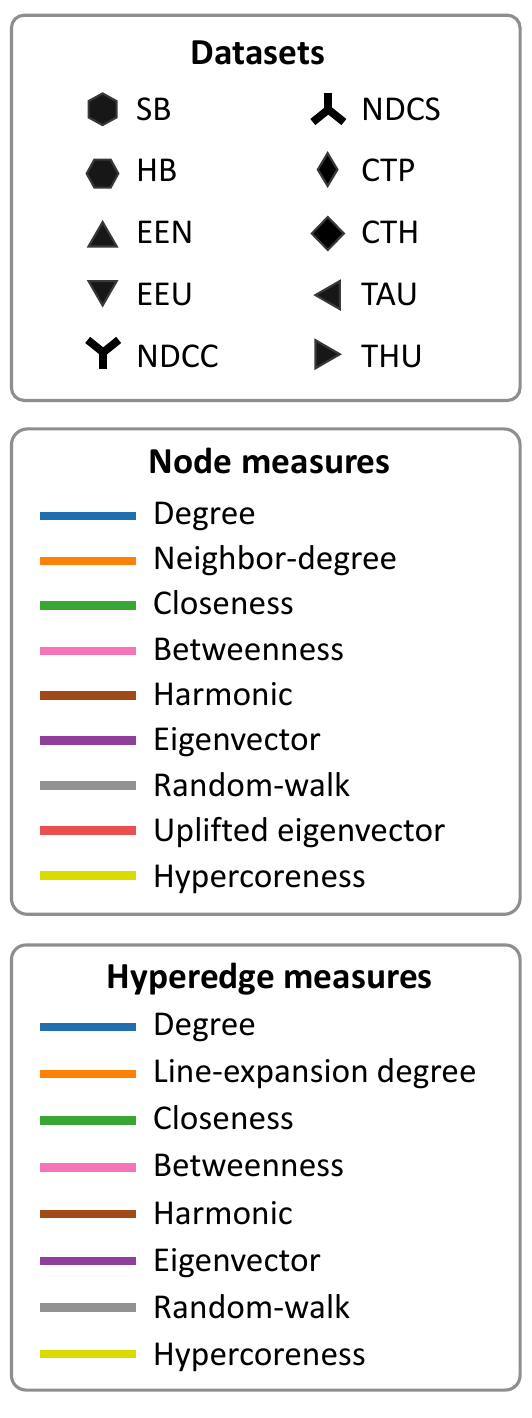}}}
\\[0.6em]

\subcaptionbox{Scalability w.r.t. hyperedge count \\ (node measures) \label{fig:runtime:node_edge}}[0.32\textwidth]{%
  \centering\includegraphics[width=\linewidth]{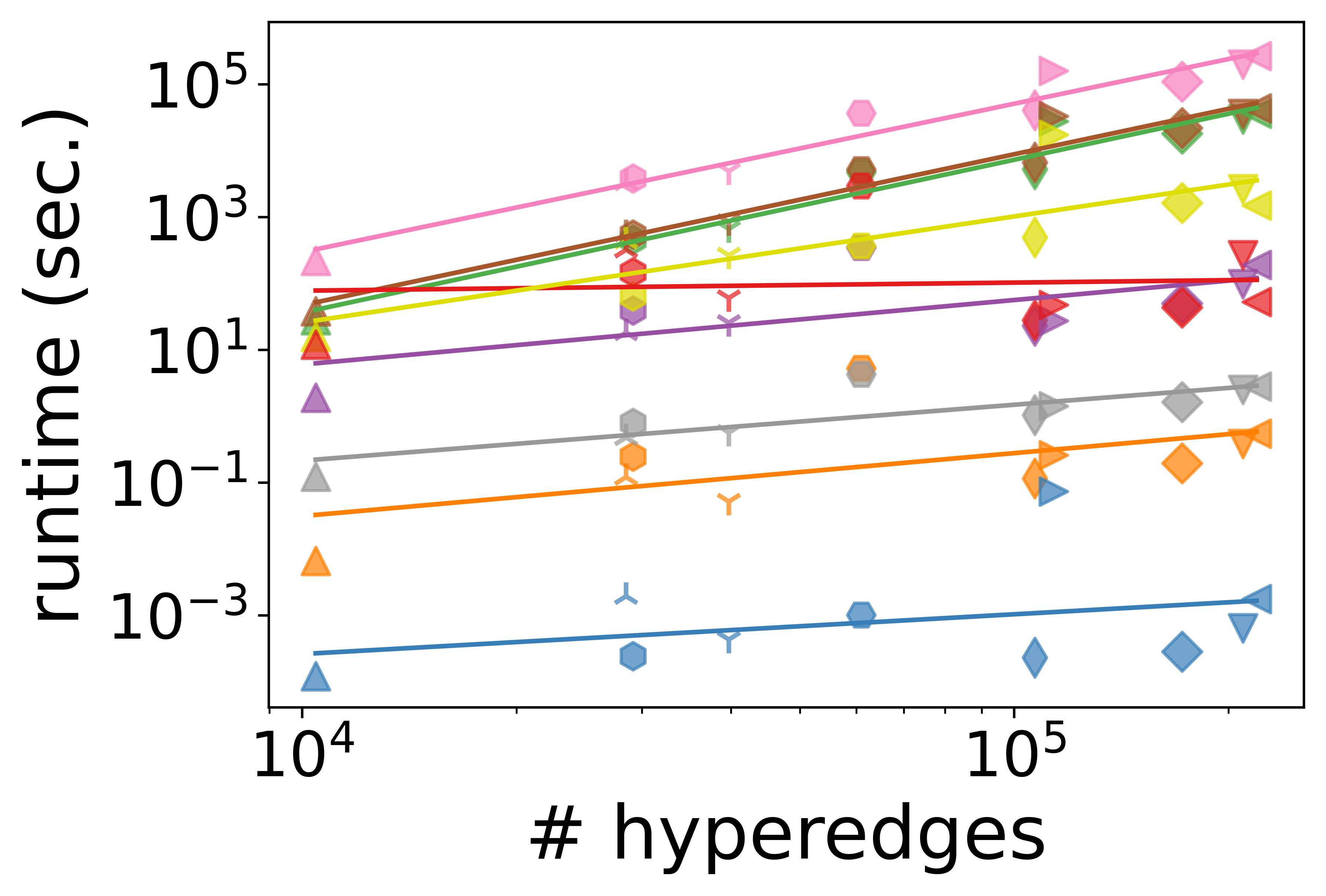}} &
\subcaptionbox{Scalability w.r.t. hyperedge count \\ (hyperedge measures)\label{fig:runtime:edge_edge}}[0.32\textwidth]{%
  \centering\includegraphics[width=\linewidth]{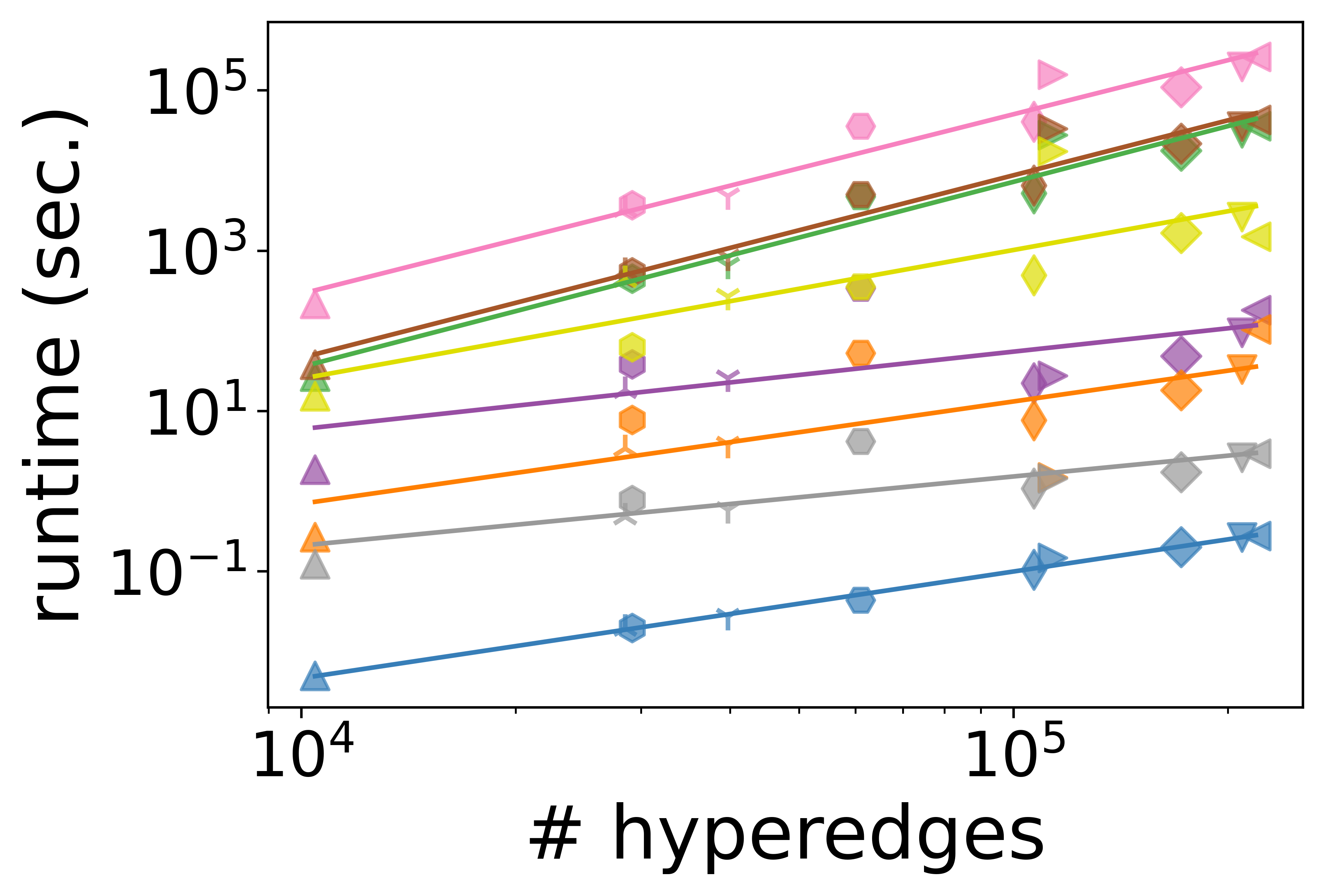}} &
\\[0.6em]

\subcaptionbox{Scalability w.r.t. degree sum \\ (node measures) \label{fig:runtime:node_degree}}[0.32\textwidth]{%
  \centering\includegraphics[width=\linewidth]{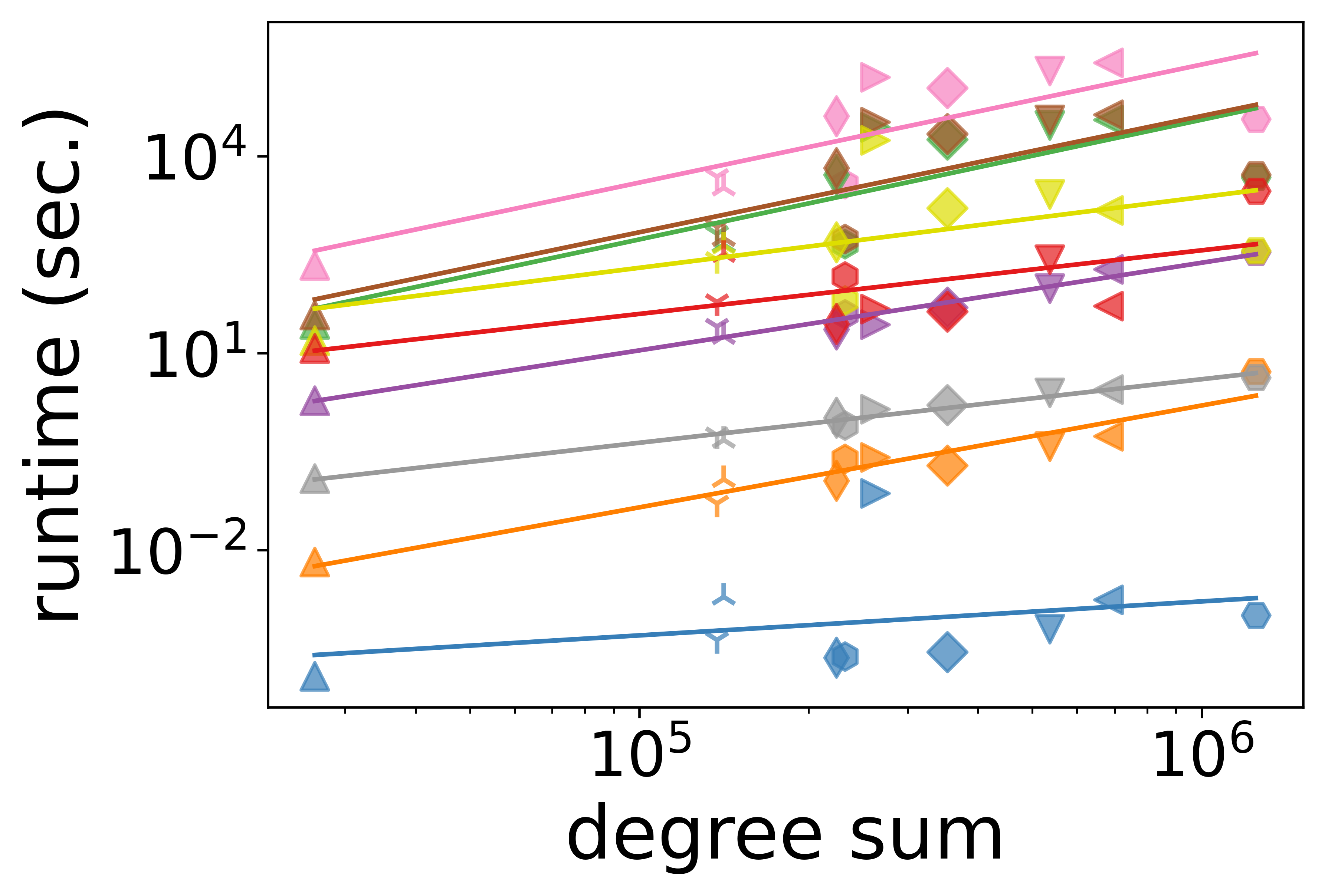}} &
\subcaptionbox{Scalability w.r.t. degree sum \\ (hyperedge measures)\label{fig:runtime:edge_degree}}[0.32\textwidth]{%
  \centering\includegraphics[width=\linewidth]{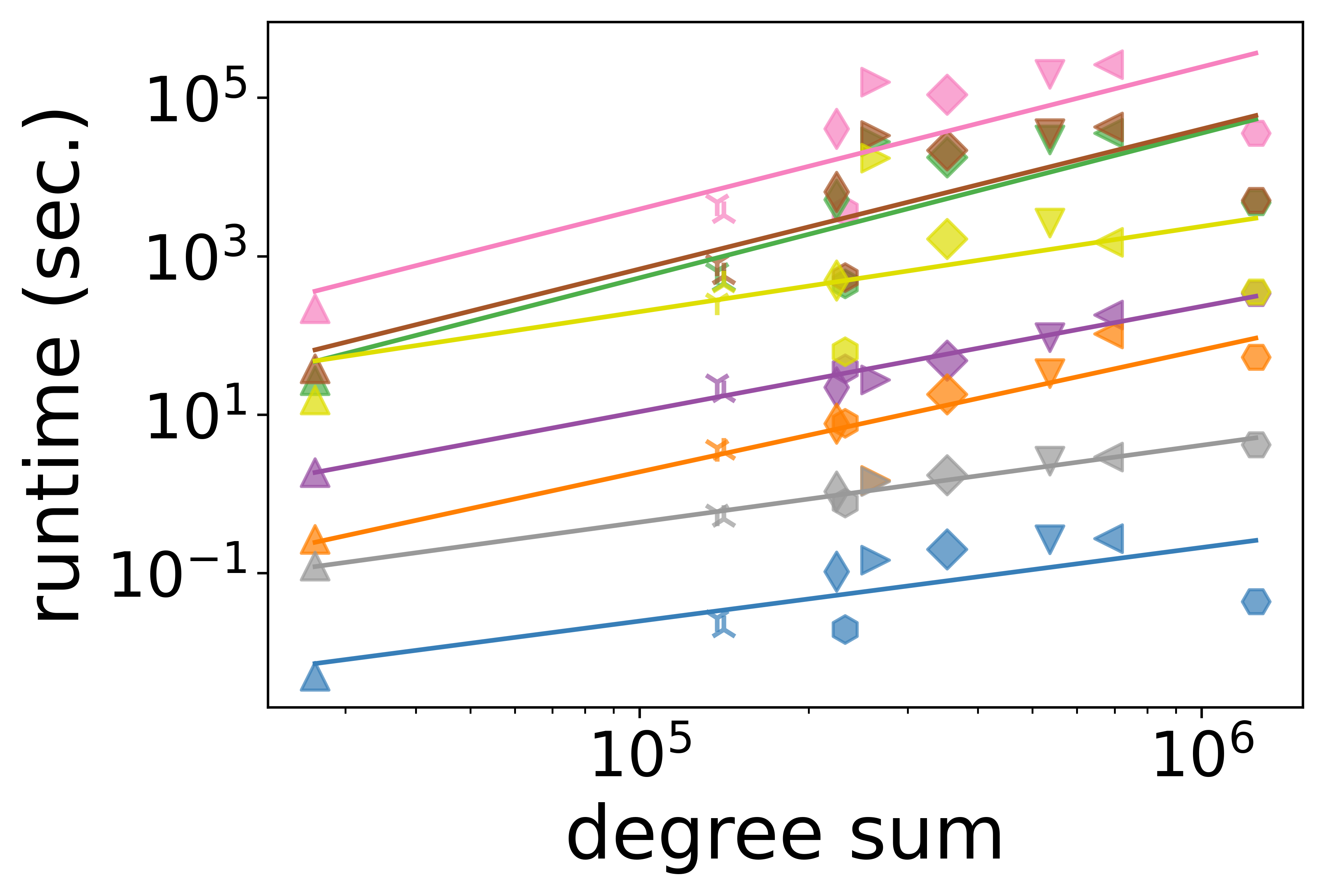}} &
\\
\end{tabular}
\vspace{-2mm}
\caption{Computation time of node and hyperedge measures with respect to different structural quantities on a log-log scale. 
Each subfigure reports the total computation time with respect to 
(\subref{fig:runtime:node_node} \& \subref{fig:runtime:edge_node}) the number of nodes, 
(\subref{fig:runtime:node_edge} \& \subref{fig:runtime:edge_edge}) the number of hyperedges, and 
(\subref{fig:runtime:node_degree} \& \subref{fig:runtime:edge_degree}) the total degree sum, 
for node and hyperedge measures, respectively.}
\label{fig:runtime}
\end{figure}
}
\smallsection{{Empirical computation time decreases from path-based to walk- and degree-based measures.}}
The computation time of a measure strongly depends on the structural element it utilizes, as categorized in our proposed taxonomy. 
In general, the computational cost in terms of time follows the order:
\[
\text{path-based} > \text{subhypergraph-based} > \{\text{walk-based \& degree-based}\}
\]
Moreover, based on the slopes of the regression lines in~\cref{fig:runtime}, we further examine the empirical scalability of each measure. 
For every measure, we compute the slope of the regression line between computation time and three structural quantities---the number of nodes, the number of hyperedges, and the degree sum---and take the largest slope among them as its representative scaling factor. 
We classify scalability into three categories according to this slope: \textit{sublinear} ($<1$), \textit{subquadratic} ($1 \leq \text{slope} < 2$), and \textit{superquadratic} ($\geq 2$). 
The resulting empirical scalability, grouped by the utilized structural element, is summarized as follows:
\begin{itemize}
    \item \textbf{Path-based (superquadratic).} 
    Path-based measures---closeness, betweenness, and harmonic centralities---exhibit the highest computational complexity due to their reliance on all-pairs or multi-step distance computations. 
    These measures show superquadratic scaling, indicating rapidly increasing computation time with hypergraph size.

    \item \textbf{Subhypergraph-based (subquadratic).} 
    The subhypergraph-based measure, hypercoreness, demonstrates moderately high computational cost, as it iteratively evaluates dense substructure membership. 
    Its scaling slope remains below~2, suggesting subquadratic growth.

    \item \textbf{Degree- and walk-based (sublinear to subquadratic).} 
    Among all measures, degree centrality is the most scalable, exhibiting sublinear growth on the log--log regression line. 
    Walk-based measures---eigenvector, uplifted eigenvector, and random-walk centralities---are more efficient than path- and subhypergraph-based measures but still require iterative convergence. 
    Empirically, they display subquadratic scaling, with slopes below approximately~1.4. 
    Other degree-based measures, including neighbor-degree and line-expansion degree centralities, also fall within the subquadratic range, with slopes around~1.6.
\end{itemize}

\smallsection{Subhypergraph- and path-based measures can become computationally infeasible at scale.}
While the previous analysis characterizes the empirical scalability of measures in terms of their asymptotic growth with hypergraph size, here we examine the practical implications in terms of absolute computation time. 
We focus on the largest dataset, \textit{tags-ask-ubuntu (TAU)}, which contains the greatest number of hyperedges among all datasets. 
In practice, degree- and walk-based measures, which exhibit sublinear to subquadratic scaling, complete computation within a few seconds to a few minutes even on TAU.
The subhypergraph-based measure, hypercoreness, requires moderately more computation, completing in approximately four hours. 
In contrast, path-based measures---closeness, betweenness, and harmonic centralities---display superquadratic scaling and require from ten hours to several days of computation on the same dataset. 
These observations suggest that for million- or billion-scale hypergraphs, subhypergraph- and path-based measures may become computationally infeasible.

\smallsection{Degree-based measures can serve as cheaper proxies for expensive measures.}
By jointly considering the similarity analysis in~\cref{sec:insight:correlation} and the computation time results in this section, we identify practical substitutes for computationally expensive measures. 
For node measures, closeness and harmonic centralities show strong correlation with neighbor-degree centrality, while the latter is at least 905~times faster across datasets (the minimum speedup is observed on the \textit{house-bills} dataset).
Similarly, betweenness centrality correlates well with degree centrality, with degree centrality being at least 1.7~million times faster across datasets (the minimum speedup is observed on \textit{ndc-substances}). 
Moreover, random-walk centrality correlates strongly with degree centrality while being more expensive, {taking at least 8.6 times longer to compute} across datasets (minimum observed on \textit{contact-high-school}). 
Overall, these findings demonstrate that degree-based measures---in particular, degree and neighbor-degree centralities---can serve as scalable, low-cost proxies for more complex path- and walk-based measures, offering a practical balance between interpretability and computational efficiency.

\section{Real-World Applications}\label{sec:applications}

Centrality and importance measures for hypergraphs provide powerful tools for analyzing systems in which interactions occur among groups rather than just pairs. Across diverse domains, such measures have been employed to identify influential individuals in social networks, key molecules in biological systems, and critical nodes in transportation and infrastructure networks. By quantifying how both nodes and hyperedges contribute to the organization, efficiency, and resilience of higher-order systems, hypergraph-based centrality and importance measures yield actionable insights that extend beyond the capabilities of traditional pairwise analysis.
In this section, we review representative applications across social, biological, and transportation domains, illustrating how hypergraph formulations uncover central structures in complex real-world systems.

\subsection{Social Networks}\label{sec:applications:social}

In social systems where individuals often engage in group interactions---such as event co-attendance, group chats, or multi-author collaborations---modeling the structure as a hypergraph enables a richer view of influence dynamics~\citep{xiao2024information}.
By applying centrality and importance measures designed for hypergraphs, one can better identify those individuals who are influential in group-based processes.
The appropriate measure depends on how influence operates: core-based measures capture repeated participation in cohesive groups, distance-based measures capture potential reach across groups, and perturbation-based measures capture the structural damage caused by removing an individual.

\citet{mancastroppa2023hyper} and \citet{bu2023hypercore} showed that nodes with high hypercoreness (see \cref{cent:core}) are effective seeds for spreading and consensus, as inner hypercores sustain repeated higher-order reinforcement.
More recently, \citet{zhang2025locating} used distance-based fuzzy centrality (see \cref{cent:fuzzy_collective_influence}) to capture both local proximity and diversified reach across higher-order distance layers.
In contrast, \citet{hu2023identifying} used higher-order Von Neumann entropy (see \cref{cent:higher_order_von_neumann}) to quantify the structural effects of hyperedge removal, aggregating hyperedge-level changes into node scores.
For lurker detection, \citet{amato2018centrality} considered degree (see \cref{cent:degree}), closeness (see \cref{cent:closeness}), betweenness (see \cref{cent:betweenness}), and neighborhood centralities (see \cref{cent:neighborhood}). These measures provide complementary signals of direct participation, global reach, brokerage, and local exposure, which help characterize silent users through their structural position even when they contribute little content.

\smallsection{Discussion.}
Contextual measures that consider external labels or features remain underexplored, while in many real-world social networks, e.g., social media platforms, side information from text, images, video, and profiles can complement topology by signaling credibility, affect, or topic alignment~\citep{poshtiban2023identification,amato2019hypergraph}.

\subsection{Biological Systems}\label{sec:applications:biological}
Many biological interactions are inherently polyadic: protein complexes involve several subunits, metabolic reactions require multiple substrates and enzymes, and gene regulation can depend on combinations of transcription factors and co-factors. Representing these as hypergraphs preserves the integrity of such multi-component interactions~\citep{murgas2022hypergraph}.
Hypergraph-based centrality and importance measures then help us spot which molecules play a key role in these multi-way systems, e.g., which protein's removal would cause major disruption, or which gene strongly influences a cellular response.
Accordingly, path-based measures are useful for finding molecules that mediate between biological contexts, whereas spectral measures are attractive when importance should be reinforced by connections to other important genes, proteins, complexes, or disease sets.

\citet{feng2021hypergraph} modeled genes as hyperedges linking the experimental conditions in which they were significantly perturbed by viral infection and used betweenness centrality (see \cref{cent:betweenness}) to highlight genes that bridge otherwise distinct perturbation patterns across viruses or conditions.
Likewise, \citet{barton2023hypergraphs} used line-graph-based degree (see \cref{cent:degree}), closeness (see \cref{cent:closeness}), betweenness (see \cref{cent:betweenness}), and eigenvector centralities (see \cref{cent:eigenvector}) to capture complementary aspects of gene salience, including feature overlap, proximity, brokerage, and association with other salient genes.
The line graph also makes these measures practical on a dense gene-expression hypergraph while retaining overlaps among multi-feature gene profiles.
Moreover, \citet{lawson2024application} applied nonlinear eigenvector centrality (see \cref{cent:node_edge_nonlinear}) to a yeast protein-complex hypergraph, using mutually reinforcing node--hyperedge scores to jointly rank proteins and complexes and examine the centrality--lethality relationship.
In a related biomedical application, \citet{rafferty2021ranking} used weighted eigenvector centrality (see \cref{cent:eigenvector}) for multimorbidity. 
Disease sets are ranked by their overlap with other well-connected sets while incorporating both disease prevalence and joint occurrence.

\smallsection{Discussion.}
In many biological systems, the interactions may have special structures, e.g., they may be directed or require exact ratios of participants~\citep{chang2024hypergraph}.
Ideal measures for biological systems should take these facts into consideration.

\subsection{Neuroscience Networks}\label{sec:applications:neuroscience}

The human brain is a complex system where cognitive functions---such as multisensory integration, decision-making, and high-level abstraction---emerge from the coordinated activity of multiple neural populations rather than simple pairwise communication. While traditional graph theory models the brain via dyadic links, this approach often fails to capture the synergy and redundancy inherent in neural computation~\citep{luppi2024quantifying}. Hypergraphs provide a rigorous framework for modeling these polyadic interactions, and the corresponding hypergraph centrality and importance measures offer novel insights into brain organization and pathology.
Here, measures that aggregate participation or reinforcement across hyperedges are especially relevant because the object of interest is a region's recurring role in coordinated sets of connections, rather than the strength of any single pairwise link.

\citet{gu2017functional} grouped functional connections that covaried across subjects into hyperedges and used degree-like incidence counts (see \cref{cent:degree}) to identify brain regions that repeatedly participate in functional cores, localized drivers, and connectors that region-based pairwise analyses can miss.
Furthermore, \citet{santos2023emergence} applied eigenvector centrality (see \cref{cent:eigenvector}) to 3-uniform hypergraphs constructed from multivariate interaction information. The spectral score identifies triplets embedded in mutually reinforcing higher-order connectivity and distinguishes synergistic integration hubs from redundant robustness hubs.

\smallsection{Discussion.}
While most current models are static, neural processing is fluid. 
{Future research should develop} dynamic weighted hypergraph approaches that can track centrality fluctuations in real time~\citep{zheng2024rich}.

\subsection{Transportation Networks}\label{sec:applications:transportation}

Transportation systems---including maritime shipping, rail networks, and multimodal transit---often involve group-wise interactions (e.g., multiple vessels calling at a hub port, multi-stop high-speed-rail services, or multimodal transfer hubs) that are naturally represented by hypergraphs~\citep{prakash2017finding,harrod2011modeling}.
Hypergraph-based centrality and importance measures then help identify critical nodes or hubs whose disruption would significantly impair system connectivity and flow efficiency.
The measures have direct operational interpretations: degree and hyperdegree describe the number of services or routes available at a hub, closeness describes network-wide accessibility, and betweenness describes dependence on a hub for transfers, transshipments, or other intermediate movements.

\citet{tocchi2022hypergraph} used hyperpath-aware betweenness (see~\cref{cent:betweenness}) to quantify how strongly ports mediate alternative service sequences and transshipment routes in worldwide maritime container networks.
Likewise, \citet{yin2026hypergraph} combined degree and hyperdegree (see~\cref{cent:degree}), betweenness (see~\cref{cent:betweenness}), closeness (see~\cref{cent:closeness}), and a neighbor-sensitive weighted hyperdegree for the Chinese high-speed rail system. Their combination captures complementary aspects of station importance, including route volume, transfer brokerage, global accessibility, and interactions with nearby hubs.
Moreover, \citet{yin2025resilience} used betweenness (see~\cref{cent:betweenness}) in a multimodal highway--railway--aviation hypergraph. Hubs lying on many cross-modal paths are natural points of transfer dependence, since their failure could disconnect transportation modes or regions.

\smallsection{Discussion.}
{Many transportation networks are inherently spatial and temporal~\citep{wang2023gmhann}.} Ideal measures for transportation networks should be able to incorporate both spatial and temporal information.

\subsection{Broader Discussions}

Across these domains, the applications reveal that \textit{centrality} or \textit{importance} is not a single, monolithic concept but is highly problem-dependent.
A useful selection principle is therefore to match the measure to the mechanism of interest: degree for direct participation, closeness for accessibility, betweenness for brokerage, hypercoreness for cohesive reinforcement, eigenvector-based measures for mutual reinforcement, and perturbation-based measures for vulnerability.
A clear trend has emerged toward shifting from purely structural measures to learned, task-specific ones. In many state-of-the-art applications---particularly in data-rich areas such as traffic forecasting~\citep{luo2022directed}, recommender systems~\citep{liu2025aspect}, and logistics (e.g., resource allocation and task scheduling)~\citep{singh2025ranking}---learning-based methods, especially hypergraph neural networks, have become increasingly prevalent. We expect this evolution to continue, with more contextual measures being used in real-world scenarios.

\color{black}

\section{Conclusions and Future Directions}\label{sec:conclusion}

In this survey, we systematized 39 hypergraph centrality and importance measures (see \cref{fig:taxonomy_elements}) into a unified, three-way taxonomy: \emph{structural} (\cref{sec:measures:structural}), \emph{functional} (\cref{sec:measures:functional}), and \emph{contextual} (\cref{sec:measures:contextual}) measures.
Beyond this taxonomy, we provided an empirical comparison of their similarities and computation time characteristics (\cref{sec:insight}), and reviewed representative applications across social, biological, and transportation domains (\cref{sec:applications}).
This integrated view is intended to help researchers and practitioners select measures that align with their modeling assumptions, and to reveal principled avenues for new designs. 
Below, we outline several concrete opportunities that emerge from this survey.

\smallsection{Axiomatic foundations.} 
There have been efforts to develop axioms for centrality and importance measures on pairwise graphs~\citep{boldi2014axioms,bloch2023centrality}. Researchers may also develop axioms for measures on hypergraphs. 
Axiomatic foundations would be helpful for clarifying the fundamental principles that a measure should satisfy, enabling formal comparisons among different measures, and guiding the design of new ones that adhere to desirable properties, bridging theoretical rigor and practical interpretability.

\smallsection{Unified framework for paths and distances.}
As discussed in \cref{sec:prelim:connectivity,sec:measures:structural:path}, the notions of paths and distances have been generalized in various ways on hypergraphs~\citep{vasilyeva2023distances,preti2024hyper}, which gives rise to different path-based measures.
Researchers may try to provide a meta-framework that parameterizes and unifies different choices of definitions of paths and distances.
This would allow practitioners to systematically select or tune the most suitable notion for their task, compare measures under a common lens, and analyze how different modeling assumptions (e.g., transition rules, walk constraints, or hyperedge weights) influence centrality outcomes, facilitating fair evaluations and theoretical connections among diverse path-based measures.

\smallsection{Truly non-uniform spectral measures.} 
Although \citet{contreras2023uplifting} have proposed uplifted eigenvector centrality (see \cref{cent:genralized_eigenvector}) that enables spectral analysis on non-uniform hypergraphs by introducing auxiliary nodes to achieve uniformity, this approach effectively reduces the problem to a uniform case rather than handling non-uniformity natively.
Researchers may aim to develop truly non-uniform spectral operators that directly incorporate variable hyperedge sizes into the spectral definition itself, ensuring well-posedness, convergence guarantees, and scalable solvers for joint node-hyperedge importance estimation.

\smallsection{Measures for generalized hypergraphs.}
Real-world systems are often better modeled as generalized hypergraphs, e.g., temporal~\citep{fischer2020visual,lee2023temporal}, directed~\citep{ausiello2017directed,moon2023four}, and multilayer~\citep{wang2024identification,qian2025modeling} ones.
Researchers may extend and propose measures for such generalized settings.
In turn, such measures would allow practitioners to better capture the multi-aspect dynamics of real-world systems---from time-sensitive coordination in temporal networks to interdependent processes in multilayer infrastructures---broadening both the explanatory power and practical applicability of hypergraph-based analysis.

\smallsection{Fast approximation for expensive measures.} 
Based on our empirical analyses in \cref{sec:insight}, some measures (e.g., path-based measures) are computationally expensive in practice.
While several studies have proposed efficient approximation techniques for analogous measures on pairwise graphs~\citep{riondato2014fast,riondato2018abra,alghamdi2017benchmark}, similar efforts for hypergraphs remain limited.
Future research may design approximation schemes that preserve accuracy guarantees while drastically reducing computation time and memory costs. Leveraging hardware acceleration, such as GPUs, and distributed computing frameworks~\citep{katz2008all,djidjev2014efficient} can further scale these methods to massive real-world datasets.
Such advances would make {computationally expensive} measures more accessible in practice, enabling their integration into real-time analytics, dynamic monitoring, and large-scale machine learning pipelines.

\smallsection{Standardized and comprehensive benchmarking.} 
Our empirical analyses in \cref{sec:insight} provide initial insights into the behaviors and relationships among various measures, but they remain limited in scope and preliminary in scale.
Similar benchmarking efforts have been actively conducted for pairwise graphs~\citep{saqr2022curious,iacobucci2017social}.
Future research may pursue more standardized and comprehensive benchmarking efforts that systematically evaluate a wide range of measures on hypergraphs across diverse datasets, domains, and hypergraph structures.
Such benchmarking frameworks---ideally supported by open repositories, unified implementations, and reproducible protocols---would enable fairer comparisons, reveal domain-dependent strengths and weaknesses, and ultimately guide the development of more robust and practically useful centrality and importance measures.

\color{black}

\begin{acks}
This work was partly supported by the National Research Foundation of Korea (NRF) grant funded by the Korea government (MSIT) (No. RS-2024-00406985).
This work was partly supported by Institute of Information \& Communications Technology Planning \& Evaluation (IITP) grant funded by the Korea government (MSIT) (No. RS-2024-00438638, EntireDB2AI: Foundations and Software for Comprehensive Deep Representation Learning and Prediction on Entire Relational Databases) (No. RS-2019-II190075, Artificial Intelligence Graduate School Program (KAIST)).
\end{acks}

\bibliographystyle{ACM-Reference-Format}
\bibliography{reference}

@String{Computing = "Computing" }

@String{Computer = "{IEEE} Computer" }

@String{Psychometrika = "Psychometrika" }

@String{Springer = "Springer-Verlag" }

@inproceedings{sun2008hypergraph,
  title={Hypergraph spectral learning for multi-label classification},
  author={Sun, Liang and Ji, Shuiwang and Ye, Jieping},
  booktitle={KDD},
  year={2008}
}

@article{zien2002multilevel,
  title={Multilevel spectral hypergraph partitioning with arbitrary vertex sizes},
  author={Zien, Jason Y and Schlag, Martine DF and Chan, Pak K},
  journal={IEEE Transactions on computer-aided design of integrated circuits and systems},
  volume={18},
  number={9},
  pages={1389--1399},
  year={1999},
  publisher={IEEE}
}

@inproceedings{inariba2017random,
  title={Random-radius ball method for estimating closeness centrality},
  author={Inariba, Wataru and Akiba, Takuya and Yoshida, Yuichi},
  booktitle={AAAI},
  year={2017}
}

@article{dangalchev2006residual,
  title={Residual closeness in networks},
  author={Dangalchev, Chavdar},
  journal={Physica A: Statistical Mechanics and its Applications},
  volume={365},
  number={2},
  pages={556--564},
  year={2006},
  publisher={Elsevier}
}

@incollection{whitney1992congruent,
  title={Congruent graphs and the connectivity of graphs},
  author={Whitney, Hassler},
  booktitle={Hassler Whitney Collected Papers},
  pages={61--79},
  year={1992},
  publisher={Springer}
}

@article{bermond1977line,
  title={Line graphs of hypergraphs I},
  author={Bermond, Jean-Claude and Heydemann, Marie-Claude and Sotteau, Dominique},
  journal={Discrete Mathematics},
  volume={18},
  number={3},
  pages={235--241},
  year={1977},
  publisher={Elsevier}
}

@article{boldi2014axioms,
  title={Axioms for centrality},
  author={Boldi, Paolo and Vigna, Sebastiano},
  journal={Internet Mathematics},
  volume={10},
  number={3-4},
  pages={222--262},
  year={2014},
  publisher={Taylor \& Francis}
}

@article{faust1997centrality,
  title={Centrality in affiliation networks},
  author={Faust, Katherine},
  journal={Social Networks},
  volume={19},
  number={2},
  pages={157--191},
  year={1997},
  publisher={Elsevier}
}

@incollection{koschutzki2005centrality,
  title={Centrality indices},
  author={Kosch{\"u}tzki, Dirk and Lehmann, Katharina Anna and Peeters, Leon and Richter, Stefan and Tenfelde-Podehl, Dagmar and Zlotowski, Oliver},
  booktitle={Network analysis: methodological foundations},
  pages={16--61},
  year={2005},
  publisher={Springer}
}

@article{saxena2020centrality,
  title={Centrality measures in complex networks: A survey},
  author={Saxena, Akrati and Iyengar, Sudarshan},
  journal={arXiv preprint arXiv:2011.07190},
  year={2020}
}

@article{bavelas1948mathematical,
  title={A mathematical model for group structures},
  author={Bavelas, Alex},
  journal={Human Organization},
  volume={7},
  number={3},
  pages={16--30},
  year={1948},
  publisher={Taylor \& Francis}
}

@article{shaw1954group,
  title={Group structure and the behavior of individuals in small groups},
  author={Shaw, Marvin E},
  journal={The Journal of Psychology},
  volume={38},
  number={1},
  pages={139--149},
  year={1954},
  publisher={Taylor \& Francis}
}

@article{estrada2006subgraph,
  title={Subgraph centrality and clustering in complex hyper-networks},
  author={Estrada, Ernesto and Rodr{\'\i}guez-Vel{\'a}zquez, Juan A},
  journal={Physica A: Statistical Mechanics and its Applications},
  volume={364},
  pages={581--594},
  year={2006},
  publisher={Elsevier}
}

@article{amato2018centrality,
  title={Centrality in heterogeneous social networks for lurkers detection: An approach based on hypergraphs},
  author={Amato, Flora and Moscato, Vincenzo and Picariello, Antonio and Piccialli, Francesco and Sperl{\'\i}, Giancarlo},
  journal={Concurrency and Computation: Practice and Experience},
  volume={30},
  number={3},
  pages={e4188},
  year={2018},
  publisher={Wiley Online Library}
}

@inproceedings{yin2017local,
  title={Local higher-order graph clustering},
  author={Yin, Hao and Benson, Austin R and Leskovec, Jure and Gleich, David F},
  booktitle={KDD},
  year={2017}
}

@article{xiao2016node,
  title={A node Importance Measuring Method Based on Hypernetwork},
  author={Xiao, Quan and Yang, Fangli and Luo, Song and Zhang, Lihong and Zhao, Hua and Shu, Wei},
  journal={International Journal of Future Generation Communication and Networking},
  volume={9},
  number={12},
  pages={187--196},
  year={2016}
}

@article{xiao2013research,
  title={A Method for Measuring Node Importance in Hypernetwork Model},
  author={Xiao, Quan},
  journal={Research Journal of Applied Sciences, Engineering and Technology},
  volume={5},
  number={2},
  pages={568--573},
  year={2013}
}

@article{aksoy2020hypernetwork,
  title={Hypernetwork science via high-order hypergraph walks},
  author={Aksoy, Sinan G and Joslyn, Cliff and Marrero, Carlos Ortiz and Praggastis, Brenda and Purvine, Emilie},
  journal={EPJ Data Science},
  volume={9},
  number={1},
  pages={16},
  year={2020},
  publisher={Springer Berlin Heidelberg}
}

@article{chodrow2021generative,
  title={Generative hypergraph clustering: From blockmodels to modularity},
  author={Chodrow, Philip S and Veldt, Nate and Benson, Austin R},
  journal={Science Advances},
  volume={7},
  number={28},
  pages={eabh1303},
  year={2021},
  publisher={American Association for the Advancement of Science}
}

@article{veldt2023combinatorial,
  title={Combinatorial characterizations and impossibilities for higher-order homophily},
  author={Veldt, Nate and Benson, Austin R and Kleinberg, Jon},
  journal={Science Advances},
  volume={9},
  number={1},
  pages={eabq3200},
  year={2023},
  publisher={American Association for the Advancement of Science}
}

@article{benson2019three,
  title={Three hypergraph eigenvector centralities},
  author={Benson, Austin R},
  journal={SIAM Journal on Mathematics of Data Science},
  volume={1},
  number={2},
  pages={293--312},
  year={2019},
  publisher={SIAM}
}

@article{tudisco2021node,
  title={Node and edge nonlinear eigenvector centrality for hypergraphs},
  author={Tudisco, Francesco and Higham, Desmond J},
  journal={Communications Physics},
  volume={4},
  number={1},
  pages={201},
  year={2021},
  publisher={Nature Publishing Group UK London}
}

@inproceedings{chitra2019random,
  title={Random walks on hypergraphs with edge-dependent vertex weights},
  author={Chitra, Uthsav and Raphael, Benjamin},
  booktitle={ICML},
  year={2019}
}

@article{kovalenko2022vector,
  title={Vector centrality in hypergraphs},
  author={Kovalenko, Kirill and Romance, Miguel and Vasilyeva, Ekaterina and Aleja, David and Criado, Regino and Musatov, Daniil and Raigorodskii, Andrei M and Flores, Julio and Samoylenko, Ivan and Alfaro-Bittner, Karin and others},
  journal={Chaos, Solitons \& Fractals},
  volume={162},
  pages={112397},
  year={2022},
  publisher={Elsevier}
}

@article{bu2023hypercore,
  title={Hypercore decomposition for non-fragile hyperedges: concepts, algorithms, observations, and applications},
  author={Bu, Fanchen and Lee, Geon and Shin, Kijung},
  journal={Data Mining and Knowledge Discovery},
  volume={37},
  number={6},
  pages={2389--2437},
  year={2023},
  publisher={Springer}
}

@article{puzis2013augmented,
  title={Augmented betweenness centrality for environmentally aware traffic monitoring in transportation networks},
  author={Puzis, Rami and Altshuler, Yaniv and Elovici, Yuval and Bekhor, Shlomo and Shiftan, Yoram and Pentland, Alex},
  journal={Journal of Intelligent Transportation Systems},
  volume={17},
  number={1},
  pages={91--105},
  year={2013},
  publisher={Taylor \& Francis}
}

@article{hu2023identifying,
  title={Identifying vital nodes in hypergraphs based on Von Neumann entropy},
  author={Hu, Feng and Tian, Kuo and Zhang, Zi-Ke},
  journal={Entropy},
  volume={25},
  number={9},
  pages={1263},
  year={2023},
  publisher={MDPI}
}

@article{zhang2025locating,
  title={Locating influential nodes in hypergraphs via fuzzy collective influence},
  author={Zhang, Su-Su and Yu, Xiaoyan and Sun, Gui-Quan and Liu, Chuang and Zhan, Xiu-Xiu},
  journal={Communications in Nonlinear Science and Numerical Simulation},
  volume={142},
  pages={108574},
  year={2025},
  publisher={Elsevier}
}

@article{piao2025identifying,
  title={Identifying important nodes of hypergraph: An improved PageRank algorithm},
  author={Piao, Yu-Hao and Wang, Jun-Yi and Li, Ke-Zan},
  journal={Chinese Physics B},
  year={2025},
  month = {apr},
publisher = {Chinese Physical Society and IOP Publishing Ltd},
volume = {34},
number = {4},
pages = {048902},
}

@inproceedings{liu2024influential,
  title={Influential-nodes Identification in Hypergraphs Based on Degree-heterogeneous Hierarchical Spherical Algorithm},
  author={Liu, Daxi and Zhang, Su-Su and Liu, Chuang and Pei, Xin and Zhan, Xiu-Xiu},
  booktitle={SocialMeta},
  year={2024}
}

@inproceedings{li2025hypergraph,
  title={When hypergraph meets heterophily: New benchmark datasets and baseline},
  author={Li, Ming and Gu, Yongchun and Wang, Yi and Fang, Yujie and Bai, Lu and Zhuang, Xiaosheng and Lio, Pietro},
  booktitle={AAAI},
  year={2025}
}

@inproceedings{tejaswi2024identifying,
  title={Identifying Influential Nodes in Hypergraph Using Isolating Centrality},
  author={Tejaswi, Anupoju and Enduri, Murali Krishna and Tokala, Srilatha},
  booktitle={BTS-I2C},
  year={2024}
}

@article{wang2024identification,
  title={Identification of important nodes in multi-layer hypergraphs based on fuzzy gravity model and node centrality distribution characteristics},
  author={Wang, Peng and Ling, Guang and Zhao, Pei and Pan, Wenqiu and Ge, Ming-Feng},
  journal={Chaos, Solitons \& Fractals},
  volume={188},
  pages={115503},
  year={2024},
  publisher={Elsevier}
}

@article{xie2023vital,
  title={Vital node identification in hypergraphs via gravity model},
  author={Xie, Xiaowen and Zhan, Xiuxiu and Zhang, Zike and Liu, Chuang},
  journal={Chaos: An Interdisciplinary Journal of Nonlinear Science},
  volume={33},
  number={1},
  year={2023},
  publisher={AIP Publishing}
}

@article{dubey2025influential,
  title={Influential nodes in ray cluster hypergraph networks},
  author={Dubey, Vivek Kumar and Samanta, Sovan},
  journal={Expert Systems with Applications},
  volume={275},
  pages={127014},
  year={2025},
  publisher={Elsevier}
}

@article{vasilyeva2024matrix,
  title={Matrix centrality for annotated hypergraphs},
  author={Vasilyeva, E and Samoylenko, I and Kovalenko, K and Musatov, D and Raigorodskii, AM and Boccaletti, S},
  journal={Chaos, Solitons \& Fractals},
  volume={186},
  pages={115256},
  year={2024},
  publisher={Elsevier}
}

@article{contreras2023uplifting,
  title = {Uplifting edges in higher-order networks: Spectral centralities for non-uniform hypergraphs},
journal = {AIMS Mathematics},
volume = {9},
number = {11},
pages = {32045--32075},
year = {2024},
issn = {2473-6988},
author = {Gonzalo Contreras-Aso and Cristian Pérez-Corral and Miguel Romance},
}

@inproceedings{chen2024hyperedge,
  title={Hyperedge Importance Estimation via Identity-aware Hypergraph Attention Network},
  author={Chen, Yin and Wang, Xiaoyang and Chen, Chen},
  booktitle={CIKM},
  year={2024}
}

@inproceedings{chen2023identifying,
  title={Identifying Vital Nodes in Hypernetworks Based on Improved PageRank Algorithm and Information Entropy},
  author={Chen, Junjie and Wei, Liang and Li, Pengyue and Ding, Haiping and Li, Faxu and Wang, Defang},
  booktitle={ICIVIS},
  year={2023}
}

@inproceedings{li2023important,
  title={Important hyperedge sorting method based on grounded Laplacian matrix},
  author={Li, Mingda and Wu, Yinghan and Bai, Libing and Hu, Feng},
  booktitle={ICCAIS},
  year={2023}
}

@article{huang2024clique,
  title={Clique based centrality measure in hypergraphs},
  author={Huang, Ruolin and Tur, Anna},
  journal={Contributions to Game Theory and Management},
  volume={17},
  pages={25--37},
  year={2024}  
}

@article{lee2024hypergraph,
  title={Hypergraph motifs and their extensions beyond binary},
  author={Lee, Geon and Yoon, Seokbum and Ko, Jihoon and Kim, Hyunju and Shin, Kijung},
  journal={The VLDB Journal},
  volume={33},
  number={3},
  pages={625--665},
  year={2024},
  publisher={Springer}
}

@article{kim2025estimating,
  title={Estimating simplet counts via sampling},
  author={Kim, Hyunju and Moon, Heechan and Bu, Fanchen and Ko, Jihoon and Shin, Kijung},
  journal={The VLDB Journal},
  volume={34},
  eid={15},
  pages={1--26},
  year={2025},
  publisher={Springer}
}

@article{aksoy2024scalable,
  title={Scalable Tensor Methods for Nonuniform Hypergraphs},
  author={Aksoy, Sinan G and Amburg, Ilya and Young, Stephen J},
  journal={SIAM Journal on Mathematics of Data Science},
  volume={6},
  number={2},
  pages={481--503},
  year={2024},
  publisher={SIAM}
}

@inproceedings{brandes2005centrality,
  title={Centrality measures based on current flow},
  author={Brandes, Ulrik and Fleischer, Daniel},
  booktitle={Annual symposium on theoretical aspects of computer science},
  pages={533--544},
  year={2005},
  organization={Springer}
}

@article{torres2021and,
  title={The why, how, and when of representations for complex systems},
  author={Torres, Leo and Blevins, Ann S and Bassett, Danielle and Eliassi-Rad, Tina},
  journal={SIAM Review},
  volume={63},
  number={3},
  pages={435--485},
  year={2021},
  publisher={SIAM}
}

@article{klamt2009hypergraphs,
  title={Hypergraphs and cellular networks},
  author={Klamt, Steffen and Haus, Utz-Uwe and Theis, Fabian},
  journal={PLoS Computational Biology},
  volume={5},
  number={5},
  pages={e1000385},
  year={2009},
  publisher={Public Library of Science San Francisco, USA}
}

@article{klimm2021hypergraphs,
  title={Hypergraphs for predicting essential genes using multiprotein complex data},
  author={Klimm, Florian and Deane, Charlotte M and Reinert, Gesine},
  journal={Journal of Complex Networks},
  volume={9},
  number={2},
  pages={cnaa028},
  year={2021},
  publisher={Oxford University Press}
}

@article{miyashita2025clustering,
  title={Clustering coefficient reflecting pairwise relationships within hyperedges},
  author={Miyashita, Rikuya and Hironaka, Shiori and Shudo, Kazuyuki},
  journal={Scientific Reports},
  volume={15},
  number={1},
  pages={20729},
  year={2025},
  publisher={Nature Publishing Group UK London}
}

@inproceedings{roy2015measuring,
  title={Measuring network centrality using hypergraphs},
  author={Roy, Sanjukta and Ravindran, Balaraman},
  booktitle={CODS},
  year={2015}
}

@inproceedings{kapoor2013weighted,
  title={Weighted node degree centrality for hypergraphs},
  author={Kapoor, Komal and Sharma, Dhruv and Srivastava, Jaideep},
  booktitle={IEEE Network Science Workshop},
  year={2013}
}

@article{benson2018simplicial,
  title={Simplicial closure and higher-order link prediction},
  author={Benson, Austin R and Abebe, Rediet and Schaub, Michael T and Jadbabaie, Ali and Kleinberg, Jon},
  journal={Proceedings of the National Academy of Sciences},
  volume={115},
  number={48},
  pages={E11221--E11230},
  year={2018},
  publisher={National Academy of Sciences}
}

@inproceedings{kim2023transitive,
  title={How transitive are real-world group interactions?-measurement and reproduction},
  author={Kim, Sunwoo and Bu, Fanchen and Choe, Minyoung and Yoo, Jaemin and Shin, Kijung},
  booktitle={KDD},
  year={2023}
}

@techreport{page1999pagerank,
  title={The PageRank citation ranking: Bringing order to the web.},
  author={Page, Lawrence and Brin, Sergey and Motwani, Rajeev and Winograd, Terry},
  year={1999},
  institution={Stanford infolab}
}

@article{seidman1983network,
  title={Network structure and minimum degree},
  author={Seidman, Stephen B},
  journal={Social Networks},
  volume={5},
  number={3},
  pages={269--287},
  year={1983},
  publisher={Elsevier}
}

@article{katz1953new,
  title={A new status index derived from sociometric analysis},
  author={Katz, Leo},
  journal={Psychometrika},
  volume={18},
  number={1},
  pages={39--43},
  year={1953},
  publisher={Springer-Verlag}
}

@article{seeley1949net,
  title={The net of reciprocal influence. A problem in treating sociometric data},
  author={Seeley, John R},
  journal={Canadian Journal of Psychology},
  volume={3},
  number={4},
  pages={234--240},
  year={1949},
  publisher={Canadian Psychological Association}
}

@article{freeman1977set,
  title={A set of measures of centrality based on betweenness},
  author={Freeman, Linton C},
  journal={Sociometry},
  pages={35--41},
  year={1977},
  volume={40},
  number={1},
  publisher={JSTOR}
}

@article{freeman1978centrality,
  title={Centrality in social networks conceptual clarification},
  author={Freeman, Linton C},
  journal={Social Networks},
  volume={1},
  number={3},
  pages={215--239},
  year={1978},
  publisher={North-Holland}
}

@article{zhou2011properties,
  title={Properties of metabolic graphs: biological organization or representation artifacts?},
  author={Zhou, Wanding and Nakhleh, Luay},
  journal={BMC bioinformatics},
  volume={12},
  number={1},
  pages={132},
  year={2011},
  publisher={Springer}
}

@article{pena2012bipartite,
  title={Bipartite graphs as models of population structures in evolutionary multiplayer games},
  author={Pena, Jorge and Rochat, Yannick},
  journal = {PLOS ONE},
  year={2012},  
  month = {09},
  volume = {7},
  eid={e44514},
  pages = {1-13},
  number = {9},
  publisher = {Public Library of Science},
}

@inproceedings{le2005clustering,
  title={Clustering in p2p exchanges and consequences on performances},
  author={Le Blond, Stevens and Guillaume, Jean-Loup and Latapy, Matthieu},
  booktitle={International Workshop on Peer-to-Peer Systems},
  year={2005}
}

@inproceedings{gallagher2013clustering,
  title={Clustering coefficients in protein interaction hypernetworks},
  author={Gallagher, Suzanne Renick and Goldberg, Debra S},
  booktitle={BCB},
  year={2013}
}

@article{chun2024random,
  title={Random walk with restart on hypergraphs: fast computation and an application to anomaly detection},
  author={Chun, Jaewan and Lee, Geon and Shin, Kijung and Jung, Jinhong},
  journal={Data Mining and Knowledge Discovery},
  volume={38},
  number={3},
  pages={1222--1257},
  year={2024},
  publisher={Springer}
}

@inproceedings{limnios2021hcore,
  title={Hcore-init: Neural network initialization based on graph degeneracy},
  author={Limnios, Stratis and Dasoulas, George and Thilikos, Dimitrios M and Vazirgiannis, Michalis},
  booktitle={ICPR},
  year={2021}
}

@article{arafat2023neighborhood,
  title={Neighborhood-based Hypergraph Core Decomposition},
  author={Arafat, Naheed Anjum and Khan, Arijit and Rai, Arpit Kumar and Ghosh, Bishwamittra},
  journal={Proceedings of the VLDB Endowment},
  volume={16},
  number={9},
  pages={2061--2074},
  year={2023},
  publisher={VLDB Endowment}
}

@inproceedings{luo2024hierarchical,
  title={Hierarchical structure construction on hypergraphs},
  author={Luo, Qi and Zhang, Wenjie and Yang, Zhengyi and Wen, Dong and Wang, Xiaoyang and Yu, Dongxiao and Lin, Xuemin},
  booktitle={CIKM},
  year={2024}
}

@inproceedings{choe2023classification,
  title={Classification of edge-dependent labels of nodes in hypergraphs},
  author={Choe, Minyoung and Kim, Sunwoo and Yoo, Jaemin and Shin, Kijung},
  booktitle={KDD},
  year={2023}
}

@article{leng2013m,
  title={An $O(m)$ algorithm for cores decomposition of undirected hypergraph},
  author={Leng, Ming and SUN, Ling-yu and Bian, Ji-nian and MA, Yu-chun},
  journal={Journal of Chinese Computer Systems},
  volume={34},
  number={11},
  pages={2568--2573},
  year={2013}
}

@inproceedings{kim2023exploring,
  title={Exploring cohesive subgraphs in hypergraphs: The (k, g)-core approach},
  author={Kim, Dahee and Kim, Junghoon and Lim, Sungsu and Jeong, Hyun Ji},
  booktitle={CIKM},
  year={2023}
}

@article{kim2025beyond,
  title={Beyond trivial edges: A fractional approach to cohesive subgraph detection in hypergraphs},
  author={Kim, Hyewon and Shin, Woocheol and Kim, Dahee and Kim, Junghoon and Lim, Sungsu and Jeong, Hyun Ji},
  journal={Knowledge-Based Systems},
  pages={113472},
  year={2025},
  publisher={Elsevier}
}

@inproceedings{wang2022efficient,
  title={Efficient truss computation for large hypergraphs},
  author={Wang, Xinzhou and Chen, Yinjia and Zhang, Zhiwei and Qiao, PengPeng and Wang, Guoren},
  booktitle={WISE},
  year={2022}
}

@inproceedings{takai2020hypergraph,
  title={Hypergraph clustering based on pagerank},
  author={Takai, Yuuki and Miyauchi, Atsushi and Ikeda, Masahiro and Yoshida, Yuichi},
  booktitle={KDD},
  year={2020}
}

@inproceedings{zhou2006learning,
  title={Learning with hypergraphs: Clustering, classification, and embedding},
  author={Zhou, Dengyong and Huang, Jiayuan and Sch{\"o}lkopf, Bernhard},
  booktitle={NeurIPS},
  year={2006}
}

@article{estrada2005subgraph,
  title={Subgraph centrality in complex networks},
  author={Estrada, Ernesto and Rodriguez-Velazquez, Juan A},
  journal={Physical Review E—Statistical, Nonlinear, and Soft Matter Physics},
  volume={71},
  number={5},
  pages={056103},
  year={2005},
  publisher={APS}
}

@article{chodrow2020annotated,
  title={Annotated hypergraphs: models and applications},
  author={Chodrow, Philip and Mellor, Andrew},
  journal={Applied Network Science},
  volume={5},
  number={1},
  pages={9},
  year={2020},
  publisher={Springer}
}

@article{ni2025structural,
title = {Structural-aware key node identification in hypergraphs via representation learning and fine-tuning},
journal = {Engineering Applications of Artificial Intelligence},
volume = {169},
pages = {114108},
year = {2026},
issn = {0952-1976},
author = {Xiaonan Ni and Guangyuan Mei and Su-Su Zhang and Yang Chen and Xin Xu and Chuang Liu and Xiu-Xiu Zhan}
}

@article{puzis2013betweenness,
  title={Betweenness computation in the single graph representation of hypergraphs},
  author={Puzis, Rami and Purohit, Manish and Subrahmanian, VS},
  journal={Social Networks},
  volume={35},
  number={4},
  pages={561--572},
  year={2013},
  publisher={Elsevier}
}

@inproceedings{bader2007approximating,
  title={Approximating betweenness centrality},
  author={Bader, David A and Kintali, Shiva and Madduri, Kamesh and Mihail, Milena},
  booktitle={WAW},
  year={2007}
}

@inproceedings{baglioni2012fast,
  title={Fast exact computation of betweenness centrality in social networks},
  author={Baglioni, Miriam and Geraci, Filippo and Pellegrini, Marco and Lastres, Ernesto},
  booktitle={ASONAM},
  year={2012}
}

@article{cohen2008trusses,
  title={Trusses: Cohesive subgraphs for social network analysis},
  author={Cohen, Jonathan},
  journal={National Security Agency Technical Report},
  volume={16},
  number={3.1},
  pages={1--29},
  year={2008},
  publisher={Citeseer}
}

@article{kostochka2013hypergraph,
  title={Hypergraph Ramsey numbers: triangles versus cliques},
  author={Kostochka, Alexandr and Mubayi, Dhruv and Verstraete, Jacques},
  journal={Journal of Combinatorial Theory, Series A},
  volume={120},
  number={7},
  pages={1491--1507},
  year={2013},
  publisher={Elsevier}
}

@article{li2022chromatic,
  title={The chromatic number of triangle-free hypergraphs},
  author={Li, Lina and Postle, Luke},
  journal={arXiv preprint arXiv:2202.02839},
  year={2022}
}

@article{lotito2022higher,
  title={Higher-order motif analysis in hypergraphs},
  author={Lotito, Quintino Francesco and Musciotto, Federico and Montresor, Alberto and Battiston, Federico},
  journal={Communications Physics},
  volume={5},
  number={1},
  pages={79},
  year={2022},
  publisher={Nature Publishing Group UK London}
}

@article{stehle2011high,
  title={High-resolution measurements of face-to-face contact patterns in a primary school},
  author={Stehl{\'e}, Juliette and Voirin, Nicolas and Barrat, Alain and Cattuto, Ciro and Isella, Lorenzo and Pinton, Jean-Fran{\c{c}}ois and Quaggiotto, Marco and Van den Broeck, Wouter and R{\'e}gis, Corinne and Lina, Bruno and others},
  journal={PloS One},
  volume={6},
  number={8},
  pages={e23176},
  year={2011},
  publisher={Public Library of Science San Francisco, USA}
}

@article{webber2010similarity,
  title={A similarity measure for indefinite rankings},
  author={Webber, William and Moffat, Alistair and Zobel, Justin},
  journal={ACM Transactions on Information Systems},
  volume={28},
  number={4},
  pages={1--38},
  year={2010},
  publisher={ACM New York, NY, USA}
}

@article{mastrandrea2015contact,
  title={Contact patterns in a high school: a comparison between data collected using wearable sensors, contact diaries and friendship surveys},
  author={Mastrandrea, Rossana and Fournet, Julie and Barrat, Alain},
  journal={PloS One},
  volume={10},
  number={9},
  pages={e0136497},
  year={2015},
  publisher={Public Library of Science San Francisco, CA USA}
}

@article{fowler2006legislative,
  title={Legislative cosponsorship networks in the US House and Senate},
  author={Fowler, James H},
  journal={Social Networks},
  volume={28},
  number={4},
  pages={454--465},
  year={2006},
  publisher={Elsevier}
}

@article{juul2024hypergraph,
  title={Hypergraph patterns and collaboration structure},
  author={Juul, Jonas L and Benson, Austin R and Kleinberg, Jon},
  journal={Frontiers in Physics},
  volume={11},
  pages={1301994},
  year={2024},
  publisher={Frontiers Media SA}
}

@article{fowler2006connecting,
  title={Connecting the congress: A study of cosponsorship networks},
  author={Fowler, James H},
  journal={Political Analysis},
  volume={14},
  number={4},
  pages={456--487},
  year={2006},
  publisher={Cambridge University Press}
}

@article{laita2011graph,
  title={Graph-theoretic connectivity measures: what do they tell us about connectivity?},
  author={Laita, A and Kotiaho, JS and M{\"o}nkk{\"o}nen, M},
  journal={Landscape Ecology},
  volume={26},
  number={7},
  pages={951--967},
  year={2011},
  publisher={Springer}
}

@article{freitas2022graph,
  title={Graph vulnerability and robustness: A survey},
  author={Freitas, Scott and Yang, Diyi and Kumar, Srijan and Tong, Hanghang and Chau, Duen Horng},
  journal={IEEE Transactions on Knowledge and Data Engineering},
  volume={35},
  number={6},
  pages={5915--5934},
  year={2023},
  publisher={IEEE}
}

@article{das2018study,
  title={Study on centrality measures in social networks: a survey},
  author={Das, Kousik and Samanta, Sovan and Pal, Madhumangal},
  journal={Social Network Analysis and Mining},
  volume={8},
  number={1},
  pages={13},
  year={2018},
  publisher={Springer}
}

@article{wan2021survey,
  title={A survey on centrality metrics and their network resilience analysis},
  author={Wan, Zelin and Mahajan, Yash and Kang, Beom Woo and Moore, Terrence J and Cho, Jin-Hee},
  journal={IEEE Access},
  volume={9},
  pages={104773--104819},
  year={2021},
  publisher={IEEE}
}

@article{grando2018machine,
  title={Machine learning in network centrality measures: Tutorial and outlook},
  author={Grando, Felipe and Granville, Lisandro Z and Lamb, Luis C},
  journal={ACM Computing Surveys},
  volume={51},
  number={5},
  pages={1--32},
  year={2018},
  publisher={ACM New York, NY, USA}
}

@article{antelmi2023survey,
  title={A survey on hypergraph representation learning},
  author={Antelmi, Alessia and Cordasco, Gennaro and Polato, Mirko and Scarano, Vittorio and Spagnuolo, Carmine and Yang, Dingqi},
  journal={ACM Computing Surveys},
  volume={56},
  number={1},
  pages={1--38},
  year={2023},
  publisher={ACM New York, NY}
}

@inproceedings{kim2024survey,
  title={A survey on hypergraph neural networks: an in-depth and step-by-step guide},
  author={Kim, Sunwoo and Lee, Soo Yong and Gao, Yue and Antelmi, Alessia and Polato, Mirko and Shin, Kijung},
  booktitle={KDD},
  year={2024}
}

@article{lee2025survey,
  title={A survey on hypergraph mining: Patterns, tools, and generators},
  author={Lee, Geon and Bu, Fanchen and Eliassi-Rad, Tina and Shin, Kijung},
  journal={ACM Computing Surveys},
  volume={57},
  number={8},
  pages={1--36},
  year={2025},
  publisher={ACM New York, NY}
}

@article{gao2020hypergraph,
  title={Hypergraph learning: Methods and practices},
  author={Gao, Yue and Zhang, Zizhao and Lin, Haojie and Zhao, Xibin and Du, Shaoyi and Zou, Changqing},
  journal={IEEE Transactions on Pattern Analysis and Machine Intelligence},
  volume={44},
  number={5},
  pages={2548--2566},
  year={2022},
  publisher={IEEE}
}

@article{bian2019identifying,
  title={Identifying top-k nodes in social networks: a survey},
  author={Bian, Ranran and Koh, Yun Sing and Dobbie, Gillian and Divoli, Anna},
  journal={ACM Computing Surveys},
  volume={52},
  number={1},
  pages={1--33},
  year={2019},
  publisher={ACM New York, NY, USA}
}

@article{chen2025critical,
AUTHOR = {Duxin Chen and Jiawen Chen and Xiaoyu Zhang and Qinghan Jia and Xiaolu Liu and Ye Sun and Linyuan Lü and Wenwu Yu},
TITLE = {Critical nodes identification in complex networks: a survey},
JOURNAL = {Complex Engineering Systems},
VOLUME = {5},
YEAR = {2025},
NUMBER = {3},
eid = {11},
}

@article{hafiene2020influential,
  title={Influential nodes detection in dynamic social networks: A survey},
  author={Hafiene, Nesrine and Karoui, Wafa and Romdhane, Lotfi Ben},
  journal={Expert Systems with Applications},
  volume={159},
  pages={113642},
  year={2020},
  publisher={Elsevier}
}

@article{sarkar2016survey,
  title={Survey of influential nodes identification in online social networks},
  author={Sarkar, Dhrubasish and Kole, Dipak K and Jana, Premananda},
  journal={International Journal of Virtual Communities and Social Networking},
  volume={8},
  number={4},
  pages={57--69},
  year={2016},
  publisher={IGI Global Scientific Publishing}
}

@article{ait2023influential,
  title={Influential nodes identification in complex networks: a comprehensive literature review},
  author={Ait Rai, Khaoula and Machkour, Mustapha and Antari, Jilali},
  journal={Beni-Suef University Journal of Basic and Applied Sciences},
  volume={12},
  number={1},
  pages={18},
  year={2023},
  publisher={Springer}
}

@inproceedings{zhao2023survey,
  title={A survey on identification of critical nodes in ad hoc network},
  author={Zhao, Jinyang and Fang, Zhen and Ge, Jiangtao and Huang, Jun and Niu, Zhao},
  booktitle={ITNEC},
  year={2023}
}

@article{wang2022mini,
  title={A mini review of node centrality metrics in biological networks},
  author={Wang, Mengyuan and Wang, Haiying and Zheng, Huiru},
  journal={International Journal of Network Dynamics and Intelligence},
  volume={1},
  number={1},
  pages={99--110},
  year={2022}
}

@inproceedings{fischer2021towards,
  title={Towards a survey on static and dynamic hypergraph visualizations},
  author={Fischer, Maximilian T and Frings, Alexander and Keim, Daniel A and Seebacher, Daniel},
  booktitle={VIS},
  year={2021}
}

@article{ausiello2017directed,
  title={Directed hypergraphs: introduction and fundamental algorithms—a survey},
  author={Ausiello, Giorgio and Laura, Luigi},
  journal={Theoretical Computer Science},
  volume={658},
  pages={293--306},
  year={2017},
  publisher={Elsevier}
}

@inproceedings{wang2022survey,
  title={Survey of hypergraph neural networks and its application to action recognition},
  author={Wang, Cheng and Ma, Nan and Wu, Zhixuan and Zhang, Jin and Yao, Yongqiang},
  booktitle={CICAI},
  year={2022}
}

@article{yang2025recent,
  title={Recent advances in hypergraph neural networks},
  author={Yang, Mu-Rong and Xu, Xin-Jian},
  journal={Journal of the Operations Research Society of China},
  pages={1--37},
  year={2025},
  publisher={Springer}
}

@article{ccatalyurek2023more,
  title={More recent advances in (hyper) graph partitioning},
  author={{\c{C}}ataly{\"u}rek, {\"U}mit and Devine, Karen and Faraj, Marcelo and Gottesb{\"u}ren, Lars and Heuer, Tobias and Meyerhenke, Henning and Sanders, Peter and Schlag, Sebastian and Schulz, Christian and Seemaier, Daniel and others},
  journal={ACM Computing Surveys},
  volume={55},
  number={12},
  pages={1--38},
  year={2023},
  publisher={ACM New York, NY}
}

@article{chen2023survey,
  title={A survey on hyperlink prediction},
  author={Chen, Can and Liu, Yang-Yu},
  journal={IEEE Transactions on Neural Networks and Learning Systems},
  volume={35},
  number={11},
  pages={15034--15050},
  year={2024},
  publisher={IEEE}
}

@inproceedings{liu2022survey,
  title={A survey of recommender systems based on hypergraph neural networks},
  author={Liu, Canwei and He, Tingqin and Zhu, Hangyu and Li, Yanlu and Xie, Songyou and Hosam, Osama},
  booktitle={SmartCom},
  year={2022}
}

@article{qin2025truss,
  title={Truss decomposition in hypergraphs},
  author={Qin, Hongchao and Zeng, Guang and Li, Rong-Hua and Lin, Longlong and Yuan, Ye and Wang, Guoren},
  journal={Proceedings of the VLDB Endowment},
  volume={18},
  number={7},
  pages={2185--2197},
  year={2025},
  publisher={VLDB Endowment}
}

@inbook{shapley1951notes,
  title={A value for n-person games},
  booktitle = {Contributions to the Theory of Games, Volume II},
  author={Shapley, Lloyd S},
  year={1953},  
publisher = {Princeton University Press},
pages = {307--317},
}

@article{chang2025structure,
  title={Structure-and-embedding-based centrality on network fragility in hypergraphs},
  author={Chang, Lanlan and Qiu, Tian and Chen, Guang},
  journal={Chaos: An Interdisciplinary Journal of Nonlinear Science},
  volume={35},
  number={3},
  year={2025},
  publisher={AIP Publishing}
}

@article{landry2023xgi,
  title={XGI: A Python package for higher-order interaction networks},
  author={Landry, Nicholas W and Lucas, Maxime and Iacopini, Iacopo and Petri, Giovanni and Schwarze, Alice and Patania, Alice and Torres, Leo},
  journal={Journal of Open Source Software},
  volume={8},
  number={85},
  pages={5162},
  year={2023}
}

@inproceedings{Bu2025Anchor,
  title={Identifying Group Anchors in Real-World Group Interactions Under Label Scarcity}, 
  author={Bu, Fanchen and Lee, Geon and Choe, Minyoung and Shin, Kijung},
  booktitle={ICDM},
  year={2025}
}

@article{nortier2025higher,
  title = {Higher-order shortest paths in hypergraphs},
  author = {Nortier, Bern\'e L. and Dobson, Simon and Battiston, Federico},
  journal = {Phys. Rev. E},
  volume = {112},
  issue = {5},
  pages = {054302},
  numpages = {9},
  year = {2025},
  month = {Nov},
  publisher = {American Physical Society}
}

@article{vasilyeva2023distances,
  title={Distances in higher-order networks and the metric structure of hypergraphs},
  author={Vasilyeva, Ekaterina and Romance, Miguel and Samoylenko, Ivan and Kovalenko, Kirill and Musatov, Daniil and Raigorodskii, Andrey Mihailovich and Boccaletti, Stefano},
  journal={Entropy},
  volume={25},
  number={6},
  pages={923},
  year={2023},
  publisher={MDPI}
}

@article{carletti2020random,
  title={Random walks on hypergraphs},
  author={Carletti, Timoteo and Battiston, Federico and Cencetti, Giulia and Fanelli, Duccio},
  journal={Physical Review E},
  volume={101},
  number={2},
  pages={022308},
  year={2020},
  publisher={APS}
}

@article{carletti2021random,
  title={Random walks and community detection in hypergraphs},
  author={Carletti, Timoteo and Fanelli, Duccio and Lambiotte, Renaud},
  journal={Journal of Physics: Complexity},
  volume={2},
  number={1},
  pages={015011},
  year={2021},
  publisher={IOP Publishing}
}

@inproceedings{tian2025representing,
  title={Representing Higher-Order Networks with Spectral Moments},
  author={Tian, Hao and Jin, Shengmin and Zafarani, Reza},
  booktitle={PAKDD},
  year={2025}
}

@article{battiston2020networks,
  title={Networks beyond pairwise interactions: Structure and dynamics},
  author={Battiston, Federico and Cencetti, Giulia and Iacopini, Iacopo and Latora, Vito and Lucas, Maxime and Patania, Alice and Young, Jean-Gabriel and Petri, Giovanni},
  journal={Physics Reports},
  volume={874},
  pages={1--92},
  year={2020},
  publisher={Elsevier}
}

@article{lu2013high,
  title={High-order random walks and generalized laplacians on hypergraphs},
  author={Lu, Linyuan and Peng, Xing},
  journal={Internet Mathematics},
  volume={9},
  number={1},
  pages={3--32},
  year={2013},
  publisher={Taylor \& Francis}
}

@article{larock2023encapsulation,
  title={Encapsulation structure and dynamics in hypergraphs},
  author={LaRock, Timothy and Lambiotte, Renaud},
  journal={Journal of Physics: Complexity},
  volume={4},
  number={4},
  pages={045007},
  year={2023},
  publisher={IOP Publishing}
}

@inproceedings{lee2021hyperedges,
  title={How do hyperedges overlap in real-world hypergraphs?-patterns, measures, and generators},
  author={Lee, Geon and Choe, Minyoung and Shin, Kijung},
  booktitle={WWW},
  year={2021}
}

@article{lotito2024hyperlink,
  title={Hyperlink communities in higher-order networks},
  author={Lotito, Quintino Francesco and Musciotto, Federico and Montresor, Alberto and Battiston, Federico},
  journal={Journal of Complex Networks},
  volume={12},
  number={2},
  pages={cnae013},
  year={2024},
  publisher={Oxford University Press}
}

@article{zhou2022topological,
  title={Topological simplifications of hypergraphs},
  author={Zhou, Youjia and Rathore, Archit and Purvine, Emilie and Wang, Bei},
  journal={IEEE Transactions on Visualization and Computer Graphics},
  volume={29},
  number={7},
  pages={3209--3225},
  year={2023},
  publisher={IEEE}
}

@article{esposito2022venture,
  title={Venture capital investments through the lens of network and functional data analysis},
  author={Esposito, Christian and Gortan, Marco and Testa, Lorenzo and Chiaromonte, Francesca and Fagiolo, Giorgio and Mina, Andrea and Rossetti, Giulio},
  journal={Applied Network Science},
  volume={7},
  number={1},
  pages={42},
  year={2022},
  publisher={Springer}
}

@article{watts1998collective,
  title={Collective dynamics of `small-world' networks},
  author={Watts, Duncan J and Strogatz, Steven H},
  journal={Nature},
  volume={393},
  number={6684},
  pages={440--442},
  year={1998},
  publisher={Nature Publishing Group}
}

@article{zhou2021identification,
  title={Identification methods of vital nodes based on k-shell in hypernetworks},
  author={Zhou, Lina and Li, Faxu and Gong, Yunchao and Hu, Feng},
  journal={Complex Systems and Complexity Science},
  volume={18},
  number={03},
  pages={15--22},
  year={2021},  
}

@article{wu2023multi,
  title={A multi-attribute decision-making method based on entropy to identify important nodes in hypernetworks},
  author={Wu, Yinghan and Li, Mingda and Hu, Feng},
  journal={Complex Systems and Complexity Science},
  volume={20},
  number={04},
  pages={40--46},
  year={2023},
  publisher = {Complex Systems and Complexity Science},
}

@inproceedings{guo2025method,
  title={Method for identifying vital nodes in the scientific-citation double-layer hypernetwork based on the h-index},
  author={Guo, Lei and Liu, Wei and Zhang, Xin and Hu, Feng},
  booktitle={ICCAIS},
  year={2024}
}

@article{amburg2021planted,
  title={Planted hitting set recovery in hypergraphs},
  author={Amburg, Ilya and Kleinberg, Jon and Benson, Austin R},
  journal={Journal of Physics: Complexity},
  volume={2},
  number={3},
  pages={035004},
  year={2021},
  publisher={IOP Publishing}
}

@article{tudisco2023core,
  title={Core-periphery detection in hypergraphs},
  author={Tudisco, Francesco and Higham, Desmond J},
  journal={SIAM Journal on Mathematics of Data Science},
  volume={5},
  number={1},
  pages={1--21},
  year={2023},
  publisher={SIAM}
}

@inproceedings{papachristou2022core,
  title={Core-periphery models for hypergraphs},
  author={Papachristou, Marios and Kleinberg, Jon},
  booktitle={KDD},
  year={2022}
}

@inproceedings{jeh2003scaling,
  title={Scaling personalized web search},
  author={Jeh, Glen and Widom, Jennifer},
  booktitle={WWW},
  year={2003}
}

@article{chai2024hypergraph,
  title={Hypergraph modeling and hypergraph multi-view attention neural network for link prediction},
  author={Chai, Lang and Tu, Lilan and Wang, Xianjia and Su, Qingqing},
  journal={Pattern Recognition},
  volume={149},
  pages={110292},
  year={2024},
  publisher={Elsevier}
}

@inproceedings{saifuddin2023hygnn,
  title={Hygnn: Drug-drug interaction prediction via hypergraph neural network},
  author={Saifuddin, Khaled Mohammed and Bumgardner, Briana and Tanvir, Farhan and Akbas, Esra},
  booktitle={ICDE},
  year={2023}
}

@article{bai2021hypergraph,
  title={Hypergraph convolution and hypergraph attention},
  author={Bai, Song and Zhang, Feihu and Torr, Philip HS},
  journal={Pattern Recognition},
  volume={110},
  pages={107637},
  year={2021},
  publisher={Elsevier}
}

@article{li2023hypergraph,
  title={Hypergraph transformer neural networks},
  author={Li, Mengran and Zhang, Yong and Li, Xiaoyong and Zhang, Yuchen and Yin, Baocai},
  journal={ACM Transactions on Knowledge Discovery from Data},
  volume={17},
  number={5},
  pages={1--22},
  year={2023},
  publisher={ACM New York, NY}
}

@article{lu2016vital,
  title={Vital nodes identification in complex networks},
  author={L{\"u}, Linyuan and Chen, Duanbing and Ren, Xiao-Long and Zhang, Qian-Ming and Zhang, Yi-Cheng and Zhou, Tao},
  journal={Physics Reports},
  volume={650},
  pages={1--63},
  year={2016},
  publisher={Elsevier}
}

@article{liu2016evaluating,
  title={Evaluating the importance of nodes in complex networks},
  author={Liu, Jun and Xiong, Qingyu and Shi, Weiren and Shi, Xin and Wang, Kai},
  journal={Physica A: Statistical Mechanics and its Applications},
  volume={452},
  pages={209--219},
  year={2016},
  publisher={Elsevier}
}

@article{leskovec2007graph,
  title={Graph evolution: Densification and shrinking diameters},
  author={Leskovec, Jure and Kleinberg, Jon and Faloutsos, Christos},
  journal={ACM Transactions on Knowledge Discovery from Data},
  volume={1},
  number={1},
  articleno={2},
  numpages = {41},
  year={2007},
  publisher={ACM New York, NY, USA}
}

@article{benson2016higher,
  title={Higher-order organization of complex networks},
  author={Benson, Austin R and Gleich, David F and Leskovec, Jure},
  journal={Science},
  volume={353},
  number={6295},
  pages={163--166},
  year={2016},
  publisher={American Association for the Advancement of Science}
}

@book{bianconi2021higher,
  title={Higher-order networks},
  author={Bianconi, Ginestra},
  year={2021},
  publisher={Cambridge University Press}
}

@article{bick2023higher,
  title={What are higher-order networks?},
  author={Bick, Christian and Gross, Elizabeth and Harrington, Heather A and Schaub, Michael T},
  journal={SIAM Review},
  volume={65},
  number={3},
  pages={686--731},
  year={2023},
  publisher={SIAM}
}

@article{battiston2021physics,
  title={The physics of higher-order interactions in complex systems},
  author={Battiston, Federico and Amico, Enrico and Barrat, Alain and Bianconi, Ginestra and Ferraz de Arruda, Guilherme and Franceschiello, Benedetta and Iacopini, Iacopo and K{\'e}fi, Sonia and Latora, Vito and Moreno, Yamir and others},
  journal={Nature Physics},
  volume={17},
  number={10},
  pages={1093--1098},
  year={2021},
  publisher={Nature Publishing Group UK London}
}

@book{shen2011spectral,
  title={Spectral methods: algorithms, analysis and applications},
  author={Shen, Jie and Tang, Tao and Wang, Li-Lian},
  volume={41},
  year={2011},
  publisher={Springer Science \& Business Media}
}

@article{jeong2001lethality,
  title={Lethality and centrality in protein networks},
  author={Jeong, Hawoong and Mason, Sean P and Barab{\'a}si, A-L and Oltvai, Zoltan N},
  journal={Nature},
  volume={411},
  number={6833},
  pages={41--42},
  year={2001},
  publisher={Nature Publishing Group UK London}
}

@inproceedings{myers2012information,
  title={Information diffusion and external influence in networks},
  author={Myers, Seth A and Zhu, Chenguang and Leskovec, Jure},
  booktitle={KDD},
  year={2012}
}

@book{slikker2012social,
  title={Social and economic networks in cooperative game theory},
  author={Slikker, Marco and Van den Nouweland, Anne},
  volume={27},
  year={2012},
  publisher={Springer Science \& Business Media}
}

@article{mancastroppa2023hyper,
  title={Hyper-cores promote localization and efficient seeding in higher-order processes},
  author={Mancastroppa, Marco and Iacopini, Iacopo and Petri, Giovanni and Barrat, Alain},
  journal={Nature Communications},
  volume={14},
  number={1},
  pages={6223},
  year={2023},
  publisher={Nature Publishing Group UK London}
}

@article{xiao2024information,
  title={Information propagation in hypergraph-based social networks},
  author={Xiao, Hai-Bing and Hu, Feng and Li, Peng-Yue and Song, Yu-Rong and Zhang, Zi-Ke},
  journal={Entropy},
  volume={26},
  number={11},
  pages={957},
  year={2024},
  publisher={MDPI}
}

@article{feng2021hypergraph,
  title={Hypergraph models of biological networks to identify genes critical to pathogenic viral response},
  author={Feng, Song and Heath, Emily and Jefferson, Brett and Joslyn, Cliff and Kvinge, Henry and Mitchell, Hugh D and Praggastis, Brenda and Eisfeld, Amie J and Sims, Amy C and Thackray, Larissa B and others},
  journal={BMC Bioinformatics},
  volume={22},
  number={1},
  pages={287},
  year={2021},
  publisher={Springer}
}

@article{murgas2022hypergraph,
  title={Hypergraph geometry reflects higher-order dynamics in protein interaction networks},
  author={Murgas, Kevin A and Saucan, Emil and Sandhu, Romeil},
  journal={Scientific Reports},
  volume={12},
  number={1},
  pages={20879},
  year={2022},
  publisher={Nature Publishing Group UK London}
}

@article{barton2023hypergraphs,
  title={Hypergraphs and centrality measures identifying key features in gene expression data},
  author={Barton, Samuel and Broad, Zoe and Ortiz-Barrientos, Daniel and Donovan, Diane and Lefevre, James},
  journal={Mathematical Biosciences},
  volume={366},
  pages={109089},
  year={2023},
  publisher={Elsevier}
}

@article{lawson2024application,
  title={An application of node and edge nonlinear hypergraph centrality to a protein complex hypernetwork},
  author={Lawson, Sarah and Donovan, Diane and Lefevre, James},
  journal={PloS One},
  volume={19},
  number={10},
  pages={e0311433},
  year={2024},
  publisher={Public Library of Science San Francisco, CA USA}
}

@article{prakash2017finding,
  title={Finding the most reliable strategy on stochastic and time-dependent transportation networks: A hypergraph based formulation},
  author={Prakash, A Arun and Srinivasan, Karthik K},
  journal={Networks and Spatial Economics},
  volume={17},
  number={3},
  pages={809--840},
  year={2017},
  publisher={Springer}
}

@article{harrod2011modeling,
  title={Modeling network transition constraints with hypergraphs},
  author={Harrod, Steven},
  journal={Transportation Science},
  volume={45},
  number={1},
  pages={81--97},
  year={2011},
  publisher={INFORMS}
}

@article{yin2026hypergraph,
  title={Hypergraph-based high-speed rail hypernetwork analysis and node importance evaluation using operational data: A case study of china},
  author={Yin, Mengmeng and Tang, Kun and Xu, Tian and Ding, Jinhong and Guo, Tangyi},
  journal={Journal of Transportation Engineering, Part A: Systems},
  volume={152},
  number={1},
  pages={04025113},
  year={2026},
  publisher={American Society of Civil Engineers}
}

@inproceedings{yin2025resilience,
  title={Resilience Assessment of Multimodal Transportation Networks: A Hypergraph-Based Modeling Framework},
  author={Yin, Mengmeng and Tang, Kun and Ding, Jinghong and Guo, Tangyi},
  booktitle={FASTA},
  year={2025}
}

@article{tocchi2022hypergraph,
  title={Hypergraph-based centrality metrics for maritime container service networks: A worldwide application},
  author={Tocchi, Daniela and Sys, Christa and Papola, Andrea and Tinessa, Fiore and Simonelli, Fulvio and Marzano, Vittorio},
  journal={Journal of Transport Geography},
  volume={98},
  pages={103225},
  year={2022},
  publisher={Elsevier}
}

@article{poshtiban2023identification,
  title={Identification of Influential nodes in social networks based on profile analysis},
  author={Poshtiban, Zeinab and Ghanbari, Elham and Jahangir, Mohammadreza},
  journal={Journal of AI and Data Mining},
  volume={11},
  number={4},
  pages={535--545},
  year={2023},
  publisher={Journal of AI and Data Mining}
}

@article{chang2024hypergraph,
  title={Hypergraph: A unified and uniform definition with application to chemical hypergraph and more},
  author={Chang, Daniel T},
  journal={arXiv preprint arXiv:2405.12235},
  year={2024}
}

@article{wang2023gmhann,
  title={{GMHANN}: A novel traffic flow prediction method for transportation management based on spatial-temporal graph modeling},
  author={Wang, Qing and Liu, Weiping and Wang, Xiumei and Chen, Xinghong and Chen, Guannan and Wu, Qingxiang},
  journal={IEEE Transactions on Intelligent Transportation Systems},
  volume={25},
  number={1},
  pages={386--401},
  year={2024},
  publisher={IEEE}
}

@article{amato2019hypergraph,
  title={A hypergraph data model for expert-finding in multimedia social networks},
  author={Amato, Flora and Cozzolino, Giovanni and Sperl{\`\i}, Giancarlo},
  journal={Information},
  volume={10},
  number={6},
  pages={183},
  year={2019},
  publisher={MDPI}
}

@article{rafferty2021ranking,
  title={Ranking sets of morbidities using hypergraph centrality},
  author={Rafferty, James and Watkins, Alan and Lyons, Jane and Lyons, Ronan A and Akbari, Ashley and Peek, Niels and Jalali-Najafabadi, Farideh and Dhafari, Thamer Ba and Pate, Alexander and Martin, Glen P and others},
  journal={Journal of Biomedical Informatics},
  volume={122},
  pages={103916},
  year={2021},
  publisher={Elsevier}
}

@article{singh2025ranking,
  title={A Ranking Framework for Network Resource Allocation and Scheduling via Hypergraphs},
  author={Singh, Rajpreet and Bo{\v{s}}kov, Novak and Gudal, Aditya and Khan, Manzoor A},
  journal={arXiv preprint arXiv:2506.01571},
  year={2025}
}

@article{luo2022directed,
  title={Directed hypergraph attention network for traffic forecasting},
  author={Luo, Xiaoyi and Peng, Jiaheng and Liang, Jun},
  journal={IET Intelligent Transport Systems},
  volume={16},
  number={1},
  pages={85--98},
  year={2022},
  publisher={Wiley Online Library}
}

@inproceedings{liu2025aspect,
  title={An Aspect Performance-aware Hypergraph Neural Network for Review-based Recommendation},
  author={Liu, Junrui and Li, Tong and Wu, Di and Tang, Zifang and Fang, Yuan and Yang, Zhen},
  booktitle={WSDM},
  year={2025}
}

@article{bloch2023centrality,
  title={Centrality measures in networks},
  author={Bloch, Francis and Jackson, Matthew O and Tebaldi, Pietro},
  journal={Social Choice and Welfare},
  volume={61},
  number={2},
  pages={413--453},
  year={2023},
  publisher={Springer}
}

@article{fischer2020visual,
  title={Visual analytics for temporal hypergraph model exploration},
  author={Fischer, Maximilian T and Arya, Devanshu and Streeb, Dirk and Seebacher, Daniel and Keim, Daniel A and Worring, Marcel},
  journal={IEEE Transactions on Visualization and Computer Graphics},
  volume={27},
  number={2},
  pages={550--560},
  year={2021},
  publisher={IEEE}
}

@article{lee2023temporal,
  title={Temporal hypergraph motifs},
  author={Lee, Geon and Shin, Kijung},
  journal={Knowledge and Information Systems},
  volume={65},
  number={4},
  pages={1549--1586},
  year={2023},
  publisher={Springer}
}

@article{moon2023four,
author={Moon, Heechan and Kim, Hyunju and Kim, Sunwoo and Shin, Kijung},
journal={IEEE Transactions on Knowledge \& Data Engineering },
title={Four-Set Hypergraphlets for Characterization of Directed Hypergraphs },
year={2026},
volume={38},
number={06},
ISSN={1558-2191},
pages={3640-3654},
publisher={IEEE},
}

@article{qian2025modeling,
  title={Modeling and analysis of cascading failures in multilayer higher-order networks},
  author={Qian, Cheng and Zhao, Dandan and Zhong, Ming and Peng, Hao and Wang, Wei},
  journal={Reliability Engineering \& System Safety},
  volume={253},
  pages={110497},
  year={2025},
  publisher={Elsevier}
}

@article{preti2024hyper,
  title={Hyper-distance oracles in hypergraphs},
  author={Preti, Giulia and De Francisci Morales, Gianmarco and Bonchi, Francesco},
  journal={The VLDB Journal},
  volume={33},
  number={5},
  pages={1333--1356},
  year={2024},
  publisher={Springer}
}

@inproceedings{riondato2014fast,
  title={Fast approximation of betweenness centrality through sampling},
  author={Riondato, Matteo and Kornaropoulos, Evgenios M},
  booktitle={WSDM},
  year={2014}
}

@article{riondato2018abra,
  title={Abra: Approximating betweenness centrality in static and dynamic graphs with rademacher averages},
  author={Riondato, Matteo and Upfal, Eli},
  journal={ACM Transactions on Knowledge Discovery from Data},
  volume={12},
  number={5},
  pages={1--38},
  year={2018},
  publisher={ACM New York, NY, USA}
}

@inproceedings{katz2008all,
  title={All-pairs shortest-paths for large graphs on the GPU},
  author={Katz, Gary J and Kider, Joseph T},
  booktitle={GH},
  year={2008}
}

@inproceedings{djidjev2014efficient,
  title={Efficient multi-GPU computation of all-pairs shortest paths},
  author={Djidjev, Hristo and Thulasidasan, Sunil and Chapuis, Guillaume and Andonov, Rumen and Lavenier, Dominique},
  booktitle={IPDPS},
  year={2014}
}

@inproceedings{alghamdi2017benchmark,
  title={A benchmark for betweenness centrality approximation algorithms on large graphs},
  author={AlGhamdi, Ziyad and Jamour, Fuad and Skiadopoulos, Spiros and Kalnis, Panos},
  booktitle={SSDBM},
  year={2017}
}

@article{saqr2022curious,
  title={The curious case of centrality measures: A large-scale empirical investigation},
  author={Saqr, Mohammed and L{\'o}pez-Pernas, Sonsoles},
  journal={Journal of Learning Analytics},
  volume={9},
  number={1},
  pages={13--31},
  year={2022}
}

@article{iacobucci2017social,
  title={In social network analysis, which centrality index should I use? Theoretical differences and empirical similarities among top centralities},
  author={Iacobucci, Dawn and McBride, Rebecca and Popovich, Deidre and Rouziou, Maria},
  journal={Journal of Methods and Measurement in the Social Sciences},
  volume={8},
  number={2},
  pages={72--99},
  year={2017}
}

@inproceedings{kolda2006tophits,
  title={The TOPHITS model for higher-order web link analysis},
  author={Kolda, Tamara and Bader, Brett},
  booktitle={LACTS},
  year={2006}
}

@article{kolda2009tensor,
  title={Tensor decompositions and applications},
  author={Kolda, Tamara G and Bader, Brett W},
  journal={SIAM Review},
  volume={51},
  number={3},
  pages={455--500},
  year={2009},
  publisher={SIAM}
}

@article{stephan2024sparse,
  title={Sparse random hypergraphs: Non-backtracking spectra and community detection},
  author={Stephan, Ludovic and Zhu, Yizhe},
  journal={Information and Inference: A Journal of the IMA},
  volume={13},
  number={1},
  pages={iaae004},
  year={2024},
  publisher={Oxford University Press}
}

@article{chodrow2023nonbacktracking,
  title={Nonbacktracking spectral clustering of nonuniform hypergraphs},
  author={Chodrow, Philip and Eikmeier, Nicole and Haddock, Jamie},
  journal={SIAM Journal on Mathematics of Data Science},
  volume={5},
  number={2},
  pages={251--279},
  year={2023},
  publisher={SIAM}
}

@article{luppi2024quantifying,
  title={Quantifying synergy and redundancy between networks},
  author={Luppi, Andrea I and Olbrich, Eckehard and Finn, Conor and Su{\'a}rez, Laura E and Rosas, Fernando E and Mediano, Pedro AM and Jost, J{\"u}rgen},
  journal={Cell Reports Physical Science},
  volume={5},
  number={4},  
  pages = {101892},
  year={2024},
  publisher={Elsevier}
}

@article{gu2017functional,
  title={Functional hypergraph uncovers novel covariant structures over neurodevelopment},
  author={Gu, Shi and Yang, Muzhi and Medaglia, John D and Gur, Ruben C and Gur, Raquel E and Satterthwaite, Theodore D and Bassett, Danielle S},
  journal={Human Brain Mapping},
  volume={38},
  number={8},
  pages={3823--3835},
  year={2017},
  publisher={Wiley Online Library}
}

@article{santos2023emergence,
  title={Emergence of high-order functional hubs in the human brain},
  author = {Santos, Fernando A.N. and Tewarie, Prejaas K.B. and Baudot, Pierre and Luchicchi, Antonio and Barros de Souza, Danillo and Girier, Guillaume and Milan, Ana P. and Broeders, Tommy and Centeno, Eduarda G.Z. and Cofre, Rodrigo and Rosas, Fernando E and Carone, Davide and Kennedy, James and Stam, Cornelis J. and Hillebrand, Arjan and Desroches, Mathieu and Rodrigues, Serafim and Schoonheim, Menno and Douw, Linda and Quax, Rick},
  journal={bioRxiv},  
    year = {2023},
    doi = {10.1101/2023.02.10.528083},
    publisher = {Cold Spring Harbor Laboratory},
}

@article{zheng2024rich,
  title={Rich-club organization of whole-brain spatio-temporal multilayer functional connectivity networks},
  author={Zheng, Jianhui and Cheng, Yuhao and Wu, Xi and Li, Xiaojie and Fu, Ying and Yang, Zhipeng},
  journal={Frontiers in Neuroscience},
  volume={18},
  pages={1405734},
  year={2024},
  publisher={Frontiers Media SA}
}

@inproceedings{hu2017maintaining,
  title={Maintaining densest subsets efficiently in evolving hypergraphs},
  author={Hu, Shuguang and Wu, Xiaowei and Chan, TH Hubert},
  booktitle={CIKM},
  year={2017}
}

@article{brandes2001faster,
  title={A faster algorithm for betweenness centrality},
  author={Brandes, Ulrik},
  journal={Journal of mathematical sociology},
  volume={25},
  number={2},
  pages={163--177},
  year={2001},
  publisher={Taylor \& Francis}
}

%





\end{document}